UNIVERSITÀ DEGLI STUDI DI PADOVA

Sede Amministrativa: Università degli Studi di Padova

Dipartimento di Matematica

**SCUOLA DI DOTTORATO DI RICERCA IN SCIENZE MATEMATICHE**

**INDIRIZZO: MATEMATICA**

**CICLO XX**

# FROM QUANTUM METALANGUAGE TO THE LOGIC OF QUBITS

**Direttore della Scuola :** Ch.mo Prof. Paolo Dai Pra

**Coordinatore d'indirizzo:** Ch.mo Prof. Franco Cardin

**Supervisori** :Ch.mo Prof. Giovanni Sambin

Ch.mo Prof. Pieralberto Marchetti

**Dottorando :** Paola Zizzi

*To the memory of my parents*

# CONTENTS:

## Summary of the thesis

The logic *Lq*, introduced within this thesis, is a logic for quantum information.
The purpose was, in fact, to describe logically the qubit structure (that is, the intrinsic quantum superposition of a two-level quantum state) and the maximal quantum entanglement of two qubits.
The logic *Lq* is obtained via Sambin's reflection principle of Basic logic, by which the metalinguistic links among assertions reflect (by solving a definitional equation) into logical connectives among propositions.

However, while in Basic logic the metalanguage is classical, in our case it is quantum.
In the quantum metalanguage each atomic assertion carries along an assertion degree, a complex number, which is interpreted as a probability amplitude. It is just the presence of assertion degrees that allows the introduction of the connective "quantum superposition" in *Lq*. This connective is a generalization of the logical conjunction "and". It is labelled by complex numbers indicating the weight by which each proposition contributes to the compound proposition.

The truth-values (or truth-degrees) are the squared modules of the assertion degrees, and their range is the real interval [0,1]. Then, the logic *Lq* is many-valued. Differently from fuzzy logics, however, the truth-degrees are interpreted here as quantum-mechanical probabilities.

The logic *Lq* keeps the three main properties of Basic logic, namely symmetry, reflection and visibility. This choice has been dictated by the following considerations:

1) The no-cloning and no-erase theorems of quantum information do not allow the corresponding logic to have the structural rules of weakening and contraction, with which they disagree. This fact rules out, in the search of a logic for quantum information, every kind of structural logic.
2) Choosing Basic logic instead of Linear logic (the other main sub-structural logic) was due to the fact that without visibility the connective "quantum entanglement" cannot be introduced.

Furthermore, we looked for a logic of quantum information which was endowed with a deductive calculus, (in particular a sequent calculus).
The logic *Lq* appears, in so far, as the only one which can take into account all the above *desiderata*.

The interpretation of the assertions of the quantum metalanguage is given in terms of quantum states (the quantum metalanguage "is" the Hilbert space).
The interpretation of the propositions of *Lq* is given in terms of (non-hermitian) operators which are weak measurements.

Then, the interpretation of *Lq* is based on a generalization of the concepts already proposed by Birkhoff and von Neumann in "orthodox" quantum logic. The difference stands in the fact that the interpretation of *Lq* is not given in terms of projectors, but in terms of weak measurements, which do not give rise to an abrupt collapse of quantum wave functions. This allows a logical description of quantum superposition, because the latter is not destroyed. The possibility of interpreting propositions as weak measurements is due to the fact that we introduced a quantum metalanguage. In fact, in the interpretation of propositions, the complex factors multiplying the projection operators are nothing else than the assertion degrees.

Some results of this thesis are:
a) The adoption of a new kind of metalanguage, the quantum metalanguage, where the metalinguistic links are quantum correlations, and assertions have a complex assertion-degree.
b) The introduction, through the reflection principle, of new (quantum) connectives, like "quantum superposition", and "quantum entanglement".
c) The introduction of a new dual operation (which is a generalization of Sambin-Girard logical duality) to take into account the dual Hilbert space occurring in the interpretation.
d) A quantum cut rule, which is interpreted as a quantum projective measurement. As the cut is a meta-rule, it follows that a quantum machine cannot perform a self-measurement and destroy itself.
e) A new meta-rule, not equivalent to the cut, named "EPR rule" (to remind the Einstein-Podolsky-Rosen paradox). This rule allows to prove simultaneously two entangled theorems.
f) The formulation of the "qubit theorem", which is the logical description of the preparation of the optical qubit state.
g) The lattice of propositions of *Lq* is, in the case of two qubits, orthomodular and non-distributive. Then, *Lq* is a quantum logic.

It should be noticed that *Lq* is the first logic which is sub-structural, many-valued and quantum at the same time.

## Riassunto della tesi

La logica introdotta in questa tesi, detta *Lq*, è una logica dell' informazione quantistica. Lo scopo, infatti, era quello di descrivere logicamente la struttura del qubit (cioè, la sovrapposizione quantistica intrinseca di uno stato quantico a due livelli) e l'intreccio (entanglement) quantistico massimale di due qubits.

La logica *Lq* è ottenuta tramite il principio di riflessione di Sambin della logica di Base, secondo il quale i legami metalinguistici tra asserzioni si riflettono (risolvendo un' equazione definitoria) in connettivi logici tra proposizioni.

Comunque, mentre nella logica di Base il metalinguaggio è classico, nel nostro caso è quantistico.
Nel metalinguaggio quantistico, ciascuna asserzione atomica è dotata di un grado di asserzione, un numero complesso che viene interpretato come un' ampiezza di probabilità. E' proprio la presenza dei gradi di asserzione che permette l'introduzione del connettivo logico di "sovrapposizione quantistica" in *Lq*. Quest'ultimo è una generalizzazione del connettivo di congiunzione "and" dotato di indici complessi indicanti con quale "peso" ciascuna proposizione contribuisce alla formazione della proposizione composta.

I valori (o gradi) di verità sono i moduli quadrati dei gradi di asserzione, con un *range* che è l'intervallo reale [0,1]. Pertanto, la logica *Lq* è polivalente. I gradi di verità, differentemente dalle logiche fuzzy, sono qui interpretati come probabilità quantistiche.

Nella logica *Lq* si mantengono le tre importanti proprietà della logica di Base, cioè simmetria, riflessione e visibilità. Questa scelta è stata dettata dalle seguenti considerazioni:

1) I teoremi di *no-cloning* e *no-erase* dell'informazione quantistica non permettono di avere, nella logica corrispondente, le regole strutturali di indebolimento e contrazione, che sono in antitesi con i suddetti teoremi. Pertanto, nella ricerca di una logica dell' informazione quantistica, ogni logica strutturale deve essere esclusa a priori.
2) La scelta tra le due più importanti logiche sub-strutturali, cioè la logica di Base e la logica Lineare, in favore della prima, è dovuta al fatto che, in assenza di visibilità, il connettivo logico "*quantum entanglement*" non può essere introdotto.

Inoltre, si è cercata una logica dell' informazione quantistica che avesse un calcolo deduttivo (in particolare il calcolo dei sequenti).
La logica *Lq* sembra essere, finora, l' unica logica dell' informazione quantistica che possa soddisfare questi *desiderata*.

L’ interpretazione delle asserzioni del metalinguaggio quantistico è data in termini di stati quantistici (il metalinguaggio quantistico “è” lo spazio di Hilbert).
L’ interpretazione delle proposizioni di *Lq* è data in termini di operatori non-hermitiani, che sono misure deboli.
L’ interpretazione di *Lq* si basa su una generalizzazione dei concetti già proposti da Birkhoff e von Neumann nella logica quantistica “ortodossa”, dove le proposizioni sono interpretate come operatori di proiezione. La differenza consiste nel fatto che in *Lq* le proposizioni sono interpretate invece come misure deboli, che, diversamente dalle misure proiettive, non danno luogo ad un brusco collasso della funzione d’onda. Questo permette una descrizione logica della sovrapposizione quantistica, perché essa non viene distrutta. La possibilità di interpretare le proposizioni come misure deboli, è dovuta al fatto che abbiamo introdotto un metalinguaggio quantistico. Infatti, il grado di asserzione si riflette, nell’ interpretazione delle proposizioni, con la presenza un fattore moltiplicativo complesso sui proiettori.

Alcuni risultati di questa tesi sono:

a) L’ adozione di un nuovo tipo di metalinguaggio, il metalinguaggio quantistico, dove i legami metalinguistici sono correlazioni quantistiche, e le asserzioni hanno un grado di asserzione complesso.
b) L’ introduzione, tramite il principio di riflessione, di nuovi connettivi logici “quantistici”, quali la “sovrapposizione quantistica” e l’ “*entanglement*”.
c) L’ introduzione di una nuova operazione duale, che è una generalizzazione della dualità logica di Sambin-Girard, che tiene conto, nell’ interpretazione, dello spazio duale di Hilbert.
d) Una regola del taglio quantistica, che viene interpretata come misura quantistica proiettiva. Poiché il taglio è una meta-regola, ne consegue che una macchina quantistica non può effettuare una auto-misura e quindi auto-distruggersi.
e) Una nuova meta-regola, non equivalente al taglio, detta regola EPR (rifacentesi al paradosso di Einstein-Podolsky-Rosen). Questa regola permette di dimostrare simultaneamente due teoremi *entanglati*.
f) La formulazione del “teorema del qubit”, che è la descrizione logica della preparazione dello stato quantistico del qubit ottico.
g) Il fatto che il reticolo delle proposizioni di *Lq* nel caso di due qubits è orto-modulare non-distributivo. Quindi *Lq* è una logica quantistica.

E’ da notare il fatto che *Lq* è la prima logica ad essere contemporaneamente sub-strutturale, a molti valori di verità, e quantistica.

**Aknowledgements**

I am very grateful to my advisors, Prof. G. Sambin and Prof. P. Marchetti, and also to Prof. E. Pessa, Prof. S. Valentini, Prof. G. Gerla, Prof. G. Vitiello, Prof. M. Rasetti, Prof. P. Odifreddi, Prof. G. Priest, Prof. H. Stapp and Prof. R. Penrose for useful discussions and advices.

I wish to thank the AIRS, in particular, the President, Prof. G. Minati, for the award of a research prize for “Research on quantum control”, the Department of Psychology of the University of Pavia, in particular the Dean, Prof E. Pessa, for the research grant on “Quantum decision making”, and IBIOCAT, for the research grant on “The logic of the unconscious”, in particular Prof. M. Pregnolato (Head of the research group Cognitive.2009-I).

I also wish to thank to the Director of the Doctorate School, Prof. F. Cardin, for his kind interest and encouragement.

Finally, I thank my sister Patty and many friends and colleagues for moral support.

## Introduction

The main aim of this thesis is to look for a logical deductive calculus (we will adopt sequent calculus, originally introduced in Gentzen, 1935), which could describe quantum information and its properties (for a review of quantum information and quantum computation see, for example, Nielsen and Chuang, 2000).

More precisely, we intended to describe in logical terms the formation of the qubit (the unit of quantum information) which is a particular linear superposition of the two classical bits 0 and 1. To do so, we had to introduce the new connective "quantum superposition", in the logic of one qubit, *Lq*, as the classical conjunction cannot describe this quantum link.

The logic of *n* qubits, *Lnq* drastically depends on the number *n* of qubits considered. A consequence is that the connective of conjunction (as well as disjunction) can be used only for a finite number of times, which depends on the number *n* of qubits. In a sense, this is a logic with "quantum jumps" as it is characterized by discrete and sudden changes. The origin of such "quantum jumps" stands in a constraint (called Meta Data) encapsulated in the quantum metalanguage, from which the connective of quantum superposition is derived. We are aware that this is a quite new situation in logic, where in general, connectives can be used for an unlimited number of times. However, since now, logicians were not confronted with the very peculiar issue of quantum superposition, where this new feature arises. In this thesis, we make the first step in this direction.

Also, we will give a logical calculus for the case of two qubits, in the case they are maximally entangled, namely Bell's states (Bell, 1987). Quantum entanglement is a pure quantum correlation, with no analogous in classical physics. Two quantum states $\psi_A$ and $\psi_B$ are said entangled when the bipartite state $\psi_{AB}$ cannot be written as a tensor product of the two. We needed therefore to introduce the new connective "quantum entanglement".

When two states are entangled, every action (measurement) performed on one of them instantaneously modifies the other, even if the two are spatially separated. This is the so-called EPR paradox (the original reference is Einstein *et al.*, 1935; many textbooks, both specialized and popular, discuss its implications; see, for instance, Bell, 1987; Selleri, 1988; Aczel, 2003). We then introduced the logical EPR rule, which describes the EPR paradox in logical terms.

Quantum information is basically different from classical information (see, in this regard, Vedral, 2006). Also, the quantum computational process itself is different from classical computation, as it obeys quantum mechanical laws (on the mathematical analysis of the difference between classical and quantum computation see Kitaev *et al.*, 2002). Quantum computation exploits quantum superposition and quantum entanglement, which together give rise to the so-called "massive parallelism". Massive parallelism allows quantum computers to perform some computational tasks much more efficiently than classical computers.

Quantum information is different from classical information under three main aspects:

i) While the unit of classical information is the bit (0 and 1), the unit of quantum information is the qubit, which can be 0, 1 or a particular linear superposition of 0 and 1, with complex coefficients called probability amplitudes.

ii) While the bit can be copied or erased, it is impossible to copy or erase a qubit in a (unknown) superposed state. These impossibilities result from two no-go theorems: the no-cloning (Wootters and Zurek, 1982; Dieks, 1982) and no-erase (Pati and Braunstein, 2000) theorems.

iii) Two or more units of quantum information can be in the so-called *entangled* state. Entanglement (Bengtsson and Zyczkowski, 2006) is a particular kind of superposition, where the two (or more) original qubits form a non separable bipartite (multipartite) state. The two (many)-qubits state cannot be written as a tensor product of the original states. In a sense, in an entangled state the two (or more) qubits lose their individuality. Entanglement is a very peculiar and strong quantum correlation, with no classical analogous.

The main features of the quantum computational process are:

iv) Quantum computation, being a quantum process, has a dynamical evolution described by unitary operators. Therefore, quantum computation is reversible. These unitary operators are the quantum logical gates, considered in the quantum network approach (in this regard see, besides the references already quoted for quantum computing, even Mahler and Weberruß, 1998).

v) Quantum superposition and quantum entanglement lead to the so-called "massive quantum superposition" which allows an exponential speed-up of the quantum computational calculus, with respect to classical computation.

vi) A projective measurement stops the quantum computational process. In fact a measurement is an irreversible operation, and destroys quantum superposition, upon which quantum computation is based.

This point is very important for the search of the logic of quantum computing. In fact, if one uses orthodox quantum logic, where propositions are interpreted as projectors, that is, if one makes yes/no questions by using projective measurements, he would have at his disposal only the result of the quantum computation, the proof having been destroyed.

All these peculiar features of quantum information and quantum computation restrict the choice of the logical framework, leading to consider sub-structural logics. The restriction to sub-structural logics is due to the fact that the absence of the two structural rules of contraction and weakening correspond in quantum computing to the no-cloning theorem and to the no-erase theorem respectively. Moreover, sub-structural logics have more room for connectives and meta-rules. This is what allowed us to introduce the new quantum connective "entanglement" (discussed in Sect. 11) in the framework of the weakest logic, Basic logic (Sambin *et al*., 2000). In Linear logic (Girard, 1987), instead, the connective of quantum entanglement collapses to a multiplicative disjunction because of the distributive property, which is absent in Basic logic.

In this thesis, we will limit ourselves to the cases of one qubit state in the 2-dimensional complex Hilbert space, and of (maximally) entangled bipartite states (Bell states) in the 4-dimensional Hilbert space. In the first case, we are faced with the intrinsic quantum superposition of the qubit, which requires the introduction of a new logical connective, which turns out to be a (non commutative) conjunction, labelled by complex numbers called “assertion-degrees” interpreted as probability amplitudes. In the second case, entanglement requires the introduction of the connective “entanglement”, which is a mixture of additive conjunction and multiplicative disjunction (the latter being a peculiar connective of sub-structural logics like Linear logic and Basic logic).

In general, our aim is to investigate about the logical meaning of the specific properties of quantum information, induced by the physical model of Quantum Mechanics (QM).

To obtain such an interpretation, we will exploit the modular environment offered by Basic logic that allows the formulation of several quantum logical calculi. The intrinsic properties of the calculus of Basic logic, such as linearity and visibility, are surprisingly close to the computational consequences of quantum superposition, no-cloning and entanglement.

Also, to achieve our scope, we adopted a constructivist approach. In the philosophy of mathematics, constructivism asserts that it is necessary to find (or "construct") a mathematical object to prove that it exists. When one assumes that an object does not exist and derives a contradiction from that assumption, one still has not found the object and therefore not proved its existence, according to constructivists (the number of references on constructivism is very high; here we will limit ourselves to quote Heyting, 1971; Dummett, 1977; Martin-Löf, 1984; Beeson, 1985; Troelstra & Van Dalen, 1988; Feferman, 1998; for the history of constructivism a good reference is Troelstra, 1991).

Under some aspects, intuitionism could be viewed as a particular form of constructivism. However, intuitionism maintains that the foundations of mathematics lie in the individual mathematician's intuition, thereby making mathematics into an intrinsically subjective activity. On the contrary, constructivism is not based on this view of intuition, and is entirely consonant with an objective view of mathematics.

Constructivist mathematics uses intuitionist logic, which is essentially classical logic without the law of the excluded middle. This is not to say that the law of the excluded middle is denied entirely; special cases of the law will be provable. It is just that the general law is not assumed as an axiom. The algebraic structure underlying Intuitionist logic is Heyting algebra.

In order to reach the aim of this thesis the main problem has been to understand which are the propositions which can lead, up to an interpretation, to the logical description of quantum superposition.

The logical approach to quantum computation based on quantum networks of logical gates has suggested new forms of quantum logic that have been called quantum computational logics (see, in this regard, Dalla Chiara *et al.*, 2003). However, the quantum logical gates are unitary operators, that is, they operate in a reversible way inside the computer, with no access to an external logic.

The main difference between orthodox quantum logic (that is, the one introduced by Birkhoff and Von Neumann; their fundamental paper is Birkhoff & Von Neumann, 1936; see also the analysis of their ideas given in Redei, 2007) and quantum computational logics concerns a basic semantic question: how to represent the *meanings* of the sentences of a given language?

The answer given by Birkhoff and Von Neumann is the following: the meanings of the elementary experimental sentences of quantum theory have to be regarded as determined by convenient sets of states. Since these sets should satisfy some special closure conditions, it turns out that, in the framework of orthodox quantum logic, sentences can be adequately interpreted as closed subspaces of the Hilbert space associated to the physical systems under investigation.

The answer given in the framework of quantum computational logics is quite different. The *meaning* of a sentence is identified with a quantum information quantity: a *qubit* or a quantum register (Dalla Chiara *et al.* , 2003).

In other words, the semantics of propositions concerning quantum computing is encoded in the states of quantum information, namely, qubits and quantum registers. Propositions are not viewed as bearers of truth, but as bearers of quantum information.

As we will see in this thesis, the answer given by us, in the framework of a substructural quantum logic, is that atomic propositions can be still interpreted as closed subspaces of the Hilbert space, each closed subspace, however, being "smaller" than the one considered by Birkhoff and Von Neumann. The consequence is that, in our case, propositions are bearers of partial truths, corresponding to quantum mechanical probabilities.

In Birkhoff-Von Neumann quantum logic, the closed subspaces are in one-to-one correspondence to projector operators in the Hilbert space. As projectors are measurement operators, the propositions are called "experimental". Orthodox quantum logic takes as primitive concepts the measurements. But this is just a point of view, as it is not dictated by any theoretical principle. It just relies on some postulates, on which the empiricist, positivist interpretation of Quantum Mechanics of the Copenhagen School has been based. But, as there are several interpretations of Quantum Mechanics, there are also different choices for the underlying logic. Some examples are given in the framework of quantum structures, which are structures at the border between algebra and logic (see, in this regard, Engesser & Gabbay, 2002; Lehmann *et al.*, 2006; Engesser *et al.*, 2007; Engesser *et al.*, 2009).

The conceptual weakness of orthodox quantum logic arises when one tries to use it in the framework of quantum computing. The only propositions one is able to formulate are about the system at the end of the computational process, when a measurement is performed. The very computational process remains inaccessible: a proposition/projector would destroy the quantum superposition, and stop quantum computation. Although this looks quite acceptable in quantum physics, if one is interested mainly in the result of a measurement and not to the very (computational) process itself, this is not the case for logic: the lack of the description of the computational process means absence of a deductive calculus.

Furthermore, it is not possible to assign definite truth-values to propositions in orthodox quantum logic. This is forbidden by the Kochen-Specker theorem (Kochen & Specker, 1967; see simpler proofs of this theorem given by Mermin, 1990 and Peres, 1991) which asserts the non-existence of valuations in quantum theory, subject only to the rather plausible requirement that the value of a function of a physical quantity should be the result of applying that function to the value of the quantity. When applied to propositions, the theorem asserts the non-existence of any consistent assignment of true-false values to the propositions in quantum theory. However, although the theorem forbids any absolute assignment of truth-values, it does not exclude truth-values that are contextual. Here, ‘contextual’ means that the truth value given to a proposition depends on which other compatible (meaning ‘simultaneously measurable’) propositions are given values at the same time. Also, it does not excludes many-valued truth-values (or truth-degrees).

It is to be noticed that there are not even truth-degrees as in many-valued logics (see, for instance, Malinowski, 1993; Bergmann, 2008), which would be allowed by the above theorem. The truth-value is just indeterminate. Now a natural question arises: how can one speak about propositions with indeterminate truth-value, once one defines propositions to be truth bearers? In orthodox quantum logic propositions are not bearers of information either. The interesting thing is that Von Neumann was perfectly aware of this fallacy, as he himself formulated a theorem (Von Neumann, 1932, Chapter IV.1, 2), which thereafter was reformulated in a different context and became the Kochen-Specker theorem.

Instead, in our case, the Kochen-Specker theorem does not apply, although our logic is quantum, the reason being that our logic is also many-valued.

Another point against orthodox quantum logic is that it has the structural rules of contraction and weakening, therefore it cannot account for no-cloning and no-erase in quantum computing. This structure does not allow, as well, to introduce a connective for quantum entanglement, which requires a multiplicative connective, available only in sub-structural logics.

Another weak point in Birkhoff-Von Neumann quantum logic is the interpretation of quantum probability. In fact, to interpret the non-commutative measure space as quantum probability space with probability understood as relative frequency (the frequency theory of probability having been developed by R. Von Mises; see, for instance, Von Mises, 1964) the probability assignment needs to satisfy the superadditivity property. Instead, the latter is violated by every state for some projections. Consequently, a lattice of projections cannot be interpreted as the structure representing random events for a quantum probability space if probability is interpreted as relative frequency.

This was the main reason which induced Von Neumann wishing to reject the Hilbert space formulation of QM, and to formulate the type $\prod_1$ Von Neumann algebra (without quoting the original papers on this subject by Murray and Von Neumann, we will limit ourselves to quote a textbook on these topics such as Blackadar, 2006), which is a modular lattice. However, also in that case, not all worked as Von Neumann had hoped and, at the end, he had to abandon the relative frequency interpretation. All those problems were related to the non-commutativity of the measure space. In our case, instead, the measure space is commutative, the super-additivity property holds and we can maintain the frequentist approach to the theory of probability. It should be noticed that although in our case the measure space is commutative, the logic is still quantum: the non-distributivity of the lattice being a consequence of the fact that propositions are weak measurements, not projectors.

Going back to the approach of quantum computational structure introduced by Dalla Chiara *et al*., the inconsistency with measurement during the computational process is avoided, as propositions are not anymore projectors. Also, the indeterminacy of truth-values disappears, and actually the concept of truth value loses its main role. The semantic role is played by quantum information. Very roughly we can say that in their approach, propositions are not (projective) operators, but quantum states, namely, qubit states. However, the above approach presents some weakness too. The qubit is taken as an atomic proposition (and one can understand why, as the qubit is the unit of quantum information). This is the so-called *holistic* approach, which leads to a semantics, but not to a logical calculus (the syntax). Why this happens it due to the fact that taking the qubit as an atomic proposition, one is not concerned with the intrinsic quantum superposition of the qubit. This in turn avoids the necessity to introduce the logical connective of quantum superposition, and the related inference rules. The holistic approach reduces then to the well known quantum network approach. No sequent calculus can be provided.

Another possible approach is that of M-algebras (Engesser & Gabbay, 2002; Lehmann *et al*., 2006; Engesser *et al*., 2009), where propositions are measurements and the meaning of such propositions is given by their action on physical states. Measurements are maps from the set of physical states to itself. To these maps can be assigned truth-values 0, 1 and indeterminate. The truth value of each map is defined in term of its fixed point.

However, in the M-algebras, the fact that propositions are measurements carries along the same problems already present in Birkhoff-Von Neumann quantum logic: the quantum computational process remains inaccessible. Moreover, no explicit interpretation is given in the Hilbert space.

Therefore, none of the above approaches seems to fully satisfy the requirements dictated by quantum information and quantum computation.

We tried to solve this puzzle by providing a new kind of metalanguage, the quantum metalanguage, which consists of metalinguistic links among quantum assertions, plus a set of Meta Data (*MD*). The quantum assertions are atomic assertions carrying along an assertion-degree, which is a complex number, interpreted as a probability amplitude. The Meta Data is a constraints on the assertion degrees, which, in the interpretation, ensures that quantum-mechanical probabilities sum up to one.

The quantum metalanguage reflects properly into the object-language of the quantum computer, in the sense that it is possible to introduce the connective of quantum superposition of the qubit.

Of course, this approach is not holistic, but compositional: the qubit is a compound proposition built up by means of the logical connective of quantum superposition. This allows to get a sequent calculus, by the inference rules of the connective of quantum superposition. Therefore, considering the qubit as a compound sentence is necessary if one wants to study the syntax (the logical calculus) of quantum information, and not only the semantics. Moreover, a theorem of our logical calculus, the “qubit theorem”, logically describes the physical preparation of the qubit state in linear optics, starting from two different polarizations of the photon.
In this sense, it seems to us that the qubit state, eventually, can be interpreted as a compound proposition, as the experimental implementation suggests.
In our approach, the atomic propositions in the quantum object-language are interpreted, in the model, as weak measurements (introduced by Aharonov *et al.*, 1988), that is, measurements which do not make the quantum superposition abruptly collapse. More in detail, the weak measurement operators we consider are non-hermitian operators, which are obtained by the multiplication of a complex number times the one-dimensional projectors of the Hilbert space under consideration (the latter, being, as it is well known, hermitian).
Weak measurements are used in the theory of quantum control (see, for instance, D’Alessandro, 2008). In our case, one can argue, then, that the metalanguage itself is a quantum control, arising from our thought processes. This idea has been used with some success in the theory of control of quantum robots (Benioff, 1998; Dong *et al.*, 2006; Zizzi, 2009), although practical applications have not yet been implemented.
Nevertheless, our principal aim is, as we said already, to get a logical calculus for quantum information. We stress again the fact that, to this purpose, all structural logics are to be dismissed, and the only acceptable logic should be sub-structural. However, a logic, to be quantum, must have the algebraic structure of a non-distributive lattice (like in the Birkhoff-Von Neumann case) to satisfy the Heisenberg uncertainty principle.
We found that in our case the propositions, once interpreted, form an orthomodular lattice which is non-distributive in the least non-trivial case of two qubits. As far as one qubit is considered, instead, the lattice of this logic is distributive (as in the case of the lattice of two one-dimensional projectors). However, when the underlying set of the lattice is enlarged to comprise both the two weak measurements and the two projectors, the lattice is non-distributive also in the case of one qubit.
To conclude, the logic of quantum information that we present in this thesis allows an interpretation in the Hilbert space, has a syntax (a deductive calculus, more precisely, a sequent calculus), is sub-structural in order to fit with the no-cloning and no-erase theorems, and of course is quantum (in the sense that is a non-distributive orthomodular lattice) to satisfy the uncertainty principle. Furthermore our logic is many-valued (as truth-values are related to the quantum-mechanical probabilities).
This thesis is organized as follows.
In Sect.1, we gave a short historical review of quantum logic. This was required by the complex historical path of the ideas about the relationship between logic and quantum mechanics. As it is well known, the original proposal of Birkhoff and Von

Neumann originated a number of algebraic and logical investigations, associated to many different definitions of what is to be meant as “quantum logic”. A short mention of these conceptual developments should help to understand the sense of the proposals made within this thesis.
In Sect.2, we made a detailed survey of lattice theory and, in particular, we reviewed the non-distributive orthomodular lattice of propositions (projectors) of the orthodox quantum logic of Birkhoff-von Neumann.
In Sect.3, we recalled the basic notions on quantum computers and quantum information. In particular, we described the unit of quantum information, the qubit, a particular linear combination of the classical bits 0 and 1.We shortly described the state space of the qubit, namely the Bloch sphere. Finally we illustrated the projective measurement of the qubit state.
In Sect.4, we introduced control theory, which is related to the notion of metalanguage. In particular, we discussed quantum control in terms of a quantum metalanguage. Within this context, the best control is realized through a weak measurement (instead of a projective one) not to make the quantum state abruptly collapse. In logical terms, we found that such a quantum control is a quantum metalanguage, based on a new notion of assertion which is somehow “partial” with respect to the classical one. Then, propositions are interpreted as non-hermitian operators of the Hilbert space.
Sect. 5 has been devoted to a review of quantum computational logics. The latter constitute a serious and rigorously grounded attempt to build a complete logical theory of quantum processes, based on the knowledge related to quantum computing, mainly the one concerning the operation of quantum gates. This approach gave rise to a number of interesting results, which can be considered as the main core of a truly “logical” theory of quantum processing. However, quantum computational logics, while being very useful, do not deal with the problem of the formation of qubits themselves, which, instead, are considered as whole entities, already constituted from the beginning.
In Sect.6 we briefly recalled some notions about Gentzen’s sequent calculus, and its semantics, while in Sect.7 we will made a short review of Basic logic, and its main properties and features. In particular we enlightened the reflection property of Basic logic, which gives rise to the Reflection principle, by which the metalanguage, consisting of metalinguistic links among assertions, reflects, through a definitional equation, into the object-language, consisting of logical connectives among propositions. Moreover, we illustrated the other two properties of Basic logic: symmetry and visibility. Finally, we displayed the B-calculus.
In Sect.8, we gave some definitions and notations that are necessary for the understanding of the sequent calculus of the logic of one qubit state. In particular, we gave the formal language *Lq*, the notion of Meta Data (data about data) which is a constraint in the quantum metalanguage requiring that the truth-evaluations of sequents sum up to 1. Moreover, we introduced the *-duality, associated with an interpretation in the dual Hilbert space, and a new kind of logical duality $\perp'$ (a Sambin-Girard $\perp$-like) which takes into account the assertion-degrees. We also defined the “gluing operation” which is in fact a functional of the functions of the assertion-degrees of the antecedent and of the consequent in a sequent. We interpreted the “gluing operation” as a scalar product in the Hilbert space. Finally we generalized the cut rule to the quantum case, and showed that it can be interpreted as a projective quantum measurement.
In Sect.9 we discussed the main differences between classical and quantum metalanguages. The interesting fact is that classical and quantum metalanguages can reflect into the object-language by the same Reflection principle of Basic logic, although they have different ranges of application (a classical metalanguage does not reflect properly into a quantum object-language, and *viceversa*). While in a classical metalanguage the atomic assertions are fully asserted, in a quantum metalanguage they are asserted with “assertion-degrees”. In the physical interpretation, assertion-degrees are complex numbers referring to probability amplitudes. The quantum metalanguage, endowed with a Meta Data ensuring that probabilities sum up to 1, reflects properly into the quantum object-language, where the connective “quantum superposition” ${}_{z_0}\&_{z_1}$ (which is a non-commutative conjunction labelled by complex numbers, the assertion-degrees $z_0$ and $z_1$ of the atomic propositions) can be introduced.
In Sect.10, we presented the sequent calculus of *Lq*. First, we introduced the connective “quantum superposition ” ${}_{z_0}\&_{z_1}$, and gave its logical rules.
Then we extended the calculus to the case without Meta Data, which though not describing anymore quantum superposition, could be useful in more general logical contexts.
In the special case of physical interest, that is in presence of Meta Data, we gave a logical interpretation of the preparation of the qubit state within linear optics.
Moreover, we introduced the $\perp'$- dual of ${}_{z_0}\&_{z_1}$, that is ${}_{z_0^*}\vee_{z_1^*}$ , which is interpreted in the dual Hilbert space.
We then discussed the absence of all structural rules in *Lq*, which turns out to be the weakest sub-structural logic so far built, which, nevertheless, can describe the informational aspects of the quantum world.
In Sect.11, we presented the logic of two qubits *(L2q*). We introduced a new connective denoted by the symbol @, which is of great importance in the logical description of the quantum correlation for excellence: quantum entanglement. The connective @ is a particular combination of the connectives “quantum superposition” and “par” (the multiplicative disjunction of Linear and Basic logic). Although being a derived connective, @ has its own definitional equation, from which we derived its logical rules. We illustrated the properties of the connective @, namely, commutativity, distributivity with absorption, associativity, duality, and non-idempotence.
We introduced the $\perp'$-dual of @, namely the connective §, to be interpreted in the dual Hilbert space.
Finally, we formulated the EPR rule (where EPR stands for the EPR-paradox) which is a meta-rule like the cut rule. The EPR rule also makes a cut, but we proved that it is not equivalent to the cut rule. The EPR rule allows to prove two entangled compound propositions at once, by proving just one of them.

In Sect.12, we showed that the propositions of the logic *Lq* form an orthomodular lattice $Lq(H)$, which is distributive in the case of a two-dimensional complex Hilbert space (the case of one qubit). We also showed that the lattice $L2q(H)$ of propositions of the logic of two qubits is instead non-distributive. Moreover, there is a very special case, that we named $L_m(H)$, where "*m*" stands for "*mixed*", which is the case of a lattice whose underlying sets comprises both projectors and weak measurements on a 2-dimensional complex Hilbert space, which is non-distributive. In this case, the order relation has been defined by a "range measure", given in terms of quantum expectation values.

In Sect.13, we discussed the relationship between quantum metalanguage and the constructivist approach to logic by which logic is not given *a priori*, but is a product of the human mind. We distinguished between the macroscopic (classical) and the microscopic (quantum) points of view with respect to constructivist logic. The former, based on the phenomenology of the process of thinking, and philosophy of mind, is best realized in the framework of dynamical constructivism proposed by Sambin (*cfr.* Sambin, 2002). The latter relies on dissipative quantum processes occurring in the brain, and can be formalized in a quantum metalanguage, which plays the role of a quantum control on the formal logical language.

In Sect.14, we discussed the main interpretations of QM, and the underlying philosophical approaches, and also looked at the logical-mathematical counterparts of the latter. Furthermore, we investigated about the relations between different kinds of metalanguages and the main interpretations of QM. This led to the conclusion that the quantum metalanguage introduced in this thesis is related to a new kind of interpretation of QM, which we called "transcendental". The latter seems to avoid some inconsistencies which plague both the realist and the Copenhagen interpretations.

## 1. A short history of Quantum logic

As it is well known, Quantum logic was born in 1932, when J. Von Neumann, in Section 5 of the Chapter 3 of his celebrated book *Grundlagen der Quantenmechanik* (see for the English edition Von Neumann, 1955) interpreted projection operators defined on a Hilbert space as representing experimental propositions related to the properties of a quantum system. In the language of Von Neumann an experimental proposition was identified with the assertion that a physical quantity had a specific value. The Von Neumann's identification seems at first sight natural if we take into account that any self-adjoint operator $P$ acting on Hilbert space $H$, whose spectrum coincides with the two-elements set $\{0,1\}$, must be a projector, as it fulfils the relationship $P^2 = P$. This entails that the range of $P$ is closed, so that there is a one-to-one correspondence between these operators and the closed subspaces of $H$. Each one of these latter, on the other hand, can be identified with the set of states of a quantum system for which a particular experimental proposition is true.

The one-to-one correspondence between projectors and closed subspaces of $H$ led immediately Von Neumann to take into consideration the possibility of building a sort of "logical calculus" of projections, by relying on the analogy with traditional Propositional Calculus, whose algebraic structure is related to the Boolean lattices of subsets. Namely, if we introduce an ordering within closed subsets of $H$, based on set-inclusion, it is easy to see that they constitute a lattice, within which the meet of two subsets is their intersection, while the join is their union. This allows to speak of the meet and join operations on projectors. It is then possible to introduce the complement $P'$ of a projector $P$ through the definition $P' = 1 - P$. However, notwithstanding the apparent easiness of this way of proceeding, Von Neumann expressed some doubts both on its usefulness and the possibility of building a "logic" of experimental propositions. These doubts came from the fact that, contrarily to classical physics, Quantum Mechanics allows the existence of not simultaneously decidable experimental propositions, more precisely the ones related to pairs of observables associated with non-commuting operators (a typical case, for instance, is the one of two experimental propositions, one asserting that the coordinate $x$ of an electron has a given value, and the other asserting that its momentum $p_x$ has a given value). In these cases the meaning to be attributed to the join and the meet of these propositions (that is, of their associated projectors) appeared to Von Neumann as absolutely unclear.

In 1936 Von Neumann dealt with again with this question in a more technical paper, written jointly with Garrett Birkhoff and entitled *The Logic of Quantum Mechanics* (Birkhoff and Von Neumann, 1936). Within this paper, considered by most people as the starting point of Quantum logic, the authors showed that, while the projectors of $H$ really constitute a lattice, this lattice is no longer Boolean, but rather belongs to the class of *orthocomplemented orthomodular lattices* (these attributes were not used by Birkhoff and Von Neumann, as they were introduced later), which, in a sense, are "weaker" than Boolean ones, as they do not fulfil the distributive law.

In order to understand the significance of Birkhoff-Von Neumann results, we need here some technical notions and clarifications. First of all we stress the fact that, when dealing with a quantum system, we must take into consideration, rather than all generic closed subsets of $H$ (which obviously constitute a Boolean lattice under the usual set-theoretical operations), only a particular subclass of them, constituted by the closed *linear* subspaces of $H$. The need for closure under linear combinations arises from the fact that, in Quantum Mechanics, if the state vectors $\psi_1,...,\psi_n$ are descriptors of pure states, even their linear combination $c_1\psi_1 + ... + c_n\psi_n$ will be a pure state. This entails that the definition of meet and join operations cannot be fully done in terms only of the standard set-theoretical operations of intersection and union. Moreover, even the operation of complementation cannot be introduced by resorting only to the usual set complementation.
In this regard Birkhoff and Von Neumann start by remarking that the meet operation between closed linear subspaces, hereafter denoted by the symbol $\prod$, can still be identified with the usual set-theoretical intersection, as the intersection of two closed linear subspaces is still a closed linear subspace. When expressed in terms of projectors, then, the meet operation will correspond to the usual composition, that is the meet of the projectors $P$ and $Q$ will be given by $PQ$ (we remind that the order of the composition matters, as in general $PQ \neq QP$). Unfortunately the join of two closed linear subspaces cannot be defined only in terms of their union, as the latter, in general, does not give rise to a closed linear subspace. Here the join of two closed linear subspaces, hereafter denoted by the symbol $\coprod$, must necessarily be defined in terms of the smallest closed linear subspace including both. As regards the complement (corresponding to the logical *negation*) Birkhoff and Von Neumann resort to the *orthogonal complement* of a closed linear subspace. More precisely, given a subspace $X$ of $H$, its orthogonal complement $X'$ is defined as the set of all vectors of $H$ which are orthogonal to all vectors of $X$. In other terms:

$$X' \textit{ is the orthogonal complement of } X \textit{ iff } (\psi,\varphi) = 0 \textit{ for any } \psi \in X, \varphi \in X' \tag{1.1}$$

Here $(\psi,\varphi)$ denotes the inner product between the vectors $\psi$ and $\varphi$.
Now, once introduced as ordering relation the usual set-theoretical inclusion, and denoting by ***1*** and ***0***, respectively, the whole Hilbert space and the null subspace, and by $CL(H)$ the set of closed linear subspaces of $H$, it can be shown that the structure $C(H)$ defined by:

$$C(H) \equiv \{CL(H), \subseteq, \prod, \coprod, ', 1, 0\} \tag{1.2}$$

is a lattice. Unfortunately it is not a Boolean lattice, as it fails to satisfy the distributive law. Namely, in general:

$$X \prod (Y \coprod Z) \neq (X \prod Y) \coprod (X \prod Z) \tag{1.3}$$

It is to be remarked that the lattice $C(H)$ is isomorphic to the lattice $P(H)$ of the projectors of $H$. This isomorphism is based on the aforementioned biunivocal correspondence between projectors and closed linear subspaces.
The lack of the distributive property troubled Birkhoff and Von Neumann, who tried to show that $C(H)$ or $P(H)$ were satisfying at least a weaker form of distributivity, given by the *modular law*. We remind here that, given a generic lattice $L \equiv \{\leq, \vee, \wedge, \neg\}$, the modular law can be written under the form:

$$a \leq c \Rightarrow a \vee (b \wedge c) = (a \vee b) \wedge c \qquad \textit{for each } a,b,c \in L \tag{1.4}$$

Unfortunately this attempt was unsuccessful. They were able only to show that, if two projectors commute (a case not always occurring in Quantum Mechanics), then the sub-lattice generated by the two projectors together with their respective orthogonal complements is Boolean. However in 1937 Husimi (see Husimi, 1937) discovered that projectors of $H$ were satisfying a weaker version of modularity, now called *orthomodular law*. Within the notation adopted above the latter can be written under the form:

$$a \leq b \Rightarrow a \vee (\neg a \wedge b) = b \qquad \textit{for each } a,b \in L \tag{1.5}$$

The results achieved by Birkhoff and Von Neumann raised a number of questions, such as:
1) why the lattice $P(H)$ differs so strongly from the Boolean lattice, commonly associated with classical logic? Is this a cue of the inadequacy of classical logic to describe the physical world, which should have a deeply quantum character?
2) Is the formalism of Quantum Mechanics derivable only from the algebraic properties of $P(H)$?
3) Are the experimental propositions introduced by Von Neumann physically meaningful? Are they expressing all possible situations a physicist could encounter when doing an experimental investigation?
As regards the opinions of Von Neumann (see also, for a modern account, Redei and Stoeltzner, 2001) he seemed inclined to accept the fact that within Quantum Mechanics experimental propositions cannot satisfy the laws of classical logic, which, at least in this respect, should be inadequate. What was disturbing him, however, was that this precluded from introducing a formal algebraic basis for a quantum probability theory analogous to the one already developed in the classical case by Kolmogorov and relying on a Boolean structure. For this reason he left this line of research, focussing his attention on algebraic questions which led him to introduce the celebrated Von Neumann algebras. In any case, the Second World War shifted the attention of all scientists toward the solution of military problems, and all research on questions related to Quantum logic ceased.

In the fifties there was a renewed interest in this topic, due to two kinds of contributions: on one hand, the developments in abstract algebra which led to formulate the theory of orthocomplemented orthomodular lattices, shortly denoted *ortholattices* (for a comprehensive treatment of this subject see Kalmbach, 1983), thus rediscovering the neglected results of Husimi, and the introduction by Mackey of a number of axioms about experimental propositions allowing him to undertake a complete reconstruction of Quantum Mechanics starting only from a logical basis (a full account of the contribution by Mackey can be found in Mackey, 1963). Essentially Mackey reformulates experimental questions in terms of associations between observables, states and probabilities obtained by experimental measures. Then, after characterizing the set of these questions in terms of suitable axioms endowing it with a sort of "natural" algebraic structure, he ends with a final axiom stating that the partially ordered set of all questions in Quantum Mechanics is isomorphic to the partially ordered set of all closed linear subspaces of a separable infinite-dimensional Hilbert space. The effective reconstruction, starting from Mackey's axioms, of the whole mathematical apparatus of Quantum Mechanics, including its evolutionary aspects (see, in this regard, Varadarajan, 1968), requires, however, further theorems, among the which one of the most important ones is Gleason's theorem, showing that it is possible to introduce a probability measure on $C(H)$ and that this measure must necessarily coincide with the one used in Quantum Mechanics (for an account of these topics see Dvurečenskij, 1993).
In 1964 Piron introduced a perspective reversal with respect to the framework adopted by Birkhoff and Von Neumann, as well as by Mackey and Gleason, focussing his attention first on the lattice $P(H)$, rather than on $C(H)$(for a complete description of Piron's contribution see Piron, 1976). In this regard he showed that $P(H)$ was characterized by further properties, among which we will quote the *atomicity* and the *irreducibility*. The atomicity consists in the fact that every element of $P(H)$ is the join of *atoms* (given by one-dimensional projectors), while the irreducibility can be formulated by saying the no element of $P(H)$, except ***1*** and ***0***, commutes with all other elements. Piron then proved that a lattice such as $P(H)$, endowed with these properties, may be realised as the set of orthonormal subspaces of a generalized Hilbert space. Crucial to Piron's work is his redefinition of the concept of 'experimental question'. He starts from the concept of *preparation procedure* associated to a question $\alpha$ and introduces a relationship between preparations and questions, here denoted by $P \succ \alpha$, holding when the preparation $P$ is such that the answer to the question $\alpha$ can be predicted with certainty to be yes. The experimental question associated with $\alpha$ is then defined as follows: there exists a preparation $P$ such that the relationship $P \succ \alpha$ holds. The problem raised by Piron's proposal is that the introduction of a generalized Hilbert space forces to ask what are the further conditions singling out, within the class of these spaces, the "normal" Hilbert spaces used in Quantum Mechanics. The solution of this problem was made more difficult in the last years by a number of results showing the richness of algebraic structures underlying the world of experimental propositions. Among them we will quote the discover that even classical Propositional Calculus can be modelled by a non-Boolean lattice (see Pavičić and Megill,1999). The latter circumstance made very problematic the traditional identification of Quantum logic with the study of orthocomplemented orthomodular lattices.
Contemporary with these developments an alternative approach has been introduced by D. Foulis and C. Randall (see for a synthetic overview Randall and Foulis, 1983). Within this approach, called *empirical logic,* the fundamental notion is the one of *test*, identified with a definite set of mutually exclusive alternative possible *experimental outcomes*. The Foulis-Randall theory deals with *test spaces,* that is collections of overlapping tests. A *statistical state* within a given test space is defined as a mapping, for every test, which associates each possible outcome $x_i$ with a number $\omega_i \in [0,1]$ such that $\sum_i \omega_i = 1$. Test spaces can be characterized from the logical and algebraic point of view in a number of different spaces, in agreement with the conditions dictated by the effective experimental realm. Among these structures one of most interesting is constituted by the *orthoalgebras*. In abstract term an orthoalgebra can be defined as a pair $(L,\oplus)$, where $L$ is a set and $\oplus$ a commutative, associative *partial* binary operation on $L$ satisfying the following conditions:
(a) there exists a neutral element $0 \in L$ such that, for every $p \in L$, $p \oplus 0 = p$;
(b) there exists a unit element $1 \in L$ such that, for every $p \in L$, there is a unique $q \in L$, with $p \oplus q = 1$;
(c) if $p \oplus p$ exists, then $p = 0$.
The orthoalgebras are a generalization of structures, related to ortholattices and known as *orthomodular posets*. They were studied as alternatives to traditional Quantum logic (see for a complete account Pták and Pulmannová, 1991). It can be shown that these latter can be obtained as particular cases of orthoalgebras if we introduce the further axiom of *orthocoherence*:
(d) if $p \oplus q$, $q \oplus r$ and $r \oplus p$ all exist, then the element $p \oplus q \oplus r$ also exists.
These structures are actually the subject of extensive algebraic and logic research (see, for instance, Coecke *et al.,* 2000).
All conceptual efforts described above are, however, frustrated when one tries to take into account the composite systems, which in most applications of Quantum Mechanics, are the most frequent cases. In these situations one must resort to tensor products of Hilbert spaces. Unfortunately Randall and Foulis showed that the categories of orthomodular lattices and orthomodular posets are *small* categories and do not admit a tensor product (see Randall and Foulis, 1979). In order to allow the construction of tensor products larger categories are needed, among which there are the orthoalgebras. This is one of the reasons why these structures are actually intensively studied.
The development briefly sketched above, while leaving unanswered the question concerning the structure which should be rightly identified with Quantum logic, evidence, on the other hand, that Quantum logic does not play any role in the effective description of quantum systems, nor in the development of Quantum Mechanics or of quantum computing. Even if some weak connections between these latter topics and the theory of Hilbert lattices have been discovered (see, for a review, Megill and Pavičić, 2000), so far no theorem of Quantum logic gave indications, for instance, about the formation and processing of qubits. We feel that in the future this gap should be filled. Intuitionist logic, as well as the related concept of metalanguage could help to understand why our mind frames the experimental outcomes just within the conceptual framework of Quantum Mechanics. In a sense, this is not in contradiction with the empirical logic of Foulis-Randall approach. We claim that the conceptual tools quoted above offer a way to better implement the program of empirical logic, while looking at a wider landscape, including not only Quantum Mechanics, but even Quantum Field Theory.

## 2. The lattice of projectors in Birkhoff-von Neumann Quantum logic

In this Section, we will review the lattice of projectors in Birkhoff-von Neumann Quantum logic, which, as it is well known, is a non-distributive orthomodular lattice. In this regard, in the next Subsection, we will recall some fundamental notions about lattices.
In Subsection 2.2, we will discuss the most general case, where the projectors and the Hilbert space have whatever dimensionality.
In Subsection 2.3, we will illustrate the particular case of the lattice of one-dimensional projectors of the two-dimensional complex Hilbert space $C^2$, which is a Boolean lattice.

### 2.1 Lattices and their properties

A lattice (there is a plethora of textbooks devoted to this topic; see, for instance, Davey and Priestley, 2002) is a partially ordered set (also called a poset) ($L$, $\leq$) in which subsets of any two elements have a unique supremum (the elements' least upper bound; called their join$\vee$) and an infimum (greatest lower bound; called their meet $\wedge$).
Lattices can also be characterized as algebraic structures satisfying certain axioms. The two definitions are equivalent.
An algebraic structure ($L$, $\vee$, $\wedge$), consisting of a set $L$ and two binary operations $\vee$), and $\wedge$), on $L$ is a lattice if the following axioms hold for all elements $a$, $b$, $c$ of $L$.

| **Commutative laws** | **Associative laws** | **Absorption laws** |
|---|---|---|
| $a \vee b = b \vee a$ | $a \vee (b \vee c) = (a \vee b) \vee c$ | $a \vee (a \wedge b) = a$ |
| $a \wedge b = b \wedge a$ | $a \wedge (b \wedge c) = (a \wedge b) \wedge c$ | $a \wedge (a \vee b) = a$ |

The following identity can be derived from the axioms.

**Idempotent laws**

$a \vee a = a$
$a \wedge a = a$

**Distributivity**

Distributivity of $\vee$ over $\wedge$

$$a \vee (b \wedge c) = (a \vee b) \wedge (a \vee c).$$

Distributivity of $\wedge$ over $\vee$

$$a \wedge (b \vee c) = (a \wedge b) \vee (a \wedge c).$$

A lattice that satisfies the first or, equivalently (as it turns out), the second axiom, is called a distributive lattice.

**Modularity**

A lattice (L, $\vee$, $\wedge$) is modular if, for all elements *a*, *b*, *c* of *L*, the following identity holds.
Modular identity

$$(a \wedge c) \vee (b \wedge c) = [(a \wedge c) \vee b] \wedge c.$$

This condition is equivalent to the following axiom.

**Modular law**

$a \leq c$ implies $a \vee (b \wedge c) = (a \vee b) \wedge c$.

**Bounded lattice**

A lattice L is said to be bounded from below if there is an element $0 \in L$ such that $0 \leq x$ for all $x \in L$. Dually, *L* is bounded from above if there exists an element $1 \in L$ such that $x \leq 1$ for all $x \in L$. A bounded lattice is one that is bounded both from above and below.

**Complemented lattice**

Let *L* be a bounded lattice with greatest element 1 and least element 0. Two elements *x* and *y* of *L* are complements of each other if and only if:

$$x \vee y = 1 \text{ and } x \wedge y = 0$$

In this case, we write $\neg x = y$ and equivalently, $\neg y = x$. A bounded lattice for which every element has a complement is called a complemented lattice.

**Orthocomplemented lattice**

An orthocomplemented lattice (or just ortholattice) is a bounded lattice equipped with an orthocomplementation, i.e. an order-reversing involution that maps each element to a complement. The orthocomplement of an element *a* is written as $a^{\perp}$. It satisfies the following axioms.

- Complement law: $a^{\perp} \vee a = 1$ and $a^{\perp} \wedge a = 0$.
- Involution law: $a^{\perp\perp} = a$.
- Order-reversing if $a \leq b$ then $b^{\perp} \leq a^{\perp}$.

**Orthomodular lattice**

An orthomodular lattice is defined as an orthocomplemented lattice such that for any two elements the orthomodular law

if $a \leq c$, then $a \vee (a^{\perp} \wedge c) = c$

holds.
Notice that the orthomodular law is a weaker form of the modular law, and is obtained from the latter in the particular case $b = a^{\perp}$.
The lattice *L*(*H*) of closed subspaces of a Hilbert space *H* is orthomodular. *L*(*H*) is modular iff *H* is finite dimensional.

## 2.2 The lattice *L(H)* of the projectors on the Hilbert space *H*

The quantum-probabilistic formalism, as developed by von Neumann [1932], assumes that each physical system is associated with a (separable) Hilbert space *H*, the unit vectors of which correspond to possible physical states of the system.
Each "observable" real-valued random quantity is represented by a self-adjoint operator *A* on H, the spectrum of which is the set of possible values of *A*.
If *u* is a unit vector in the domain of *A*, representing a state, then the expected value of the observable represented by *A* in this state is given by the scalar product $<Au,u>$.
The observables represented by two operators *A* and *B* are commensurable iff *A* and *B* commute, i.e., $AB = BA$.
As stressed by von Neumann, the {0,1}-valued observables may be regarded as encoding propositions about (properties of) the state of the system. It is not difficult to show that a self-adjoint operator *P* with spectrum contained in the two-point set {0,1} must be a projection; i.e., $P^2 = P$. Such operators are in one-to-one correspondence with the closed subspaces of H. Indeed, if *P* is a projection, its range is closed, and any closed subspace is the range of a unique projection. If *u* is any unit vector, then the scalar product $<Pu,u> = ||Pu||^2$ is the expected value of the corresponding observable in the state represented by *u*. Since this is {0,1}-valued, we can interpret this as the *probability* that a measurement of the observable will produce the "affirmative" answer 1. In particular, the affirmative answer will have probability 1 if and only if $Pu = u$; that is, *u* lies in the range of *P*.
Let's examine the algebraic structure of projections. Ordered by set-inclusion, the closed subspaces of *H* form a complete lattice, in which the meet (greatest lower bound) of a set of subspaces is their intersection, while their join (least upper bound) is the closed span of their union.
In order to get easily acquainted with the above algebraic structure, we will discuss some lattice properties in the context of Hilbert space *H*.

**i) Orthocomplements of closed subsets of *H*.**

If *S* is a subset of *H*, the set of vectors orthogonal to *S* is defined by:

$$S^{\perp} = \{x \in H | \langle x, s \rangle = 0 \quad \forall s \in S\}$$

$S^{\perp}$ is a closed subspace of *H* and so is itself a Hilbert space.
If *V* is a closed subspace of *H*, then $V^{\perp}$ is called the orthogonal complement of *V*.
In fact, every $x \in H$ can then be written uniquely as $x = v + w$ with $v \in V$ and $w \in V^{\perp}$.
Therefore, *H* is the internal direct sum of $V$ and $V^{\perp}$, namely: $H = V \oplus V^{\perp}$.
The linear operator $P_V : H \to H$ which maps *x* to *v* is called the orthogonal projection onto *V*.

**Theorem** 2.1

The orthogonal projection $P_V$ is a self-adjoint linear operator of $H$ of norm $\leq 1$ with the property: ${P_V}^2 = P_V$. Moreover, any self-adjoint linear operator *E* such that $E^2 = E$ is of the form $P_V$, where *V* is the range of *E*.

In view of the above-mentioned one-to-one correspondence between closed subspaces and projections, we may impose upon the set *L*(*H*) the structure of a complete orthocomplemented lattice, defining the order relation between two projectors $P_V$ and $Q_U$ :

$$P_V \le Q_U \text{ iff } V \subseteq U$$

and the complement of $P_V$ as $P_{V'} = \mathrm{I} - P_V$, so that $V' = V^{\perp}$.
The orthocomplement of a projector $P_V$ is then defined as:

$$P_V{}^{\perp} = 1 - P_v .$$

**ii) The operations meet and join in the lattice of projectors**
It is easy to verify that the order relation $P_V \le Q_U$ holds just in the case:

$$P_V Q_U = Q_U P_V = P_V$$

The above relation expresses the fact that the composition of two projectors with different ranges, projects into the smallest of the two ranges, which is in fact the set intersection of them.
In other words, this requires that $P_V$ and $Q_U$ commute: $[P_V, Q_U] = 0$.
Therefore, owing to the commutative property of the meet: $P_V \wedge Q_U = Q_U \wedge P_V$, we can define the meet in this case as the product:

$$P_V \wedge Q_U = Q_U \cdot P_V \qquad (2.1)$$

In the general case the previous definition of meet must be generalized as follows:

$$P_V \wedge Q_U = \lim_{n\to\infty} (Q_U \cdot P_V)^n \qquad (2.1\text{bis})$$

The definition of the join, which in this case pertains to closed subspaces, can be recovered from the set theoretical definition of the union, that is:
$V \cup U = V \oplus U - V \cap U$
By recalling that *U* and *V* are the ranges of the projectors *P* and *Q*, the definition of the join if the projectors commute is:

$$P_V \vee Q_U = P_V + Q_U - P_V \wedge Q_U = P_V + Q_U - P_V \cdot Q_U \qquad (2.2)$$

The properties of the orthocomplement of a projector $P_V$, defined in the previous Subsection, are:

Orthogonality relation: $P_V{}^{\perp} P_V = (\mathrm{I} - P_V)P = P - P^2 = P - P = 0$

Involution law $(P_V{}^{\perp})^{\perp} = P_V$

Order reversing: if $P_V \leq Q_U$ then $Q_U{}^{\perp} \leq P_V{}^{\perp}$

Having defined above the meet and join operations for projectors, also the complement laws hold:

$$P_V{}^{\perp} \wedge P_V = P_v{}^{\perp} P_V = 0$$
$$P_V{}^{\perp} \vee P_V = \mathrm{I}$$

**iii) Non-distributivity**

It is easy to show that when two projectors *P* and *Q* do not commute, the lattice is non-distributive. As an example, we consider the distributive law:

$$a \wedge (b \vee c) = (a \wedge b) \vee (a \wedge c) \tag{2.3}$$

in the particolar case with: $a = P$, $b = Q$, $c = R = Q^{\perp}$ and $P \wedge Q = P \wedge Q^{\perp} = 0$ In this case, the left hand side of Eq.(2.3) gives:

$$P \wedge (Q \vee Q^{\perp}) = P \wedge \mathrm{I} = P \tag{2.4}$$

while the right hand side gives:

$$(P \wedge Q) \vee (P \wedge Q^{\perp}) = 0 \vee 0 = 0 \tag{2.5}$$

If *Q* and *P* commute, it follows $P = 0$ and distributivity holds.

**iv) Boolean sublattices**

Let us now consider the case when the projectors *P* an *Q* commute.

The orthocomplement of *P*, denoted by $P^{\perp}$, is defined as. It holds: $P \cdot P^{\perp} = P \cdot (\mathrm{I} - P) = P - P^2 = P - P = 0$, and $P + P^{\perp} = \mathrm{I}$.

Therefore, in terms of *P* and its orthocomplement $P^{\perp}$, Eqs.(2.1) and (2.2) become, respectively:

$$P \wedge P^{\perp} = 0 \tag{2.6}$$

$$P \vee P^{\perp} = \mathrm{I} \tag{2.7}$$

Of course, the same relations hold for the projector *Q* and its orthocomplement $Q^{\perp}$:

$$Q \wedge Q^{\perp} = 0$$

$Q \vee Q^{\perp} = \mathrm{I}$

Each pair of commuting projectors like *P* and *Q* discussed above, together with their respective orthocomplements, form a Boolean (distributive orthocomplemented) sub-lattice of the lattice *L* (*H*) of projectors on *H*.

In a more formal way, this is expressed by the following lemma:

**Lemma 2.1**

*Let* P *and* Q *be projection operators on the Hilbert space H. The following are equivalent:*

a. $PQ = QP$
b. The sublattice of $L(H)$ generated by *P*, *Q*, $P^{\perp}$ and $Q^{\perp}$ is Boolean
c. *P*, *Q* lie in a common Boolean sub-ortholattice of $L(H)$.

Commuting observables are associated to simultaneously measurable physical quantities. As projectors are a particular kind of observables, we conclude that the members of a Boolean sub-ortholattice of $L(H)$ correspond to simultaneously testable physical quantities. This suggests that we can maintain a classical logical interpretation of the meet, join and orthocomplement as applied to commuting projections.

In the case discussed above, the Boolean sub-lattice generated by two commuting projectors *P* and *Q* and their orthocomplements $P^{\perp}$ and $Q^{\perp}$, is represented by an Hasse diagram whose form reminds the Benzene molecule. See Fig. 2.1.

**Fig. 2.1**

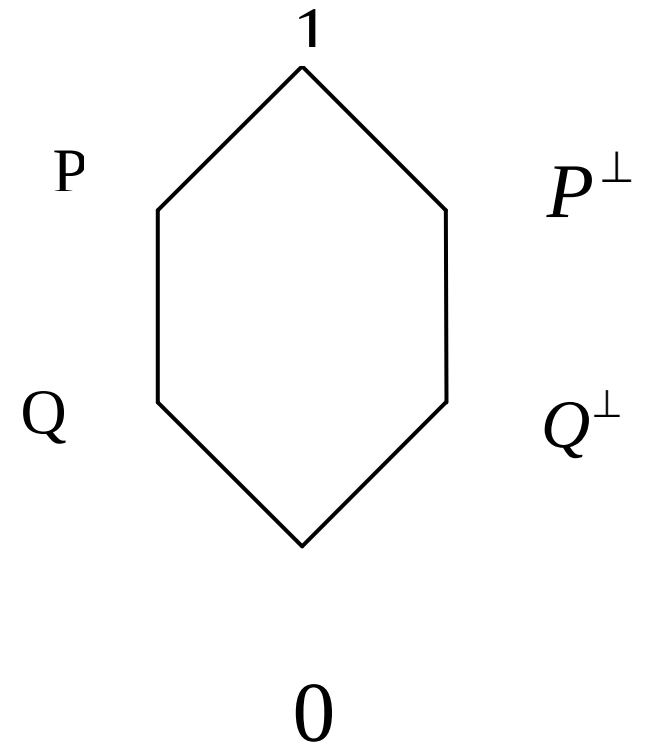

**2.3 The lattice L($C^2$) of the two one-dimensional projectors on $C^2$**

Let us consider the two one-dimensional projectors of $C^2$:

$$P_0 \equiv |0\rangle\langle 0| = \begin{pmatrix} 1 & 0 \\ 0 & 0 \end{pmatrix} \quad \text{and} \quad P_1 \equiv |1\rangle\langle 1| = \begin{pmatrix} 0 & 0 \\ 0 & 1 \end{pmatrix} \qquad (2.8)$$

They are Hermitian operators (observables), which are:

i) Idempotent:

$P_0{}^2 = P_0 \quad P_1{}^2 = P_1$

ii) Orthogonal to each other:

$$P_0 P_1 = P_0 P_1 = 0 \tag{2.9}$$

iii) Therefore, their commutator trivially vanishes:

$[P_0, P_1] = 0$

iv) They satisfy the completeness relation:

$$P_0 + P_1 = \mathrm{I} \tag{2.10}$$

The commutativity in iii) allows to define the meet operation as:

$$P_0 \wedge P_1 = P_0 P_1 \tag{2.11}$$

and the join operation as:

$$P_0 \vee P_1 = P_0 + P_1 - P_0 P_1 \tag{2.12}$$

Because of ii) we get, from ( 2.2):

$$P_0 \wedge P_1 = 0 \tag{2.13}$$

From Eq.(2.12), using ii) and the completeness relation in iv), we get:

$$P_0 \vee P_1 = \mathrm{I} \tag{2.14}$$

From iv) we can say that $P_1$ is the complement of $P_0$ (and *viceversa*, $P_0$ is the complement of $P_1$):

$$P_1 = (\mathrm{I} - P_0) \tag{2.15}$$

Moreover, because of ii), we say that $P_1$ is the orthocomplement of $P_0$ (and *viceversa*) which is generally indicated with the symbol "$\perp$":

$$P_1 = P_0{}^{\perp} \tag{2.16}$$

Therefore we get:

$$P_0{}^{\perp} = (\mathrm{I} - P_0) \tag{2.17}$$

$$P_0 \wedge P_0^{\perp} = 0 \qquad (2.18)$$

$$P_0 \vee P_0^{\perp} = \mathrm{I} \qquad (2.19)$$

The unary operation $\perp$ is involutive:

$$P_0^{\perp\perp} = P_1^{\perp} = P_0 \qquad (2.20)$$

This lattice is orthocomplemented, and distributive. Therefore it cannot be the algebraic structure of a Quantum logic aimed to describe the qubit, which is in fact a quantum state of $C^2$. The Boolean lattice discussed in this example is represented by the Hasse diagram given in Fig. 2.2.

**Fig. 2.2**

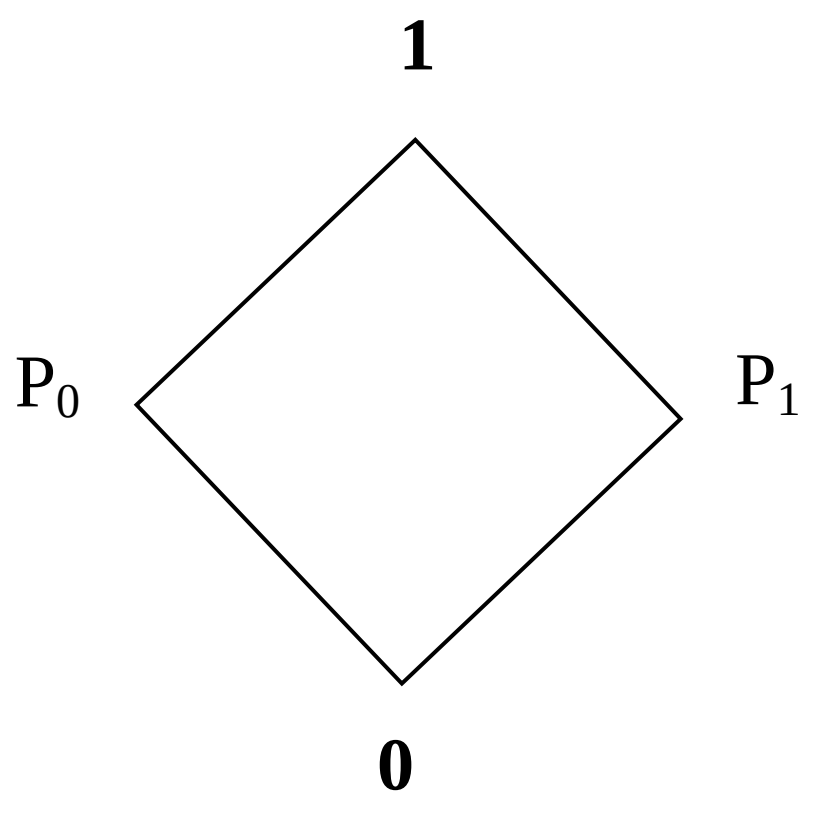

## 3. Some basics of Quantum Information and Quantum Computation

In this Section, we will give a few notions about quantum computers, and quantum information.

### 3.1 Quantum computers

The qubit is the unit of quantum information, the quantum analogous of the classical bit. But while a bit can only be either 0 or 1, the qubit can be a suitable linear superposition of 0 and 1, with complex coefficients called probability amplitudes, such that the sum of their squared modules is one (probabilities sum up to one).

In a traditional computer, information is encoded in a series of bits, which are manipulated by Boolean logic gates arranged in succession to produce an end result. Similarly, a quantum computer manipulates qubits by executing a series of quantum gates, each a unitary transformation acting on a single qubit or pair of qubits. In applying these gates in succession, a quantum computer can perform a complicated unitary transformation to a set of qubits in some initial state. The qubits can then be measured, with this measurement serving as the final computational result. This similarity in calculation between a classical and quantum computer affords that, in theory, a classical computer can accurately simulate a quantum computer. In other words, a classical computer would be able to do anything a quantum computer can.

However, although a classical computer can theoretically simulate a quantum computer, it is incredibly inefficient, so much so that a classical computer is effectively incapable of performing many tasks that a quantum computer could perform with ease. The simulation of a quantum computer on a classical one is a computationally hard problem because the correlations among qubits are qualitatively different from correlations among classical bits. For a classical computer, to simulate a quantum system would take an exponentially longer time than a quantum computer. Richard Feynman (Feynman 1982) was among the first to recognize the potential in quantum superposition for solving such problems very much faster.

Feynman asserted that a quantum computer could function as a kind of simulator for quantum physics, potentially opening the doors to many discoveries in the field.

Soon after realizing the potential of such immense computing power, the goal was on to find some useful application. Peter Shor provided such an application by devising the first quantum algorithm (Shor 1994).

Shor's algorithm utilizes the power of quantum superposition to rapidly factor very large numbers (on the order of 10200 digits and greater) in a matter of seconds. Hence, the premier application of a quantum computer capable of implementing this algorithm lies in the field of encryption, where the most common and best encryption code, (RSA), relies on the difficulty of factoring very large numbers into their primes.

Another very important quantum algorithm, although of a different kind, is Grover's algorithm (Grover 1996), which is meant to search unsorted database in the square root of the time it would take a classical computer to find an entry.

### 3. 2 The qubit

The qubit is the unit of quantum information. It is the quantum analog of the classical bit $\{0,1\}$, with the difference that the qubit can be also in a linear superposition of 0 and 1 at the same time.
We briefly introduce a few notions, which will be useful for understanding what follows. Every quantum system is associated to an Hilbert space. Each vector in the Hilbert space (apart from the origin) corresponds to a pure quantum state. In addition, two vectors that differ only by a nonzero complex scalar correspond to the same state (in other words, each pure state is a *ray* in the Hilbert space). Alternatively, many authors choose to only consider normalized vectors (vectors of norm 1) as corresponding to quantum states. In this case, the set of all pure states corresponds to the unit sphere of a Hilbert space, with the proviso that two normalized vectors correspond to the same state if they differ only by a complex scalar of absolute value 1, which is called the phase factor.
The qubit is a unit vector in the 2-dimensional complex Hilbert space $C^2$.
The expression of the qubit is:

$$|Q\rangle = \alpha|0\rangle + \beta|1\rangle \tag{3.1}$$

Where the symbol $|\ \rangle$ is the ket vector in the (bra-ket) Dirac notation in the Hilbert space.
The two kets:

$$|0\rangle = \begin{pmatrix}1\\0\end{pmatrix},\ |1\rangle = \begin{pmatrix}0\\1\end{pmatrix} \tag{3.2}$$

Form the orthonormal basis of the Hilbert space $C^2$, called the computational basis.
The coefficients $\alpha, \beta$ are complex numbers called probability amplitudes, with the constraint:

$$|\alpha|^2 + |\beta|^2 = 1 \tag{3.3}$$

to make probabilities sum up to one. (Any quantum measurement of the qubit, either gives $|0\rangle$ with probability $|\alpha|^2$, or $|1\rangle$ with probability $|\beta|^2$).
The qubit, in the standard vector notation is:

$$|Q\rangle = \begin{pmatrix}\alpha\\\beta\end{pmatrix} \tag{3.4}$$

**3. 3 The Bloch sphere**
The geometrical representation of the qubit corresponds to the Bloch sphere, which is the sphere $S^2$ with unit radius. Formally, the qubit, which is a point of a two-dimensional vector space with complex coefficients, would have four degrees of freedom, but the *completeness condition* (see next Subsection) and the impossibility to observe the *phase factor* (see next Subsection) reduce the number of degrees of freedom to two. Then, a qubit can be represented as a point on the surface of a sphere with unit radius.
The Bloch sphere is defined by:

$$S^2 = \left\{ x_i \in R^3 \middle| \sum_{i=1}^{3} x_i^2 = 1 \right\} \tag{3.5}$$

Any generic 1-qubit state in (3.1) can be rewritten as:

$$|Q\rangle = \cos\frac{\vartheta}{2}|0\rangle + e^{i\phi}\sin\frac{\vartheta}{2}|1\rangle \tag{3.6}$$

where the Euler angles $\vartheta$ and $\phi$ define a point on the unit sphere $S^2$. Thus, any 1-qubit state can be visualized as a point on the Bloch sphere, the two basis states being the poles. See Fig.3.1.
We remind that any transformation on a qubit during a computational process is a reversible operation, as it is performed by a unitary operator *U*:

$$U^{\dagger}U = \mathrm{I}\,. \tag{3.7}$$

where $U^{\dagger}$ is the hermitian conjugate of $U$.
This can be seen geometrically as follows.
Any unitary $2\times 2$ matrix $U$ on the 2-dimensional complex Hilbert space $C^2$ (which is an element of the group SU(2)) multiplied by a global phase factor):

$$U = e^{i\phi}\begin{pmatrix} \alpha & \beta \\ -\beta^* & \alpha^* \end{pmatrix} \tag{3.8}$$

(where $\alpha^*$ is the complex conjugate of $\alpha$) and $|\alpha|^2 + |\beta|^2 = 1$), can be rewritten in terms of a rotation of the Bloch sphere:

$$U_2 = e^{i\phi}R_{\hat{n}}(\theta) \tag{3.9}$$

where $R_{\hat{n}}(\theta)$ in Eq. (3.9) is the rotation matrix of the Bloch sphere by an angle $\theta$ about an axis $\hat{n}$.

**Fig.3.1 The Bloch sphere**

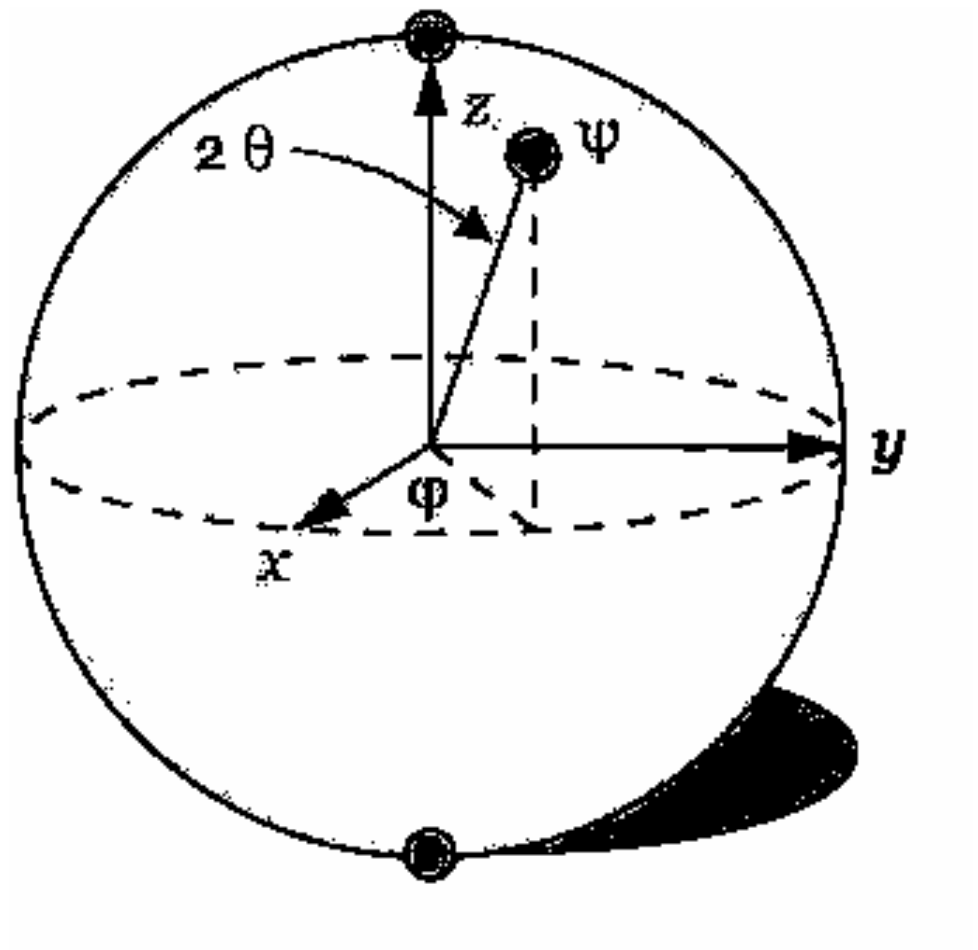

## 3.4 Projective measurement of the qubit

Let us consider a qubit in the superposed state in Eq.(3.1):

$|Q\rangle = \alpha|0\rangle + \beta|1\rangle$

The standard quantum measurement of the qubit $|Q\rangle$ gives either $|0\rangle$ with probability $|\alpha|^2$, or $|1\rangle$ with probability $|\beta|^2$.

This is achieved by the use of projector operators.

Let us consider a general superposed quantum state:

$$|\psi\rangle = \sum_{i=1}^{n} c_i |\psi_i\rangle \qquad (3.10)$$

in the Hilbert space $C^n$, with:

$$\sum_{i=1}^{n} |c_i|^2 = 1. \qquad (3.11)$$

The probability $p(i)$ of finding the state $|\psi\rangle$ in one of the basis states $|\psi_i\rangle$ is, after a projective measurement:

$$p(i) = \left| P_i |\psi\rangle \right|^2 \qquad (3.12)$$

After the measurement, the state $|\psi\rangle$ has "collapsed" to the state:

$$|\psi\rangle' = \frac{P_i|\psi\rangle}{\sqrt{p(i)}}. \qquad (3.13)$$

The n projectors $P_i \quad i = (1...n)$ where *n* is the dimension of the Hilbert space *H*, are orthogonal and idempotent:

$$P_i P_j = \delta_{ij} P_i \tag{3.14}$$

and satisfy the *completeness relation*:

$$\sum_{i=1}^{n} P_i = \mathrm{I} \tag{3.15}$$

which ensures that the sum of the probabilities is equal to 1.
In the case of the qubit, $i = (0,1)$ , we have the two one-dimensional projectors:

$$P_0 = \begin{pmatrix} 1 & 0 \\ 0 & 0 \end{pmatrix} \qquad P_1 = \begin{pmatrix} 0 & 0 \\ 0 & 1 \end{pmatrix} \tag{3.16}$$

for which it holds:

$$P_0^{\dagger} = P_0 \qquad P_1^{\dagger} = P_1 \tag{3.17}$$

$$P_0 P_1 = P_1 P_0 = 0 \tag{3.18}$$

$$P_0^2 = P_0 , \quad P_1^2 = P_1 \tag{3.19}$$

$$P_0 + P_1 = \mathrm{I} \tag{3.20}$$

The basis states $|0\rangle$ and $|1\rangle$ are eigenvectors of $P_0$ and $P_1$, respectively, with eigenvalue 1:

$$P_0|0\rangle = |0\rangle , \qquad P_1|1\rangle = |1\rangle \tag{3.21}$$

On the other hand, $|0\rangle$ and $|1\rangle$ are, respectively, eigenvectors of $P_1$ and $P_0$ with eigenvalue 0:

$$P_0|1\rangle = 0 \qquad P_1|0\rangle = 0 \tag{3.22}$$

from which it follows that the action of the projectors $P_0$ and $P_1$ on the superposed state (3.1) is, respectively:

$$P_0|Q\rangle = \alpha|0\rangle , \qquad P_1|Q\rangle = \beta|1\rangle \tag{3.23}$$

The probability of finding the qubit state (3.1) in the state $|0\rangle$ (for example) is:

$$p(0) = \left| P_0 \left| Q \right\rangle \right|^2 = \left| \alpha \left| 0 \right\rangle \right|^2 = \left| \alpha \right|^2 \tag{3.24}$$

After the measurement, the qubit in Eq.(3.1) has “collapsed" to the (normalized) state:

$$\left| Q \right\rangle' = \frac{P_0 \left| Q \right\rangle}{\sqrt{p(0)}} = \frac{\alpha \left| 0 \right\rangle}{\left| \alpha \right|} = \left| 0 \right\rangle \tag{3.25}$$

where $\frac{\alpha}{|\alpha|}$ in (3.25) is the *phase factor* which does not influence the state of the system and therefore can be neglected.

Then, a lot of quantum information that was encoded in the qubit is made hidden by the standard quantum measurement. As a projector is not a unitary transformation, the standard quantum measurement is not a reversible operation.

# 4. Quantum Metalanguage and Quantum Control

In this Section, we will briefly introduce control theory, which is strictly related to metalanguages.

We remind that in orthodox Quantum logic, propositions are interpreted as projectors $P_i$, while, in the logic *Lq* discussed in this thesis, propositions are interpreted as weak measurements $P_i O_i \equiv \lambda_i P_i$, (where $i = 0,1$) in the 2-dimensional complex Hilbert space $C^2$. Therefore, in *Lq*, propositions will be used to “control” the quantum system (the quantum computer, in particular) instead of asking yes/no questions about it, as in the case of Birkhoff-von Neumann Quantum logic.

We will refer to the metalanguage as a control over the object-language, in computer science (with particular attention to quantum computing). In the scheme of control an important role is played by representation of the controlled system in the controlling device. (Turchin, *et al.*,1996). When the controlled system in some object-language *L*, the language used to make the representations in the controlling system, let it be *L'* is referred to as a metalanguage. Control is the operation mode of a *control system* which includes two subsystems: the controlling (a controller) *C*, and the controlled, *S*. They interact, but there is a difference between the action of *C* on *S*, and the action of *S* on *C*. The controller *C* may change the state of the controlled system *S* in any way including the destruction of *S*.

The action of *S* on *C* is the formation of a *perception* of system S in the controller *C*. We see in the controller an *agent* which is responsible for its actions, and a *representation* of the controlled system, which is an object whose states we identify with perceptions. The relation between representation and agent is described as a flow of *information*: the actions of the agent depend on this flow. Thus the action of *S* on *C* is limited, in its effect, by changing only *S*'s representation in *C*, not the rest of the system. Thus the asymmetry of the control relation: *C* controls *S*, but S does not control *C*. The action of *S* on *C* is "filtered" through the representation: its effect on *C* cannot be greater than allowed by the changing state of the representation. See Fig. 4.1.

**Fig. 4.1**

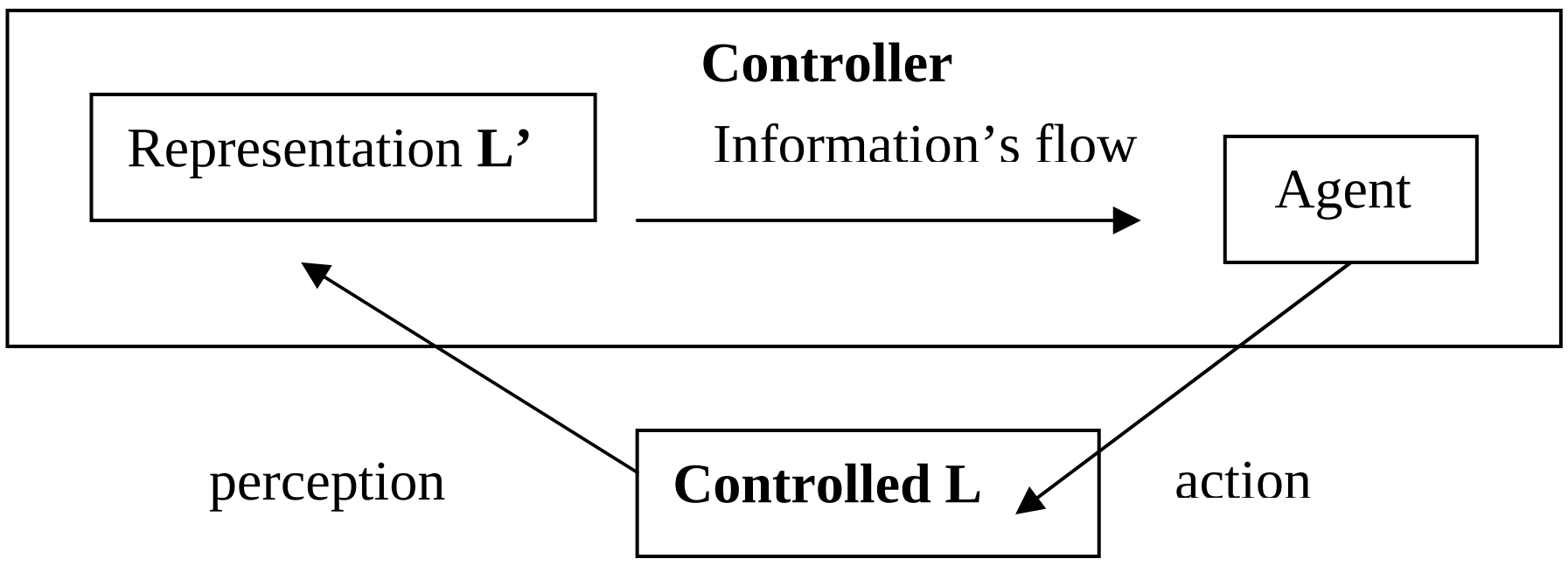

This resembles Sambin’s schema concerning the interrelations between object-language and metalanguage (Sambin 2008). See Fig. 4.2.

**Fig. 4.2**

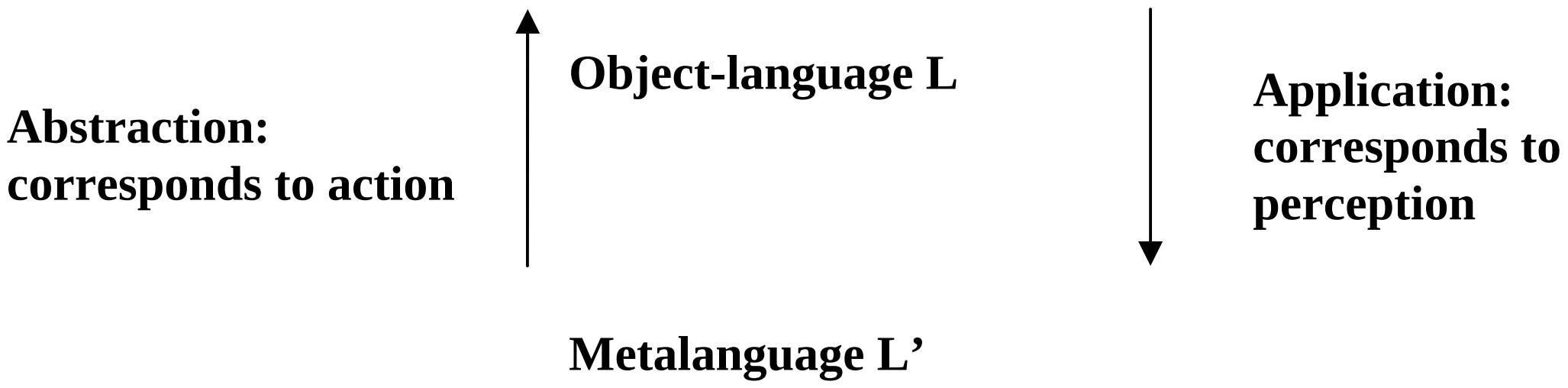

Even though the relation of control is asymmetric, it includes a closed loop. Looked from the controller, the loop starts with its action and is followed by a perception, which is an action in the opposite direction: from the controlled to the controller. This aspect of control relation is known as feedback.
In a system where a transformation occurs, there are inputs and outputs. The inputs are the result of the environment's influence on the system, and the outputs are the influence of the system on the environment. Input and output are separated by a duration of time, as in before and after, or past and present.
In every feedback loop, information about the result of a transformation or an action is sent back to the input of the system in the form of input data. See Fig. 4.3.

**Fig. 4.3**

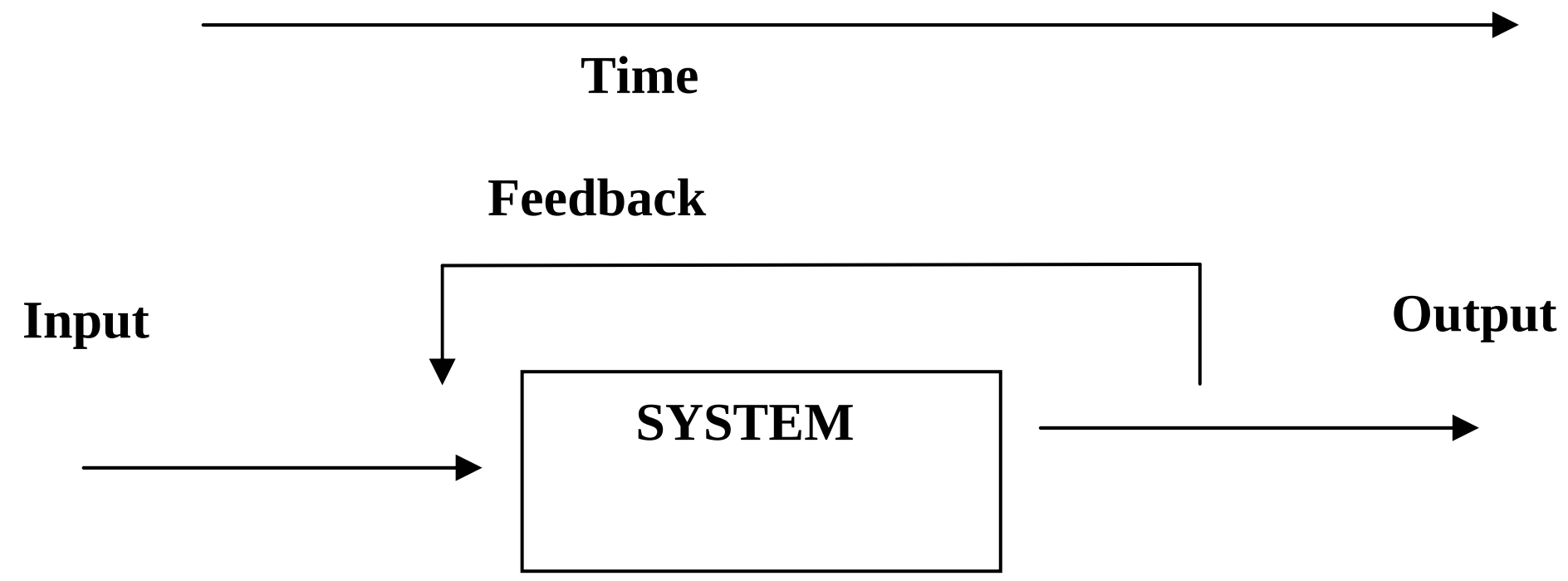

These new data facilitate and accelerate the transformation in the same direction as the preceding results, as they give a positive feedback - their effects are cumulative. If the new data produce a result in the opposite direction to previous results, they give a negative feedback - their effects stabilize the system. In the first case there is exponential growth or decline; in the second there is maintenance of the equilibrium.
In terms of metalanguage and object-language, there is a feedback from the object-language to the metalanguage. If the feedback is a positive loop, the logical theory "explodes". It would correspond to an object-language containing its own semantics. We need therefore a control in terms of negative loops. This is very difficult when the object-language is quantum, as there is no possibility of controlling the positive feedback loops by a classical control (a classical metalanguage). We need therefore a quantum control (a quantum metalanguage).

## 4.1 Quantum control

The characteristic of a quantum system, from the perspective of control theory, is that measurement causes its state to change. This property is referred to as *back action*, an inevitable consequence of the Heisenberg uncertainty principle. In the classical world, it is (in principle) possible to perform a non-invasive measurement of the system as part of the control process.
The controller continuously adjusts the system dynamics based on the outcome of these non-invasive measurements. However, in Quantum Mechanics, the very act of observation causes the state of the system to wander even farther away from the control target. In this sense, the effect of back action is to introduce a type of quantum noise into the measurement of the system's state. Fortunately, it is possible to think of quantum control in a context that is similar to classical control, but with an additional noise source due to the measurement back action. Therefore, it is theoretically possible to adapt existing control knowledge to make it apply in quantum settings.
Observation plays a central role in feedback control: there can be no feedback without checking up on the state of the system to ensure that it is doing the right thing. However, observation induces unavoidable randomness due to the projection postulate. Fortunately, quantum systems can be measured in many different ways, some of which are less intrusive than others. In the worst case scenario, a projective measurement instantaneously and completely collapses the quantum system into an eigenstate of the measurement operator. This drastic, sudden change in the quantum state appears to be a poor candidate for feedback control. Instead, weak measurements of the quantum system result in less drastic fluctuations and only gradually reduce the system to a measurement eigenstate. Most importantly, weak measurements can be performed continuously and provides the backbone of quantum feedback control.

## 4.2 Weak measurements

We will give some basics of weak measurements, which are important in quantum control.

In quantum theory, the mean value of a certain observable *A* in a (pure) quantum state $|\Psi\rangle$ is defined by :

$$\langle A \rangle_{\Psi} \equiv \langle \Psi | A | \Psi \rangle \qquad (4.1)$$

The observable *A* is an Hermitian operator on the Hilbert space of dimension n, satisfying the eigenvalue equation:

$$A|a_i\rangle = a_i|a_i\rangle \quad (i = 1,2\ldots\ldots n) \qquad (4.2)$$

The pure state $|\Psi\rangle$ in (4.1) can always be expressed by a linear superposition of the eigenstates $|a_i\rangle$ (relative to the eigenvalues $a_i$) of the observable *A*:

$$|\Psi\rangle = \sum_i \lambda_i |a_i\rangle \qquad \text{with: } \lambda_i \in C\,; \quad \sum_i |\lambda_i|^2 \qquad (4.3)$$

The notion of the mean value in Eq.(4.1) has been extended, by Aharonov, Albert and Vaidmal (see Aharonov *et al.*, 1988) to the (normalized) expression:

$$A_W = \frac{\langle f|A|i\rangle}{\langle f|i\rangle} \qquad (4.4)$$

Where the two pure states $|i\rangle$, $|f\rangle$ play the roles of the prepared initial state and post selected final state, respectively.

The statistical interpretation relies upon the concept of weak measurement. In a single weak measurement, decoherence is chosen asymptotically small, that is the coupling between the measured state and the meter is assumed asymptotically weak. The mean value (4.4) is called the (complex) weak value.

Notice that $A_W$, for a particular kind of *A*, is the equivalent of the logical assertion-degree.
To show that, we consider a projective measure (which is an observables) $P_i$, and we recall that we deal with non-hermitian operators defined as:

$$O_i \equiv \lambda_i P_i \tag{4.5}$$

The expectation value is:

$$\langle \Psi | \lambda_i P_i | \Psi \rangle = \lambda_i \langle P_i \rangle_\Psi \tag{4.6}$$

where $\lambda_i |\Psi\rangle$ plays the role of the initial state $|i\rangle$ and $|\Psi\rangle$ plays the role of the final state $|f\rangle$.
The post selection consists in the following procedure. One performs the von Neumann (ideal) measurement of the hermitian projector $|f\rangle\langle f|$, then one includes the case if the outcome is 1, and discards it if the outcome is 0. The rate of post selection is: $|\langle f \| i \rangle|^2$.
A weak measurement can be implemented through indirect observation of the quantum system where it is first entangled with a disposable meter followed by a projective measurement on only the meter. This process still involves back action because of the entanglement; however, for only moderate coupling between the system and meter, the back action is much less severe and leads to continuous, stochastic evolution.

## 5. A brief review of Quantum Computational Logic

While for a number of years, starting from the pioneering contribution of Birkhoff and Von Neumann, the study of Quantum logic was consisting in a complex mixture of algebraic and philosophical arguments, in the last times emerged the need for a closer relationship between Quantum logic and quantum computing. The advances in the latter domain, originated in the eighties from the seminal work of Benioff, Feynman, and Deutsch (see Benioff, 1980; Feynman, 1982, 1986; Deutsch, 1985), became impetuous after Shor (the original paper dates to 1994, but an expanded, easier to find, version is Shor, 1997) and Grover (1996) introduced their famous quantum algorithms, opening the way for the design of a new generation of quantum computers, much more powerful than the actual digital ones (for a widely used textbook on quantum computing see, for instance, Nielsen and Chuang, 2000; more recent treatises are Mermin, 2007; Kaye *et al.*, 2007; Yanofsky and Mannucci, 2008; a textbook describing also the physical implementations of quantum computers is Nakahara and Ohmi, 2008). This new situation called for the search of a ‘logic of quantum computation’, to be based on Quantum logic, and devoted to clarify in what sense the processes occurring within a quantum computer should be considered as implementing some sort of ‘Quantum logical reasoning’.

The latter appears as a very difficult enterprise. Namely, so far all logical reasoning, not only of mathematicians or of logicians, but even of theoreticians working on Quantum Mechanics, has been done by resorting almost exclusively to traditional classical logic. Of course, many non-classical and quantum logics have been so far introduced, but few of them found important applications in the studies on foundations of mathematics or in the description of some kind of non-classical reasoning. This circumstance can be easily explained if we take into account that proof theory and model theory are much more difficult (if not impossible) to build in non-classical logics than in traditional classical logic (from time to time this fact has been stressed by some mathematicians; see, for instance, Dieudonné, 1978). However, quantum computers are becoming reality and, in the same way in which classical computers have been described as ‘machines implementing classical logical reasoning’, we are compelled to describe quantum computers as implementing some kind of ‘quantum computational logic’.

The building of the latter can be performed by following two different, but complementary, approaches. The former, just called *quantum computational logic* (for a survey see Dalla Chiara *et al.*, 2003; a synthetic review is contained in Dalla Chiara *et al.*, 2008), starts from the basic elements of quantum computation – qubits, qumix, quregisters, quantum gates – to build a logical interpretation of them. The latter, instead, starts from the theory of automata, viewed as the primeval entities of a computation, and tries to generalize it in such a way as to introduce *quantum automata* fulfilling the rules of Quantum logic (see, for instance, Ying, 2005). Both approaches run into difficulties whose origin, in their essence, is the same of the difficulties which already troubled Birkhoff and Von Neumann. We will limit our description of these approaches only to quantum computational logic, as the theory of quantum automata is still in a rudimentary stage.

In order to understand the achievements reached within this approach, it is convenient to shortly recall some basic notions about quantum computation. In this regard we will work within a two-dimensional Hilbert space $C^2$ or *n*-times tensor products of $C^2$, which will be denoted by $\otimes^n C^2$. Let us start with the first definition, the one of *qubit*.

**Definition 5.1**
*A* qubit *is a unit vector of* $C^2$.
If we denote by $|0\rangle$ and $|1\rangle$ the canonical orthonormal basis for $C^2$, any qubit $|\psi\rangle$ can be written under the form $|\psi\rangle = a_0|0\rangle + a_1|1\rangle$, with $|a_0|^2 + |a_1|^2 = 1$.

**Definition 5.2**
*An n-*qubit *system, or n-quregister, is a unit vector of* $\otimes^n C^2$.
A *n*-quregister can be expressed in terms of a linear combination, whose squared coefficients sum up to 1, over the components of the tensor product of *n* qubits. In turn, each component of this product can be set into biunivocal correspondence with a string of *n* binary numbers $|x_1,\ldots,x_n\rangle$. Let us now introduce two projection operators $P_1^{(n)}$ and $P_0^{(n)}$, acting on *n*-quregisters in such a way that their span is constituted, respectively, by the linear combination over all components of the tensor product corresponding to strings $|x_1,\ldots,x_n\rangle$ with $x_n = 1$, and by the linear combination over all components of the tensor product corresponding to strings $|x_1,\ldots,x_n\rangle$ with $x_n = 0$. Let us now recall from standard textbooks the following:

**Definition 5.3**
*A density operator is a positive-semidefinite hermitian operator with unit trace.*
It is then possible to prove the following theorem.

**Theorem 5.4**
*The operators* $k_n P_1^{(n)}$ *and* $k_n P_0^{(n)}$, *where* $k_n = \frac{1}{2^{n-1}}$, *are density operators.*

Let us now denote by $D_n$ the set of all density operators of $\otimes^n C^2$ and let $D = \bigcup_{n=1}^{\infty} D_n$. We can now define a *qumix* as follows:

**Definition 5.5**
*A* qumix *is a density operator in* $D$.
As it is well known, a qumix corresponds to a *mixed state*, which, in general, cannot be expressed through a linear combination of pure states. According to the standard rules of Quantum Mechanics the probability associated to a qumix is defined as follows:

**Definition 5.6**
*For any qumix* $\rho$ *in* $\otimes^n C^2$ *its probability is given by* $p(\rho) = tr\left[P_1^{(n)}\rho\right]$.

Within $C^2$ a qumix can be written under the form:

$$\rho = \frac{1}{2}(I + r_1\sigma_x + r_2\sigma_y + r_3\sigma_z)$$

where $I$ is the unit matrix and $\sigma_x, \sigma_y, \sigma_z$ are the usual Pauli matrices. The real numbers $r_1, r_2, r_3$ fulfil the inequality $r_1^2 + r_2^2 + r_3^2 \leq 1$.

Besides qubits and qumixs, a quantum computation makes use of *logical gates*, identified with unitary operators acting on elements of $\otimes^n C^2$. Here it will suffice to introduce the so-called *Petri-Toffoli gate*, defined below.

**Definition 5.7**

*The* Petri-Toffoli gate $T^{(n,m,1)}$ *is the linear operator*

$$T^{(n,m,1)} : (\otimes^n C^2) \otimes (\otimes^m C^2) \otimes C^2 \to (\otimes^n C^2) \otimes (\otimes^m C^2) \otimes C^2$$

*defined for any element* $|x_1, \ldots, x_n\rangle \otimes |y_1, \ldots, y_m\rangle \otimes |z\rangle$ *of the computational basis of* $\otimes^{n+m+1} C^2$ *through the formula:*

$$T^{(n,m,1)}(|x_1, \ldots, x_n\rangle \otimes |y_1, \ldots, y_m\rangle \otimes |z\rangle) = |x_1, \ldots, x_n\rangle \otimes |y_1, \ldots, y_m\rangle \otimes |x_n y_m \oplus z\rangle$$

*where* $\oplus$ *denotes the sum modulo 2.*

The Petri-Toffoli gate allows to introduce an operation of conjunction, denoted by $AND$ and defined through:

**Definition 5.8**

*For any* $|\varphi\rangle \in \otimes^n C^2$ *and any* $|\psi\rangle \in \otimes^m C^2$ *we have that:*

$$AND(|\varphi\rangle, |\psi\rangle) = T^{(n,m,1)}(|\varphi\rangle \otimes |\psi\rangle \otimes |0\rangle)$$

As regards the *negation*, the domain of quantum computation is characterized by many different possibilities. The simplest one is given by the operator $NOT^{(n)}$ defined by:

**Definition 5.9**

*For any element* $|x_1, \ldots, x_n\rangle$ *of the computational basis of* $\otimes^n C^2$ *we have that:*

$$NOT^{(n)}(|x_1, \ldots, x_n\rangle) = |x_1, \ldots, x_{n-1}, 1 - x_n\rangle$$

In $C^2$ we have immediately that $NOT^{(1)}(|1\rangle) = |0\rangle$ and $NOT^{(1)}(|0\rangle) = |1\rangle$.

Starting from the operators $AND$ and $NOT$ it is then possible to introduce a disjunction operator, denoted by $OR$, through the De Morgan law:

**Definition 5.10**

*For any quregisters* $|\varphi\rangle$ *and* $|\psi\rangle$ *we have that:*

$$OR(|\varphi\rangle, |\psi\rangle) = NOT(AND(NOT(|\varphi\rangle), NOT(|\psi\rangle)))$$

It is, however, to be taken into account that within the domain of quantum computation it is possible to introduce another kind of negation, usually called *the square root of negation*, denoted by $\sqrt{NOT^{(n)}}$ and defined by:

**Definition 5.11**

*For every element* $|x_1, \ldots, x_n\rangle$ *of the computational basis of* $\otimes^n C^2$ *we have that:*

$$\sqrt{NOT^{(n)}}(|x_1,\ldots,x_n\rangle) = |x_1,\ldots,x_{n-1}\rangle \otimes (\frac{1+i}{2}|x_n\rangle + \frac{1-i}{2}|1-x_n\rangle)$$

*where i denotes the imaginary unit.*

In $C^2$ we obtain immediately that:

$$\sqrt{NOT^{(1)}}(|0\rangle) = \frac{1+i}{2}|0\rangle + \frac{1-i}{2}|1\rangle$$

$$\sqrt{NOT^{(1)}}(|1\rangle) = \frac{1-i}{2}|0\rangle + \frac{1+i}{2}|1\rangle$$

Among the most interesting properties of the square root of negation we have the following:

$$\sqrt{NOT^{(n)}}(\sqrt{NOT^{(n)}}(|\psi\rangle)) = NOT^{(n)}(|\psi\rangle)$$

The previous definitions of the quantum gates can be easily extended to qumixs. Without entering into further details on this point, we will limit ourselves to mention that on the set of all qumixs it is possible to introduce two interesting pre-order relationships. The first of them, denoted by the symbol $\leq$ and called *weak pre-order*, is defined by:

**Definition 5.12**

$\rho \leq \sigma$ *iff* $p(\rho) \leq p(\sigma)$

The second relationship, called *strong pre-order* and denoted by $\prec$, is defined by the following two conditions:

**Definition 5.13**

$\rho \prec \sigma$ *iff*: a) $p(\rho) \leq p(\sigma)$ , b) $p(\sqrt{NOT}(\sigma)) \leq p(\sqrt{NOT}(\rho))$.

Now, to build a quantum computational logic, it is possible to introduce a logical language in which each sentence $\alpha$ is interpreted as a suitable qumix, and the logical connectives are interpreted as operations performed by suitable gates. Moreover, logical consequence are interpreted in terms of pre-order relationships between qumixs. In this regard, a possible starting point relies on a *minimal* (*sentential*) *quantum computational language* $L$, containing a privileged atomic sentence $f$ (whose interpretation is the truth-value *Falsity*) and the following three primitive connectives: the *negation* (denoted by $\neg$), the *square root of negation* ($\sqrt{\neg}$), and the *conjunction* ($\wedge$). Latin letters $q,r,\ldots$ will denote atomic sentences while generic sentences will be denoted by Greek letters $\alpha,\beta,\ldots$ It is then possible to define a further connective, the *disjunction* ($\vee$), through the de Morgan law ($\alpha \vee \beta := \neg(\neg\alpha \wedge \neg\beta)$), as well as a privileged atomic sentence $t$ (representing the *Truth*), starting from the negation of $f$ ($t := \neg f$). Moreover, let us denote by $Form^L$ the set of all sentences of $L$.

The language introduced above allows to build a *reversible quantum computational model* (RQC-*model*) through a function $Qum : Form^L \to D$ associating to any sentence of the language a qumix and fulfilling the following rules:

$Qum(\alpha) :=$ *a density operator in* $C^2$ *if* $\alpha$ *is an atomic sentence*

$$
\begin{aligned}
&Qum(\alpha) := P_0^{(1)} \text{ if } \alpha = f \\
&Qum(\alpha) := NOT(Qum(\beta)) \text{ if } \alpha = \neg\beta \\
&Qum(\alpha) := \sqrt{NOT}(Qum(\beta)) \text{ if } \alpha = \sqrt{\neg}\beta \qquad (5.1) \\
&Qum(\alpha) := AND(Qum(\beta), Qum(\gamma)) \text{ if } \alpha = \beta \wedge \gamma
\end{aligned}
$$

The existence of two different kinds of pre-order relationships between qumixs, the strong and the weak ones, allows then to introduce two different kinds of consequence relations, the *strong* and the *weak consequence*. The definitions are reported below.

**Definition 5.14 a) (weak consequence in a RQC-model)**
*A sentence* $\beta$ *is a* weak consequence in a RQC-model *of a sentence* $\alpha$ *iff* $Qum(\alpha) \leq Qum(\beta)$.

**Definition 5.14 b) (strong consequence in a RQC-model)**
*A sentence* $\beta$ *is a* strong consequence in a RQC-model *of a sentence* $\alpha$ *iff* $Qum(\alpha) \prec Qum(\beta)$.

The existence of two different kinds of consequence relations is reflected in the possibility of introducing two different kinds of quantum computational logic: the logic $QCL$, which is based on the weak consequence relation, and the logic $\sqrt{\neg}QCL$, which is based on the strong consequence relation. In each one of these two logics it is possible to define the notions of *truth*, *logical consequence*, and *logical truth*. The related definitions are listed in the following.

**Definition 5.15 a) (Weak truth in QCL)**
*A sentence* $\alpha$ *is* weakly true *iff* $\alpha$ *is a weak consequence of* $t$ .

**Definition 5.15 b) (Weak logical consequence in QCL)**
*A sentence* $\beta$ *is a* weak logical consequence *of a sentence* $\alpha$ *iff for any choice of Qum* $\beta$ *is a weak consequence of* $\alpha$.

**Definition 5.15 c) (Logical truth in QCL)**
*A sentence* $\alpha$ *is a* weak logical truth *iff for any choice of Qum* $\alpha$ *is weakly true.*

**Definition 5.16 a) (Strong truth in $\sqrt{\neg}QCL$)**
*A sentence* $\alpha$ *is* strongly true *iff* $\alpha$ *is a strong consequence of* $t$ .

**Definition 5.16 b) (Strong logical consequence in $\sqrt{\neg}QCL$)**
*A sentence* $\beta$ *is a* strong logical consequence *of a sentence* $\alpha$ *iff for any choice of Qum* $\beta$ *is a strong consequence of* $\alpha$.

**Definition 5.16 c) (Logical truth in $\sqrt{\neg}QCL$)**
*A sentence* $\alpha$ *is a* strong logical truth *iff for any choice of Qum* $\alpha$ *is strongly true.*

From these definitions it is possible to see that the logic $\sqrt{\neg}QCL$ is a sub-logic of $QCL$. Namely the existence of a strong consequence relationship implies the existence even of a weak consequence relationship, but the converse does not hold.
It is convenient to list the most interesting rules and logical consequences holding both in $QCL$ and in $\sqrt{\neg}QCL$. In order to shorten the relevant formulae, we will introduce suitable abbreviations. More precisely, the fact that $\beta$ is a logical consequence (strong or weak) of $\alpha$ will be denoted by the symbolic expression $\alpha \Rightarrow \beta$. Moreover, the holding of both relationships $\alpha \Rightarrow \beta$ and $\beta \Rightarrow \alpha$ will be denoted by $\alpha \equiv \beta$. Finally, the formula $\textit{if } \alpha_1 \Rightarrow \beta_1, \ldots, \alpha_n \Rightarrow \beta_n \textit{ then } \gamma \Rightarrow \delta$ will be written by resorting to the usual notation $\dfrac{\alpha_1 \Rightarrow \beta_1, \ldots, \alpha_n \Rightarrow \beta_n}{\gamma \Rightarrow \delta}$.

**Theorem 5.17 (Logical consequences and rules holding in both $\sqrt{\neg}QCL$ and $QCL$)**

$\alpha \Rightarrow \alpha$
(*identity*)

$$\frac{\alpha \Rightarrow \beta, \beta \Rightarrow \gamma}{\alpha \Rightarrow \gamma}$$
(*transitivity*)

$\alpha \equiv \neg\neg\alpha$
(*double negation*)

$$\frac{\alpha \Rightarrow \beta}{\neg\beta \Rightarrow \neg\alpha}$$
(*contraposition for the negation)*

$\sqrt{\neg}\sqrt{\neg}\alpha \equiv \neg\alpha$
(*double square root of negation principle*)

$\neg\sqrt{\neg}\alpha \equiv \sqrt{\neg}\neg\alpha$
(*permutation of the negations*)

$\sqrt{\neg} f \Rightarrow \sqrt{\neg} t$

$\alpha \wedge \beta \equiv \beta \wedge \alpha \quad , \quad \alpha \vee \beta \equiv \beta \vee \alpha$
(*commutativity*)

$\alpha \wedge (\beta \wedge \gamma) \equiv (\alpha \wedge \beta) \wedge \gamma \quad , \quad \alpha \vee (\beta \vee \gamma) \equiv (\alpha \vee \beta) \vee \gamma$
(*associativity*)

$\neg(\alpha \wedge \beta) \equiv \neg\alpha \vee \neg\beta \quad , \quad \neg(\alpha \vee \beta) \equiv \neg\alpha \wedge \neg\beta$
(*de Morgan*)

$\alpha \wedge (\beta \vee \gamma) \Rightarrow (\alpha \wedge \beta) \vee (\alpha \wedge \gamma) \quad , \quad (\alpha \vee \beta) \wedge (\alpha \vee \gamma) \Rightarrow \alpha \vee (\beta \wedge \gamma)$
(*strong distributivity*)

$$\frac{\alpha \Rightarrow \sqrt{\neg}\, t}{f \Rightarrow \alpha}$$
(*Weak Duns Scoto rule*)

The list given above is partial and its complete version can be found in Dalla Chiara *et al.* (2003). Much more interesting, however, is the list of properties *not holding* both in $QCL$ and $\sqrt{\neg}QCL$. It is given by the following:

**Theorem 5.18 (Rules not holding in both $\sqrt{\neg}QCL$ and $QCL$)**

$\alpha \Rightarrow \alpha \wedge \alpha$

$t \Rightarrow \alpha \vee \neg\alpha$
(*excluded middle*)

$t \Rightarrow \neg(\alpha \wedge \neg\alpha)$
(*non contradiction principle*)

$\alpha \vee (\beta \wedge \gamma) \Rightarrow (\alpha \vee \beta) \wedge (\alpha \vee \gamma) \quad , \quad (\alpha \wedge \beta) \vee (\alpha \wedge \gamma) \Rightarrow \alpha \wedge (\beta \vee \gamma)$
(*weak distributivity*)

It is not so easy to interpret these results, as the failure of both the non contradiction principle and the excluded middle law shows that the quantum computational logics are very different from most traditional logics, included the so-called Quantum logic (in which, on the contrary, the strong distributive law fails, while the weak one holds). This seem to call for a relationship with Fuzzy logics. On the other hand, Dalla Chiara *et al.* (2008) proposed to interpret these logics in terms of an *holistic* quantum computational semantics.
One could wonder whether these paradoxical findings are deriving from the choice of qumixs as corresponding to sentences. It can be proved that it is not so. Namely one could introduce a model formulated in terms of qubits, rather than qumixs. By adopting the same symbolic apparatus used to define the RQC-model, we could formulate a *reversible qubit model* by introducing a function $Qub : Form^L \to R$ (here $R$ is the set of quregisters) associating to any sentence of the language a quregister and fulfilling the following rules:

$Qub(\alpha) := a\ qubit\ in\ C^2\ if\ \alpha\ is\ an\ atomic\ sentence$

$Qub(\alpha) := |0\rangle\ if\ \alpha = f$ (5.2)

$Qub(\alpha) := NOT(Qub(\beta))$ *if* $\alpha = \neg\beta$

$Qub(\alpha) := \sqrt{NOT}(Qub(\beta))$ *if* $\alpha = \sqrt{\neg}\beta$

$Qub(\alpha) := AND(Qub(\beta), Qub(\gamma))$ *if* $\alpha = \beta \wedge \gamma$

Within this model it is possible to introduce, like in RQC-model, the notions of consequence (strong and weak), truth, logical consequence, logical truth. It is then possible to prove the following:

**Theorem 5.19**

(i) *$\beta$ is a strong logical consequence of $\alpha$ in the logic $\sqrt{\neg}QCL$ iff $\beta$ is a strong logical consequence of $\alpha$ in the reversible qubit model.*

(ii) *$\beta$ is a weak logical consequence of $\alpha$ in the logic $QCL$ iff $\beta$ is a weak logical consequence of $\alpha$ in the reversible qubit model.*

Thus the models based on qumixs are logically equivalent to the ones based on qubits.

It could be conjectured that the problems troubling quantum computational logics, as well as the traditional Quantum logics, are a consequence of the fact that in all these cases one starts by considering entities, like the qubits, which are given as *already existing, endowed with given properties* which appear somewhat paradoxical if dealt with through the usual logic tools. If, however, we would take into consideration the *formation of these entities*, at least from a logical point of view, perhaps we should note a lesser contrast between the mental processes occurring within a "quantum reasoning" (invoked also by Dalla Chiara *et al.*, 2008) and the laws ruling the new entities (like the qubits), arising from elementary meta-logical operations. Of course, it is very difficult to individuate the physical counterparts (if any) of this new logical approach to the quantum world.

## 6. An introduction to sequent calculus

One way to classify different styles of deduction systems is to look at the form of judgments in the system, i.e., which things may appear as the conclusion of a (sub) proof. The simplest judgment form is used in Hilbert-style deduction systems, where a judgment has the form B, where B is any formula of first-order-logic (or whatever logic the deduction system applies to, e.g., propositional calculus or a higher-order logic). The theorems are those formulae that appear as the concluding judgment in a valid proof. The distinction between formulae and judgments is rather vague in the Hilbert-style system. In the Hilbert-style system (which has a very simple syntax) complete formal proofs tend to get extremely long. Concrete arguments about proofs in such a system almost always appeal to the deduction theorem. This leads to the idea of including the deduction theorem as a formal rule in the system, which happens in natural deduction. In natural deduction, judgments have the form:

$$A_1, A_2, ...., A_n \vdash B$$

where the $A_i$'s and B are again formulae and $n \geq 0$.A judgment consists of a list (possibly empty) of formulae on the left-hand side of a turnstile symbol "$\vdash$", with a single formula on the right-hand side. The theorems are those formulae B such that $\vdash B$ (with an empty left-hand side) is the conclusion of a valid proof.

The standard semantics of a judgment in natural deduction is that it asserts that whenever $A_1$, $A_2$, etc., are all true, B will also be true. The judgments

$A_1, A_2, ...., A_n \vdash B$ and $\vdash (A_1 \wedge ....... \wedge A_n) \to B$

are equivalent in the strong sense that a proof of either one may be extended to a proof of the other.

Sequent calculus is a deduction system for performing reasoning in first order logic (and propositional logic). It was introduced by Gentzen (Gentzen, 1935) initially for classic logic (LK) and then extended to Intuitionist logic (LJ). It is now used for several different logics, among which sub-structural logics, like Linear logic (Girard, 1987) and Basic logic (Sambin *et al.*, 2000).

### 6.1 The sequent

The following notation will be used:

The symbol $\vdash$ known as "turnstile", separates the *assumptions* on the left from the *propositions* on the right. *A* and *B* denote formulae of first-order predicate logic (one may also restrict this to propositional logic) Γ,Δ,Σ, and Π are finite (possibly empty) sequences of formulae, called contexts.

When on the *left* of the $\vdash$, the sequence of formulas is considered *conjunctively* (all assumed to hold at the same time) while on the *right* of the $\vdash$, the sequence of formulas is considered *disjunctively* (at least one of the formulas must hold for any assignment of variables).

We say that the finite list Δ of assertions $B_j$ (*j =1,2,….k*) *follows from* a finite list Γ of assertions $A_i$ (*i=1,2….n*) (or equivalently Γ *yields* Δ) and write: $\Gamma \vdash \Delta$, where $\vdash$ (“yields” or “therefore”) is a metalinguistic link between assertions. $\Gamma$ is said the antecedent, and $\Delta$ the consequent of the sequent.
The sequent $\Gamma \vdash \Delta$ can be rewritten, more extensively, as:

$A_1,..., A_n \vdash B_1,....B_k$

where the $A_i$ and $B_j$ $(i = 1,2...,n)$ $(j = 1,2...,k)$ are assertions.
Either $\Gamma$ or $\Delta$ (or both) can be empty.
If the consequent $\Delta$ is empty: $\Delta \equiv \emptyset$ this is interpreted as false, that is $\Gamma \vdash$ means that $\Gamma$ proves falsehood, and therefore it is inconsistent.
Instead, an empty antecedent $\Gamma \equiv \emptyset$ is assumed true, that is, $\vdash \Delta$ means that $\Delta$ follows without any assumption, that is, it is always true.
We say then that $\vdash \Delta$ is a logical assertion

## 6.2 Semantics of the sequent

Whenever all assertions of the list $\Gamma$ are true, then at least one assertion of the list $\Delta$ is true.
That is the left of the comma on the turnstile can be thought as an “and” $\wedge$ , and the comma on the right as an “or” $\vee$.
Therefore, we can write, in an equivalent way:

$\vdash (A_1 \wedge ....... \wedge A_n) \to (B_1 \vee ...... \vee B_k)$

where $\to$ is the material implication: “if…then”.
The rules of sequent calculus are written schematically by a list of sequents above and below a line.
This indicates that, if everything is true above the line, then everything is also true below.
A formal proof in sequent calculus is a sequence of sequents, where every sequent is derivable from the sequents that appear before in the sequence, using the rules of sequent calculus.

## 6.3 Rules of sequent calculus

The rules of sequent calculus are of three different kinds:
i) Inference rules (or logical rules)
They concern logical connectives, like: $\wedge$ , $\vee$ , $\neg$ , $\to$
Every inference rule introduces a new logical formula either on the right or on the left of the turnstile.

ii) Structural rules (weakening, contraction, exchange)
Weakening allows the addition of arbitrary elements to a sequence. Intuitively, this is allowed in the antecedent because we can always add assumptions to our proof. It is allowed in the consequent because the consequent is a disjunction of elements, so only one needs be provable at a time, and we can add additional unproven propositions.

Contraction and Permutation assure that neither the order (P) nor the multiplicity of occurrences (C) of elements of the sequences matters.

iii) The two exceptions to this general scheme are the axiom of identity (I) and the rule of (Cut), the latter being a meta-rule, which is practically useless when given to a machine, like a computer. The rule of cut states that, when a formula A can be concluded and this formula may also serve as a premise for concluding other statements, then the formula A can be "cut out" and the respective derivations are joined.
When constructing a proof bottom-up, this creates the problem of guessing A (since it does not appear at all below). This issue is addressed in the theorem of cut-elimination. The cut-elimination theorem is the central result establishing the significance of the sequent calculus. It was originally proved by Gerhard Gentzen in his paper "Investigations in Logical Deduction" (Gentzen, 1935) for the systems LJ and LK formalising intuitionist and classical logic respectively.
The cut-elimination theorem, that is, Gentzen 's *Hauptsatz* ("Main theorem") states that any judgement that has a proof in the sequent calculus that makes use of the cut rule also has a cut-free proof, that is, a proof that does not make use of the cut rule.
The theorem of cut-elimination has important consequences:
A system is inconsistent if it admits a proof of the absurd. If the system has a cut elimination theorem, then if it has a proof of the absurd, it should also have a proof of the absurd without cuts. It is typically very easy to check that there are no such proofs. Thus, once a system is shown to have a cut elimination theorem, it is normally immediate that the system is consistent.
All derivations without cuts posses the sub-formula property that all formulas occurring in a derivation are sub-formulas of the formulas from the end-sequent.

**6.4 System LK**
**i) Logical rules**

Left logical rules

$$\frac{\Gamma, A \vdash \Delta \qquad \Sigma, B \vdash \Pi}{\Gamma, \Sigma, A \vee B \vdash \Delta, \Pi}\ (\vee L)$$

$$\frac{\Gamma \vdash A, \Delta}{\Gamma, A \wedge B \vdash \Delta}\ (\wedge L_1)$$

$$\frac{\Gamma \vdash B, \Delta}{\Gamma, A \wedge B \vdash \Delta}\ (\vee L_2)$$

Right logical rules

$$\frac{\Gamma \vdash A, \Delta \qquad \Sigma \vdash B, \Pi}{\Gamma, \Sigma \vdash A \wedge B, \Delta, \Pi}\ (\wedge R)$$

$$\frac{\Gamma, A \vdash \Delta}{\Gamma \vdash A \vee B, \Delta}\ (\vee R_1)$$

$$\frac{\Gamma, B \vdash \Delta}{\Gamma \vdash A \vee B, \Delta}\ (\vee R_2)$$

$$\frac{\Gamma \vdash A, \Delta \qquad \Sigma, B \vdash \Pi}{\Gamma, \Sigma, A \to B \vdash \Delta, \Pi} (\to L)$$

$$\frac{\Gamma, A \vdash B, \Delta}{\Gamma \vdash A \to B, \Delta} (\to R)$$

$$\frac{\Gamma \vdash A, \Delta}{\Gamma, \neg A \vdash \Delta} (\neg L)$$

$$\frac{\Gamma, A \vdash \Delta}{\Gamma \vdash \neg A, \Delta} (\neg R)$$

**ii) Structural rules:**

**Left structural rules**

**Right structural rules**

**Weakening** (data can be erased)

$$\frac{\Gamma \vdash \Delta}{\Gamma, A \vdash \Delta} (WL)$$

$$\frac{\Gamma \vdash \Delta}{\Gamma \vdash A, \Delta} (WR)$$

**Contraction** (data can be copied)

$$\frac{\Gamma, A, A \vdash \Delta}{\Gamma, A \vdash \Delta} (CL)$$

$$\frac{\Gamma \vdash A, A, \Delta}{\Gamma \vdash A, \Delta} (CR)$$

**Exchange** (or permutation)

$$\frac{\Gamma_1, A, B, \Gamma_2 \vdash \Delta}{\Gamma_1, B, A, \Gamma_2 \vdash \Delta} (EL)$$

$$\frac{\Gamma \vdash \Delta_1, A, B, \Delta_2}{\Gamma \vdash \Delta_1, B, A, \Delta_2} (ER)$$

**iii) Identity axiom and cut**

**Identity axiom**: $A \vdash A$

**Cut rule:** $$\frac{\Gamma \vdash \Delta, A \qquad A, \Sigma \vdash \Pi}{\Gamma, \Sigma \vdash \Delta, \Pi}$$

This system of rules can be shown to be both sound and complete with respect to first-order logic, i.e. a statement A follows semantically from a set of premises Γ if and only if the sequent $\Gamma \vdash A$ can be derived by the above rules.

One may restrict or forbid the use of some of the structural rules. This yields a variety of sub-structural logic systems. They are generally weaker than LK (*i.e.*, they have fewer theorems), and thus not complete with respect to the standard semantics of first-order logic. However, they have other interesting properties that have led to applications in theoretical computer science and artificial intelligence. In this thesis, we present in fact a sub-structural (also many-valued and quantum ) logic which has applications in quantum computing.

**6.5 System LJ**

Surprisingly, some small changes in the rules of LK suffice in order to turn it into a proof system for intuitionist logic. To this end, one has to restrict to intuitionist sequents (i.e., sequents with exactly one formula in the consequent) and modify the rule ($\vee$L) as follows:

$$\frac{\Gamma, A \vdash C \qquad \Sigma, B \vdash C}{\Gamma, \Sigma, A \vee B \vdash C} \; (\vee L)$$

where C is an arbitrary formula. The resulting system is called LJ. It is sound and complete with respect to intuitionist logic and admits a similar cut-elimination proof.

# 7. Basic logic and the reflection principle

Basic logic (Sambin *et al.*, 2000) is the weakest possible logic (no structure, no free contexts) and was originally conceived as the common platform for all other logics (linear, intuitionist, quantum, classical etc.) which can be considered as its "*extensions*". See “The cube of logics” in Fig.7.1.

**Fig. 7.1 The cube of logics**

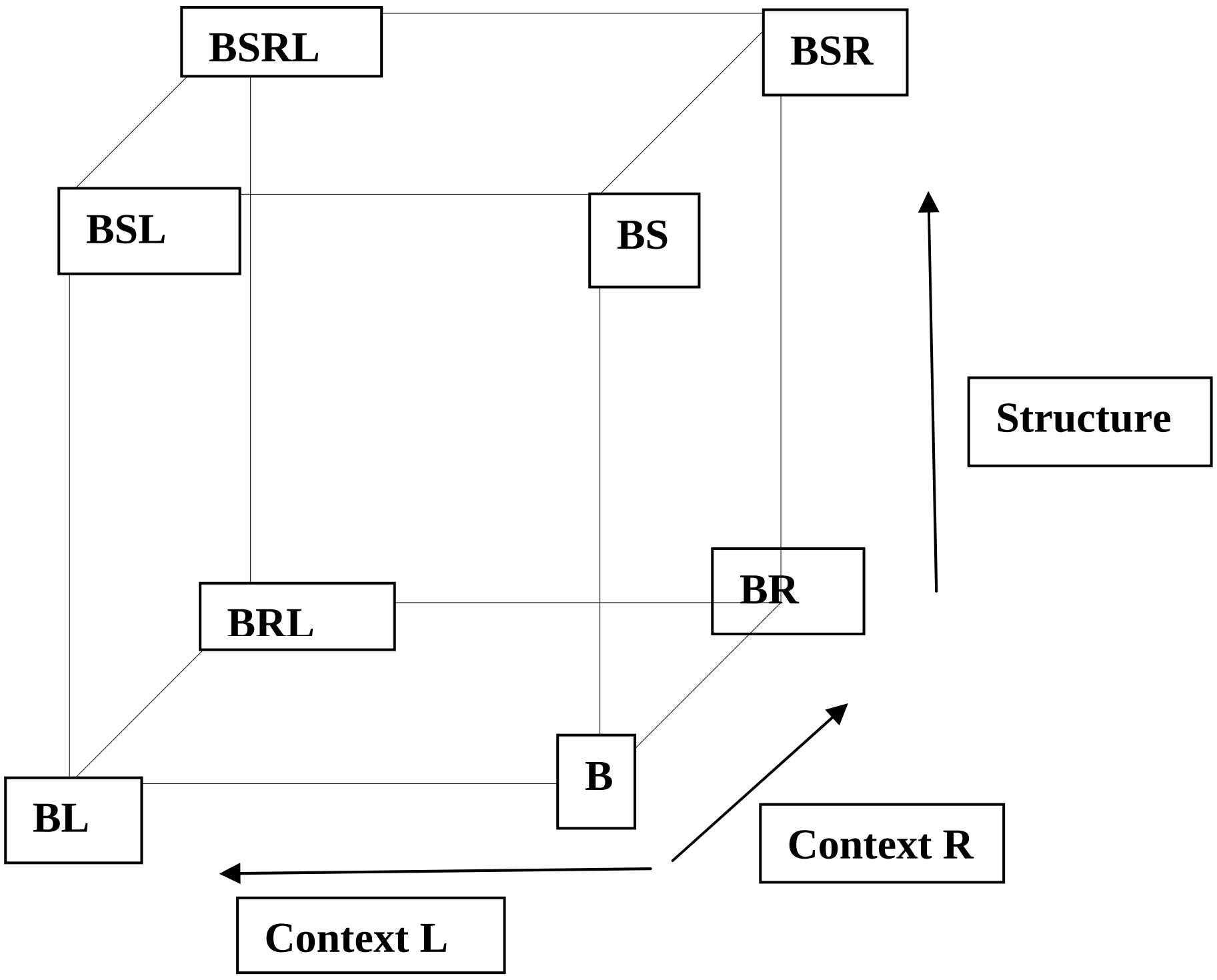

**Sub-structural logics:**
**B** = Basic logic
**BL** = Basic logic + context on the left
**BR =** Basic logic + context on the right
**BRL =** Linear logic

**Structural logics:**
**BS =** Quantum logic
**BSR** = paraconsistent logic
**BSL** = Intuitionist logic
**BSRL** = classical logic

### 7.1 The propositional calculus of Basic logic

The propositional calculus of Basic logic consists of:
**Propositional variables** (atomic formulas) which are denoted by:

$p_1, p_2, .....$

And their primitive negations:

$p_1^{\perp}, p_2^{\perp}, .....$

**Logical connectives:**

$\& = "and"$ (additive conjunction)

$\vee = "or"$ (additive disjunction) sometimes also indicated as $\oplus$ like in Linear logic.

$\otimes = "times"$ (multiplicative conjunction)

$\wp = "par"$ (multiplicative disjunction)

$\rightarrow$ (implication)

$\leftarrow$ (counter-implication)

$\neg$ (negation)

**Logical constants:**

Additive unities: T = "true" (it is the unity of &), 0 (it is the unity of $\vee$ )

Multiplicative unities: 1 (it is the unity of ⊗), $\perp$ = "false" (it is the unity of $\wp$ )

**Inference rules**

The inference rules for the connectives $\&, \vee, \neg, \rightarrow, \otimes, \wp$ are:

$$\& - form \frac{\Gamma \vdash A \qquad \Gamma \vdash B}{\Gamma \vdash A \& B} \qquad \& - refl \frac{A \vdash \Delta}{A \& B \vdash \Delta} \qquad \frac{B \vdash \Delta}{A \& B \vdash \Delta}$$

$$\vee - form \frac{A \vdash \Delta \qquad B \vdash \Delta}{A \vee B \vdash \Delta} \qquad \vee - refl \frac{\Gamma \vdash A}{\Gamma \vdash A \vee B} \qquad \frac{\Gamma \vdash B}{\Gamma \vdash A \vee B}$$

$$\neg - form \frac{A \vdash}{\vdash \neg A} \qquad \neg - rifl \frac{\vdash A}{\neg A \vdash}$$

$$\rightarrow - form \frac{A \vdash B}{\vdash A \rightarrow B} \qquad \rightarrow rifl \frac{\vdash A \quad B \vdash \Delta}{A \rightarrow B \vdash \Delta}$$

$$\otimes - form \frac{A, B \vdash \Delta}{A \otimes B \vdash \Delta} \qquad \otimes - refl \frac{\Gamma \vdash A \qquad \Gamma' \vdash B}{\Gamma, \Gamma' \vdash A \otimes B}$$

$$\wp - form \frac{\Gamma \vdash A, B}{\Gamma \vdash A \wp B} \qquad \wp - refl \frac{A \vdash \Delta \qquad B \vdash \Delta'}{A \wp B \vdash \Delta, \Delta'}$$

**Structural rules**

Basic logic has only one structural rule, the exchange rule.

**Exchange rule:**

$$exchL \frac{\Gamma, A, B, \Gamma' \vdash \Delta}{\Gamma, B, A, \Gamma' \vdash \Delta} \qquad exchR \frac{\Gamma \vdash \Delta, A, B, \Delta'}{\Gamma \vdash \Delta, B, A, \Delta'}$$

**The cut rule**:

$$\frac{\Gamma \vdash A \quad A \vdash \Delta}{\Gamma \vdash \Delta}$$

In Basic logic, the cut-elimination theorem  holds (Sambin *et.al.*, 2000, Faggian and Sambin, 1998).

**The identity axiom:**

$A \vdash A$

**7.2 The three main properties of Basic logic**

Basic logic has tree main properties:

**i) Reflection**: All the connectives of Basic logic satisfy the principle of reflection, that is, they are introduced by solving an equation (called *definitional equation*), which “reflects” meta-linguistic links between assertions into connectives between propositions in the object-language.

| **Object-language** | $\xleftrightarrow{\text{Re}\,\mathit{flection}}$ | **Meta-language** |
|---|---|---|
| Logical connectives | $\xleftrightarrow{\text{Re}\,\mathit{flection}}$ | Metalinguistic links $\vdash$; “<u>and</u>” |
| (Between propositions) | | (between assertions) |

The semantic equivalence:

*A # B true if and only if A true link B true*

which we call definitional equation for the connective #, gives all we need to know about it. A # B is semantically defined as that proposition which, when asserted true, behaves exactly as the compound assertion A true link B true. The inference rules for # are easily obtained by solving the definitional equation, and they provide an explicit definition. We then say that # is introduced according to the principle of reflection.

All logical constants of basic logic are introduced according to the principle of reflection.

**ii) Symmetry**: All the connectives are divided into "left" and "right" connectives.

A "left" connective has formation rule acting on the left, and a reflection rule acting on the right. In Basic logic, every left connective has its symmetric, a right connective, which has a formation rule acting on the right, and a reflection rule acting on the left (and vice-versa)

.

**Left connectives** $\xleftrightarrow{Symmetry}$ **Right connectives**

| | |
|---|---|
| $\vee$ ="*or*" (additive disjunction) | & ="*and*" (additive conjunction) |
| $\otimes$ ="*times*" (multiplicative conjunction) | $\wp$ =" *par*" (multiplicative disjunction) |
| $\leftarrow$ (counter-implication) | $\rightarrow$ (implication) |

**iii) Visibility**: There is a strict control on the contexts, that is, all active formulas are isolated from the contexts, and they are *visible*.

### 7.3 Some features of Basic logic

Moreover, Basic logic has the following features,

a) It is non-distributive (because of the absence of active contexts)

b) It is sub-structural, i.e., it has no structural rules like contraction:

$$\frac{\Gamma, A, A \vdash \Delta}{\Gamma, A \vdash \Delta} \quad \text{(data can be copied)}$$

and weakening:

$$\frac{\Gamma \vdash \Delta}{\Gamma, A \vdash \Delta} \quad \text{(data can be deleted)}$$

c) In Basic logic the non-contradiction principle $A \wedge \neg A = 0$ is invalidated, like in paraconsistent logic (Priest, 2002).

The contemporary logical orthodoxy has it that, from contradictory premises, anything can be inferred. To be more precise, let $\vDash$ be a relation of logical consequence, defined either semantically or proof-theoretically. Call $\vDash$ *explosive* if it validates $\{A , \neg A\} \vDash B$ for every $A$ and $B$ (*ex contradictioone quodlibet* (ECQ)). The contemporary orthodoxy, i.e., classical logic, is explosive, but also some 'non-classical' logics such as intuitionist logic and most other standard logics are explosive.

The major motivation behind *paraconsistent logic* is to challenge this orthodoxy. A logical consequence relation, $\vDash$, is said to be *paraconsistent* if it is not explosive. Thus, if $\vDash$ is paraconsistent, then even if we are in certain circumstances where the available information is inconsistent, the inference relation does not explode into *triviality*.

By symmetry, in Basic logic, the excluded middle $A \vee \neg A = 1$ is also invalidated, like in Intuitionist logic.

We will illustrate this duality in more detail in what follows.

### 7.4 Negation and falsehood

In Basic logic, negation is defined as in intuitionist logic: $\neg A \equiv A \rightarrow \perp$

In fact, “A is false” is denoted by: $A|\!-\ \bot$ , which becomes: $|\!-\ A \to \bot$

The shorthand of the above proposition: $A \to \bot$

represents the negation of A , that is: $\neg\ A$

In Linear logic (Girard, 1987) one has the primitive (or linear) negation $(.)^{\perp}$ , which is involutive:

$\left(A^{\perp}\right)^{\perp} = A$ .

This primitive negation is adopted also in Basic logic.

Girard’s primitive negation, and Sambin’s symmetry of Basic logic together give rise to “$\bot$-duality”.

The operation of $\bot$-duality on sequents where only atomic propositions appear, consists in performing the primitive negation of the atomic proposition, and inverts the sense of the sequent:

Definition: $(|\!-\ A^{\perp})^{\perp} = A|\!-$

The action of $\bot$-duality on sequents where there appear composite propositions, consists in making the primitive negation of each atomic component, taking the symmetric of the connective, and inverting the sense of the sequent:

Example: $\left(A\ \&\ B\ |\!-\right)^{\perp} = |\!-\ A^{\perp} \vee B^{\perp}$

In particular, the “dual” of the non contradiction principle is the principle of the excluded middle: $\left(A\ \&\ A^{\perp}\ |\!-\right)^{\perp} = |\!-\ A^{\perp} \vee A$ .

**7.5 Definitional equations**

A definitional equation is a correspondence between the metalanguage (formulated in terms of metalinguistic links between assertions) and the object-language (formulated in terms of connectives between propositions). By solving a definitional equation, a connective is introduced, and its inference rules are given.

The definitional equations for the connectives $\&, \vee, \neg, \to, \otimes, \wp$ are:

| | | | | |
|---|---|---|---|---|
| $\Gamma|\!-A\ \&\ B$ | iff | $\Gamma\ |\!-\ A$ | and | $\Gamma|\!-B$ |
| $A \vee B|\!-\Delta$ | iff | $A\ |\!-\ \Delta$ | and | $B\ |\!-\ \Delta$ |
| $|\!-\neg\ A$ | iff | $A|\!-$ | | |
| $|\!-\ A \to B$ | iff | $A|\!-\ B$ | | |
| $A \otimes B|\!-\Delta$ | iff | $A, B\ |\!-\Delta$ | | |
| $\Gamma|\!-A \wp B$ | iff | $\Gamma\ |\!-\ A, B$ | | |

## 8. Some preliminary definitions for the logic of the qubit (*Lq*)

In this Section, we will give some notations and definitions concerning the Language of the logic of one qubit (*Lq*), the rules for evaluating sequents, the semantics, the concept of Meta Data, and the quantum cut rule.

### 8.1 The language *Lq*

**Non-logical symbols:**

The constants $z_i$ $(i = 0,1)$

The symbol “ $^*$ ”

The symbol “ .”

The symbol $+$

The symbol $\circ$

**Logical symbols:**

The atomic propositions $p_i$ $.(i = 0,1)$

The logical connective $_{z_0}\&_{z_1}$

The logical connective $_{z_0}\vee_{z_1}$

The negation $\neg$

The *wff* $\varphi$

Logical duality (primitive negation) $\perp$

Quantum logical duality $\perp'$

### 8.2 Evaluation of sequents

We will introduce a new kind of duality, the *-duality, and the “Quantum logical duality” (a generalization of Sambin-Girard logical duality; see Sambin *et al.*, 2000) and we will show the relation among the two.

Moreover, we will define Meta Data (*MD*), to be encapsulated in the quantum metalanguage, which will be introduced in Sect.9. Meta Data (data about data) are necessary for the evaluations of sequents to sum up to 1. This constraint is required when truth-evaluations are interpreted as quantum probabilities, like in the case under consideration.

Finally we will illustrate the “gluing operation”, which is needed to give evaluations to complete sequents.

The whole procedure of performing the *-duality and the gluing operation is necessary to assign partial truth-values, without using the T-norms (Hajék, 1998) of many-valued logics (Łukasiewicz, 1920) and Fuzzy logic (Zadeh, 1965). We did not use T-norms because the latter do not satisfy the *MD* constraint, as we will see in Sect.10.

**The *-duality**

In what follows, we will consider sequents labelled by a complex number $z_i$:

$\vert\!-^{z_i}\ p_i$

denoting atomic assertions which are asserted with a grade $z_i$ ("assertion-degree"), whose physical interpretation will be given in Subsection 8.4. It should be noticed that the $\vert\!-^{z_i}\ p_i$ are a particular case of atomic assertions: they are the asserted atomic propositions, while in general atomic assertions are asserted propositions. We will come back to this distinction in this Section, and also in Sect.9.

The notion of "*-duality", takes into account the assertion-degrees $z_i$ in sequents without antecedents (or without consequents). It will turn out useful also in giving truth-evaluations to sequents having antecedent and consequent both non-empty.

The *-duality is defined as:

**-duality*: $(\vert\!-^{z_i}\ p_i)^* = p_i \vert\!-^{z_i^*}$ (8.1)

Therefore, the *-duality inverts the sense of the sequent, makes the *-operation on the assertion-degrees $z_i$, that is: $z_i \to z_i^*$ (where $z_i^*$ is the complex conjugate of $z_i$) but does not negate the atomic proposition, neither exchanges the $z_i$.

**The Quantum logical duality $\perp'$**

We also encountered the necessity of introducing a generalization of Sambin-Girard logical duality $\perp$. We called it "Quantum logical duality", denoted by $\perp'$.

The definition of $\perp'$ is:

$\perp'$*-duality*:

$$(\vert\!-^{z_i}\ p_i)^{\perp'} = p_j \vert\!-^{z_j^*} \tag{8.2}$$

where $p_j$ is the primitive negation of $p_i$, that is: $p_j = p_i^{\perp}$ for $i \neq j$, $(i, j = 0,1)$.

Therefore, the $\perp'$-duality inverts the sense of sequents, makes the primitive negation of atomic propositions, and exchanges the assertion-degrees, also making the *-operation: $z_i \to z_j^*$.

**Relation between the *-duality and the $\perp'$-duality**

The *-duality and the $\perp'$-duality are related to each other as follows:

$$(\vert\!-^{z_i}\ p_i)^* = (\vert\!-^{z_j}\ p_j)^{\perp'} = p_i \vert\!-^{z_i^*} \tag{8.3}$$

**Meta Data (*MD*)**

The *MD* are the following constraints on the assertion-degrees:

,

*MD*: $z_0^* \cdot z_0 + z_1^* \cdot z_1 = 1; \quad z_0^* \cdot z_1 + z_1^* \cdot z_0 = 0$ (8.4)

***Definition***: $v_i = z_i^* \cdot z_i$ (8.5)

**The language *Lq***

In the Language *Lq*, there are two ***atomic formulas***: $p_i \ (i = 0,1)$.

The ***logical connectives*** are: ${}_{z_0}\&_{z_1}$, ${}_{z_0^*}\vee_{z_1^*}$, $\neg$ .

The ***well formed formulas*** (*wff*) $\varphi$ are: $p_0, p_1$, $p_0 \, {}_{z_0}\&_{z_1} \, p_1$, $p_0 \, {}_{z_0^*}\vee_{z_1^*} \, p_1$.

**The metalanguage *MLq* consists of:**

***The atomic assertions***: $\vdash^{f(z_i)} \varphi$, which are asserted with a grade which is a function $f(z_i)$. We will consider in particular ***asserted atomic propositions***: $\vdash^{z_i} p_i$, which form a subset of $\vdash^{f(z_i)} \varphi$.

***The metalinguistic links***: $\vdash$ "yields", "and" .

***The sequent***: $\Gamma \vdash^{v} \varphi$, where $\Gamma$ is a list (at least empty) contained in $p_0, p_1$, $p_0 \, {}_{z_0}\&_{z_1} \, p_1$, $p_0 \, {}_{z_0^*}\vee_{z_1^*} \, p_1$, and *v* (defined in (8.5) is the evaluation of the sequent.

***The Meta Data***: *MD*: $z_0^* \cdot z_0 + z_1^* \cdot z_1 = 1; \quad z_0^* \cdot z_1 + z_1^* \cdot z_0 = 0$.

**The gluing operation**

Let us consider the two lists $\Gamma$ and $\Delta$.

In our case, the atomic propositions $p_0$ and $p_1$ are asserted with assertion-degrees $z_0$ and $z_1$ respectively. Then, in general $\Gamma$ and $\Delta$ will be asserted with assertion-degrees which are functions of $z_0$ and $z_1$, that is:

$$\vdash^{f(z_0,z_1)} \Gamma, \ \vdash^{g(z_0,z_1)} \Delta$$

The question, now, is how to form the analogous of the classical sequent $\Gamma \vdash \Delta$ by taking into account the assertion-degrees $z_0$ and $z_1$.

Let us extend the *-duality to the general case of a sequent with a list $\Gamma$ as antecedent by:

$$(\vdash^{f(z_0,z_1)} \Gamma)^* = \Gamma \vdash^{f(z_0^*, z_1^*)}$$

and define the "gluing operation" denoted by "$\circ$" as follows:

$$\Gamma \vdash^{f(z_0^*, z_1^*)} \circ \vdash^{g(z_0,z_1)} \Delta = \Gamma \vdash^{f(z_0^*, z_1^*) \circ g(z_0,z_1)} \Delta \qquad (8.6)$$

With $f(z_0^*, z_1^*) \circ g(z_0, z_1)$ defined as:

$$f(z_0^*, z_1^*) \cdot g(z_0, z_1) \tag{8.7}$$

Where the symbol $"\cdot"$ stands for the usual multiplication, and we have assumed that the $z_i$ satisfy Eq.(8.4).
Let us consider for simplicity the case with $\Delta = \Gamma$, that is:

$$\Gamma \vdash \Gamma \tag{8.8}$$

where $\Gamma$ is asserted by a function $f(z_0, z_1)$.
Eq.(8.8) should be rewritten as:

$$\Gamma \vdash^{f(z_0^*, z_1^*)} \circ \vdash^{f(z_0, z_1)} \Gamma \;=\; \Gamma \vdash^{f(z_0^*, z_1^*) \circ f(z_{0,}, z_1)} \Gamma \tag{8.9}$$

where, by applying Eq.(8.7):

$$f(z_0^*, z_1^*) \circ f(z_0, z_1) = f(z_0^*, z_1^*) \cdot f(z_0, z_1) \tag{8.10}$$

Let us now make the simplest choices for $f(z_0, z_1)$ and $f(z_0^*, z_1^*)$, namely:

$$f(z_0, z_1) = z_0 + z_1 ; \quad f(z_0^*, z_1^*) = z_0^* + z_1^* \tag{8.11}$$

From Eqs.(8.10) and (8.11), and by applying the distributive property we get:

$$f(z_0^*, z_1^*) \circ f(z_0, z_1) = z_0^* \cdot z_0 + z_1^* \cdot z_1 \tag{8.12}$$

Where we see that by the assumption of Eq.(8.4), the mixed products have disappeared.
By the definition of $v_i = z_i^* \cdot z_i$, Eq.(8.12) becomes:

$$f(z_0^*, z_1^*) \circ f(z_0, z_1) = v_0 + v_1 \tag{8.13}$$

Finally, by the use of the *MD* constraint: $z_0^* \cdot z_0 + z_1^* \cdot z_1 = 1$, we get, from Eq.(8.13):

$$f(z_0^*, z_1^*) \circ f(z_0, z_1) = 1 \tag{8.14}$$

Then, the sequent in Eq.(8.9) is a “classical” axiom, as its truth-value is 1:

$$\Gamma \vdash^{1} \Gamma$$

However, this does not hold true in general. Let us consider, for example the "classical" identity axioms $p_i \vdash p_i$ where the $p_i$ are atomic propositions.
In our case, the atomic propositions $p_i$ are asserted with assertion-degrees $z_i$.
Given the sequent $\vdash^{z_i} p_i$ and its *-dual $p_i \vdash^{z_i^*}$, we make their "gluing" by the "$\circ$" operation, and obtain:

$$\left(p_i \vdash^{z_i^*}\right) \circ \left(\vdash^{z_i} p_i\right) = p_i \vdash^{z_i^* \circ z_i} p_i \tag{8.15}$$

where, in this case, it holds:

$$z_i^* \circ z_i = z_i^* \cdot z_i = v_i \tag{8.16}$$

Therefore, in this case the identity axiom:

$$p_i \vdash^{v_i} p_i \tag{8.17}$$

is evaluated by a partial truth-value $v_i$.
From the gluing operation and the *MD*, it follows that any sequent having as antecedent a list $\Gamma$ and as consequent an atomic proposition $p_i$, will be evaluated by $v_i$, that is:

$$\Gamma \vdash^{v_i} p_i \tag{8.18}$$

The same reasoning holds for sequents having $p_i$ as antecedent and a list $\Delta$ as consequent, so that one has as well:

$$p_i \vdash^{v_i} \Delta \tag{8.19}$$

Let us just prove Eq.(8.18), the proof of Eq.(8.19) being symmetrical.
Let us consider the general case in Eq.(8.6), namely:

$$\Gamma \vdash^{f(z_0^*, z_1^*)} \circ \vdash^{g(z_0, z_1)} \Delta = \Gamma \vdash^{f(z_0^*, z_1^*) \circ g(z_0, z_1)} \Delta$$

and put, : $\Delta = p_i$ $(i = 0,1)$
Let us consider the example with $i = 0$, and let us show that the sequent having as antecedent the list $\Gamma$ and as consequent the atomic proposition $p_0$ has evaluation $v_0$, that is, we want to prove $\Gamma \vdash^{v_0} p_0$.
The assertion function $g(z_0, z_1)$ of $\Delta$, in the case $\Delta = p_0$, reduces to $g(z_0, 0)$.
We must perform the gluing operation:

$$\Gamma \vdash^{f(z_0^*, z_1^*)} \circ \vdash^{g(z_0,0)} p_0 = \Gamma \vdash^{f(z_0^*, z_1^*) \circ g(z_0,0)} p_0$$

i.e.
$f(z_0^*, z_1^*) \circ g(z_0,0) = f(z_0^*, z_1^*) \cdot g(z_0,0)$
with: $g(z_0,0) = z_0$ and $f(z_0^*, z_1^*) = z_0^* + z_1^*$
which, by using the distributive property and taking into account the constraint in Eq.(8.4), gives:

$$f(z_0^*, z_1^*) \circ g(z_0,0) = z_0^* \cdot z_0$$

Finally, by using the definition in (8.5), we get:

$$f(z_0^*, z_1^*) \circ g(z_0,0) = v_0$$

In summary, in this calculus, sequents whose antecedent and consequent are both non-empty, are all evaluated by the $v_i$ (or by their sum, which equals 1).
It should be noticed that in *Lq*, because of the use of the *-duality, the "gluing operation" substitutes the unpleasant (and problematic) notation of "nested sequents". In fact the expression:

$$\Gamma \vdash^{f(z_0^*, z_1^*)} \circ \vdash^{g(z_0,z_1)} \Delta$$

stays for:

$$(\vdash^{f} \Gamma) \vdash (\vdash^{g} \Delta)$$

Which reads: "Asserted $\Gamma$ (with assertion function $f$ ) yields asserted $\Delta$ (with assertion function *g*)".
We think that this new quantum notation of the link "yields" might be of some utility in solving the difficulty of defining the connective $\rightarrow$ (primitive implication) in Basic logic, a problem related to nested sequents (see Sambin *et al.*, 2000, pg.991).

**8.3 Semantics**
We give the $v_i$ the meaning of "partial truths" (also called truth-degrees in Fuzzy logic; see Zadeh, 1965). In what follows, we will name the $v_i$ "truth-evaluations".
The sum of the $v_i$, that is 1, is the "complete truth".
In the physical interpretation *I* the $v_i$ are quantum probabilities $p(i) = |\lambda_i|^2$ :

$$v_i \xrightarrow{I} p(i) \qquad v_i \in [0,1] \tag{8.20}$$

## 8.4 Interpretation of atomic graded assertions

Although it is not customary to give an interpretation to atomic assertions, we think that in this case (where assertions are graded) it might be of some utility to do so, for getting an intuitive understanding.

We interpret an atomic graded assertion $\vdash^{z_i} p_i$ as the action of the non hermitian operator $O_i$ on the basis state $|\psi_i\rangle$ of the Hilbert space *H*, that is: $O_i|\psi_i\rangle$.

On the other hand, we interpret the *-dual of the above sequent, namely $_i\vdash^{z_i^*}$ as the dual of the ket $O_i|\psi_i\rangle$, namely as the bra: $\langle\psi_i|O_i^\dagger$. Then, the "gluing" of $p_i\vdash^{z_i^*}$ with $\vdash^{z_i} p_i$, which gives $p_i\vdash^{v} p_i$, is interpreted as the scalar product of the bra $\langle\psi_i|O_i^\dagger$ with the ket $O_i|\psi_i\rangle$, that is: $\langle\psi_i|O_i^\dagger O_i|\psi_i\rangle = |\lambda_i|^2$, where we recall that $v_i$ is interpreted as the quantum-mechanical probability $|\lambda_i|^2$.

## 8.5 The quantum cut rule

The quantum cut rule (which is a generalization of the usual cut rule) requires that the evaluations of the two sequents (where the formula has to be cut) should be equal.

It is defined as follows:

$$\frac{\Gamma\vdash^{v}\varphi \quad \varphi\vdash^{v}\Delta}{\Gamma\vdash^{v}\Delta} \tag{8.21}$$

Notice that in Eq.(8.21), the evaluation of the conclusion is the same as those in the premises. This holds because the quantum cut rule is a generalization of the classical one, which must be recovered as a particular case. As such, the quantum cut rule acts only on the formula $\varphi$ and leaves unchanged all the rest, comprising the evaluations $v$.

Let us consider the particular case where the formula to be cut is an atomic formula, that is, let us put $\varphi \equiv p_i$ in (8.21):

$$\frac{\Gamma\vdash^{v_i} p_i \quad p_i\vdash^{v_i}\Delta}{\Gamma\vdash^{v_i}\Delta} \qquad \text{(8.21bis)}$$

By undoing the "gluing" in Eq.(8.21bis), we can rewrite the quantum cut rule as:

$$\frac{\Gamma\vdash^{z_i^*}\circ\vdash^{z_i} p_i \quad p_i\vdash^{z_i^*}\circ\vdash^{z_i^*}\Delta}{\Gamma\vdash^{v_i}\Delta} \tag{8.22}$$

We define:

$$\vdash^{z_i} p_i \quad p_i \vdash^{z_i^*} \equiv \Pi_i \tag{8.23}$$

Where $\Pi_i$ in Eq.(8.23) is a metalinguistic operator that we call "projector". Notice that the blank space between the two sequents with opposite sense in $\vdash^{z_i} p_i \quad p_i \vdash^{z_i^*}$ is the metalinguistic link "and". Therefore, we give the following:

**Definition**:
*A metalinguistic projector consists of the metalinguistic link "and" between an atomic assertion and its *-dual.*

Notice that Eq.(8.22) can be rewritten, by the use of Eq.(8.23) as:

$$\frac{\Gamma \vdash^{z_i^*} \circ \Pi_i \circ \vdash^{z_i} \Delta}{\Gamma \vdash^{v_i} \Delta} \tag{8.24}$$

Then, the quantum cut rule is equivalent to making the gluing $\Gamma \vdash^{v_i} \Delta$ of two premises with the same evaluation, $\vdash^{z_i} \Gamma$ and $\vdash^{z_i} \Delta$.
The physical interpretation of the quantum cut is that of a projective quantum measurement. In fact, by reminding the interpretation given in Subsection 8.4 to atomic graded assertions and to their *-duals, it follows the interpretation $I$ of $\Pi_i$:

$$\vdash^{z_i} p_i \quad p \vdash^{z_i^*} \xrightarrow{I} O_i^{\dagger} O_i |\psi_i\rangle\langle\psi_i| \tag{8.25}$$

where, in Eq.(8.25) $|\psi_i\rangle\langle\psi_i|$ is the one-dimensional projector $P_i$ on the Hilbert space.
Eq.(8.25) can be rewritten, in a more compact form as:

$$\Pi_i \xrightarrow{I} |\lambda_i|^2 P_i \tag{8.26}$$

The interpretation of the quantum cut rule as a quantum projective measurement just gives the probability $|\lambda_i|^2$ of measuring the state $|\psi_i\rangle$.
The quantum cut-rule is a meta-rule and (as the usual cut rule), becomes useless when given to a quantum computer. Its interpretation as a projective quantum measurement suggests the impossibility for a (quantum) machine to perform a self-measurement.

## 9. From Quantum Metalanguage to *Lq*

Metalogic is the study of the metatheory of logic. While logic is the study of the manner in which logical systems can be used to decide the correctness of arguments, metalogic studies the properties of the logical systems themselves.

In metalogic, formal languages are sometimes called object-languages. The language used to make statements about an object-language is called a metalanguage. This distinction is a key difference between logic and metalogic. While logic deals with proofs in a formal system, expressed in some formal language, metalogic deals with proofs about a formal system which are expressed in a metalanguage about some object-language.

In this Section, we will illustrate the main difference between a classical metalanguage and its quantum counterpart. It is such a difference that limits the use of a classical metalanguage, which allows, by the reflection principle, to introduce only classical logical connectives. Instead, a quantum metalanguage can be used to introduce quantum logical connectives, like the connectives of quantum superposition and quantum entanglement.

### 9.1 Classical metalanguage

We report here a very clear explanation of metalanguage directly from the source (Sambin *et.al.*, 2000):

"Assume from now on that A,B,C…. denote propositions. A proposition A must be distinct from the assertion on it. At this basic level, rather than A is true, we prefer to adopt for assertions a more neutral notation like A *is* (to recall the common form of A is true, A is available, A is measured, etc.), which shares with A is true only the fact it is an assertion, rather than a proposition. This is to recall that no specific meaning is at the moment attached to the assertion of A.

We need to consider also more complex metalinguistic statements, which are built up from atomic assertions of the form A *is* by means of some metalinguistic links and which we call compound assertions. The first discovery here is that the compound assertions used in any sequent calculus (lists $\Gamma$, sequents $\Gamma|\!-\Delta$, rules, derivations) can be seen as obtained from atomic assertions by means of only two metalinguistic links, namely "and", "yields".

A conjunction of atomic assertions *$C_1$ is and…. and$C_n$ is* is abbreviated by $C_1,\ldots C_n$, where comma takes the place both of *and* and of i*s*. Like Gentzen, we write $\Gamma$ for any conjunction of atomic assertions, either empty or $C_1,\ldots,C_n$. Similarly, for $\Delta$ and $D_1,\ldots,D_m$ and for other capital Greek letters.

The meaning of a sequent $\Gamma|\!-\Delta$ is that $\Delta$ is a consequence of $\Gamma$, that is $\Gamma$ yields $\Delta$ or (*$C_1$ is and…and$C_n$ is* ) yields (*$D_1$ is and…and$D_m$ is* ). Thus the usual sign $|\!-$` is a shorthand for yields in such a compound assertion."

As an example, we remind that the connective & is introduced, in Basic logic, by the definitional equation:

$$\Gamma|\!- A\,\&\,B \quad \underline{\text{iff}} \quad \Gamma|\!- A \quad \underline{\text{and}} \quad \Gamma|\!- B \qquad (9.1)$$

Where the compound assertion is: $\Gamma|-A$ <u>and</u> $\Gamma|-B$, built up from the atomic assertions $|-A$ and $|-B$. The metalinguistic link <u>and</u> reflects into the connective &.
We add here a few considerations, which will be useful to our scope.
In the above view, the formal system to be built is based solely on the meta-level which comes first. Of course, the assertions of the metalanguage *ML* should correspond to the propositions of the language *L* of the formal system, which is to be built. Then, those propositions should be well formed formula (*wff*) of *L*, that is, it should be possible to inductively build them from *n* atomic propositions $p_i\ (i = 0,1,...n-1)$ and from the logical connectives (the latter being those introduced by the definitional equations.

**9.2 Quantum metalanguage**

A quantum metalanguage differs from a classical metalanguage in two points:
i) the atomic assertions are asserted by a grade, in general a complex function
ii) the sequents are evaluated, and the evaluations obey a constraint (the Meta Data).
In what follows, we will give some definitions (among them, same were already given in Sect.8, nevertheless, we recall those too for more clarity).

In the language *Lq*:
i)We have two atomic propositions, $p_i\ (i = 0,1)$
ii)We have three connectives: ${}_{z_0}\&_{z_1}$, ${}_{z_0^*}\vee_{z_1^*}$, $\neg$
iii) From i) and ii) we build up the following $wff\{\varphi\} = p_0, p_1, p_0\ {}_{z_0}\&_{z_1}\ p_1, p_0\ {}_{z_0^*}\vee_{z_1^*}\ p_1$

In the quantum metalanguage:
i) We have the atomic assertions: $|-^{f(z_i)}\ \varphi$
which are asserted by an assertion grade, $f(z_i)$ which is a function of the assertion-degrees $z_i$ of the asserted atomic propositions $p_i\ (i = 0,1)$.
In particular, we will consider the asserted atomic propositions, denoted by $|-^{z_i}\ p_i$, the latter being a particular case of the atomic assertions $|-^{f(z_i)}\ \varphi$
ii) The metalinguistic links $|-$ "yields", "and" among assertions
iii) The sequent $\Gamma|-^{v}\ \varphi$
Where *v* is a partial truth-evaluation, and $\Gamma$ is a list (at least empty) contained in $p_0, p_1, p_0\ {}_{z_0}\&_{z_1}\ p_1, p_0\ {}_{z_0^*}\vee_{z_1^*}\ p$.
vi) A Meta Data (*MD*): $z_0^* \cdot z_0 + z_1^* \cdot z_1 = 1 \qquad z_0^* \cdot z_1 + z_1^* \cdot z_0 = 0$

The definitional equation of the connective ${}_{z_0}\&_{z_1}$ is:

$\Gamma|-p_0\ {}_{z_0}\&_{z_1}\ p_1$ <u>iff</u> $\Gamma|-^{v_0}\ p_0$ <u>and</u> $\Gamma|-^{v_1}\ p_1$ (9.2)
(*MD*): $z_0^* \cdot z_0 + z_1^* \cdot z_1 = 1 \qquad z_0^* \cdot z_1 + z_1^* \cdot z_0 = 0$

The definitional equation for ${}_{z_0}\&_{z_1}$ (in the case $\Gamma = \emptyset$) is :

$$\vdash p_0 \, {}_{z_0}\&_{z_1} \, p_1 \quad \underline{\text{iff}} \quad \vdash^{z_0} p_0 \quad \underline{\text{and}} \quad \vdash^{z_1} p_1 \qquad (9.3)$$

$$MD: \quad z_0^* \cdot z_0 + z_1^* \cdot z_1 = 1; \qquad z_0^{\ *} \cdot z_1 + z_1^{\ *} \cdot z_0 = 0$$

This writing is not wrong, because assertions are propositions which are always true, whatever is the antecedent.
However, in order to make the metalanguage operational in solving the definitional equation, one should put a non-empty antecedent in the sequents. This will allow getting the reflection axioms and the explicit reflection rule. Instead, getting the formation rule does not need the appearance of the antecedent in the metalanguage. We will do that in both ways. In the first one (with an empty antecedent) it is more transparent the role played by the assertion-degrees $z_0$ and $z_1$ in introducing the connective "quantum superposition".
It should be noticed that by using the classical kind of assertions of Basic logic, it would not be possible to recover the connective ${}_{z_0}\&_{z_1}$ "quantum superposition".
To show that, let us pick up, in the set of atomic assertions $\vdash A$, two of them which are asserted atomic propositions, namely $\vdash p_0$ and $\vdash p_1$, the latter being asserted with assertion 1 (because in Basic logic there is not the notion of "assertion-degree"). As a consequence, the assertion-degrees will not appear in RHS of the definitional equation for ${}_{z_0}\&_{z_1}$, while they will on the LHS:

$$\vdash p_0 \, {}_{z_0}\&_{z_1} \, p_1 \quad \underline{\text{iff}} \quad \vdash p_0 \quad \underline{\text{and}} \quad \vdash p_1$$

Therefore, a classical-like metalanguage will not help in introducing the connective "quantum superposition".

**9.3 The connective "quantum superposition"**

By solving the definitional equation (9.3) from the RHS to the LHS we get the formation rule for ${}_{z_0}\&_{z_1}$ in the case $\Gamma = \emptyset$:

$$\frac{\vdash^{z_0} p_0 \qquad \vdash^{z_1} p_1}{\vdash p_0 \, {}_{z_0}\&_{z_1} \, p_1} \qquad (9.4)$$

Let us suppose that one might want to rewrite Eq.(9.3) as:

$$\vdash^{g(z_0,z_1)} p_0 \,\&\, p_1 \quad \underline{\text{iff}} \quad \vdash^{z_0} p_0 \quad \underline{\text{and}} \quad \vdash^{z_1} p_1 \qquad (9.5)$$

$$MD: \quad z_0^* \cdot z_0 + z_1^* \cdot z_1 = 1; \qquad z_0^{\ *} \cdot z_1 + z_1^{\ *} \cdot z_0 = 0$$

Where, in Eq.(9.5), *&* is the usual additive conjunction of Basic logic, and $g(z_0, z_1)$ is a function of the assertion degrees, already discussed in Sect.8.
Pheraps, Eq.(9.5) is more explicit in expressing the fact that the compound proposition $p_0 \,\&\, p_1$ is asserted with a grade, which is a function of the assertion degrees of the atomic assertions. On the other hand, Eq.(9.3) looks more transparent in describing the formation of the connective which is to be interpreted as quantum superposition.
The expression $\vdash^{g(z_0,z_1)} p_0 \,\&\, p_1$ on the LHS of Eq.(9.5) might seem to be more appropriate in the case $\Gamma \neq \emptyset$. However, keeping this notation, will lead to an absurd conclusion as we will see in the following.
The logical connective ${}_{z_0}\&_{z_1}$ is in general non-commutative:

$$p_0 \,{}_{z_0}\&_{z_1}\, p_1 \neq p_0 \,{}_{z_1}\&_{z_0}\, p_1 \qquad (9.6)$$

In fact, the compound proposition $p_0 \,{}_{z_1}\&_{z_0}\, p_1$ would result, by reflection, from a different quantum metalanguage, namely: $\vdash^{z_1} p_0$ <u>and</u> $\vdash^{z_0} p_1$, which in fact will reflect in the quantum object-language: $\vdash p_0 \,{}_{z_1}\&_{z_0}\, p_1$, leading to the definitional equation:

$$\vdash p_0 \,{}_{z_1}\&_{z_0}\, p_1 \quad \underline{\text{iff}} \quad \vdash^{z_1} p_0 \quad \underline{\text{and}} \quad \vdash^{z_0} p_1 \qquad (9.7)$$

$$MD: \quad z_0^* \cdot z_0 + z_1^* \cdot z_1 = 1; \qquad z_0^* \cdot z_1 + z_1^* \cdot z_0 = 0$$

Notice that in the supposed alternative form, Eq.(9.7) could be rewritten as:

$$\vdash^{g(z_1,z_0)} p_0 \,\&\, p_1 \quad \underline{\text{iff}} \quad \vdash^{z_1} p_0 \quad \underline{\text{and}} \quad \vdash^{z_0} p_1 \qquad (9.8)$$

$$MD: \quad z_0^* \cdot z_0 + z_1^* \cdot z_1 = 1; \qquad z_0^* \cdot z_1 + z_1^* \cdot z_0 = 0$$

Owing to the fact that the LHS of Eq.(9.8) is different from the LHS of Eq. (9.5), it follows that the functions labelling the sequents on the RHS of Eqs.(9.8) and (9.5) are different too:

$$g(z_1, z_0) \neq g(z_0, z_1) \qquad (9.9)$$

At this point we remind that in Sect.8 we made the simplest choice for the function g, that is:

$$g(z_0, z_1) = z_0 + z_1 \qquad (9.10)$$

Then, from Eqs.(9.9) and (9.10) it follows that, if we want to use the classical *&* of Basic logic we are faced with a non-commutative sum of (complex) numbers, which is meaningless. This arises from the fact that Basic logic, although being non-classical, is not quantum either, and cannot be used to formalize quantum information.
Obviously, the connective ${}_{z_0}\&_{z_1}$ is commutative in the case $z_0 = z_1$, which, by the *MD* is furthermore restricted to $z_0 = z_1 = \frac{1}{\sqrt{2}}$ corresponding, in the physical interpretation, to the so called "cat state": $|\Psi\rangle\, cat = \frac{1}{\sqrt{2}}(|0\rangle + |1\rangle)$.

Then, in the case $\Gamma \neq \emptyset$, the definitional equation for the connective ${}_{z_0}\&_{z_1}$ is:

$$\Gamma \vdash^{v_0+v_1=1} p_0\, {}_{z_0}\&_{z_1}\, p_1 \quad \underline{\text{iff}} \quad \Gamma \vdash^{v_0} p_0 \quad \underline{\text{and}} \quad \Gamma \vdash^{v_1} p_1 \tag{9.11}$$

*MD*: $v_0 + v_1 = 1$

The *MD* means that the compound proposition is asserted with assertion-degree 1, and therefore is true, with truth-value 1.
One might argue that, by multiplying the assertion degrees $z_0, z_1$ by a phase $e^{i\varphi}$, the RHS of Eq.(9.11) is left unchanged, while the LHS is not, due to the explicit appearance of the $z_0, z_1$ in the connective ${}_{z_0}\&_{z_1}$. However, the phase is "absorbed" in the gluing operation, for the very definition of the latter, as we will see in the following. In fact, by un-gluing the sequent $\Gamma \vdash^{v_0+v_1=1} p_0\, {}_{z_0}\&_{z_1}\, p_1$ in (9.11), we get:

$$\Gamma \vdash^{g(z_0^*, z_1^*)} \circ \vdash^{g(z_0, z_1)} p_0\, {}_{z_0}\&_{z_1}\, p_1$$

Let us now multiply $z_0, z_1$ by a phase $e^{i\varphi}$:

$$\Gamma \vdash^{g(z_0^*, z_1^*)} \circ \vdash^{g(z_0, z_1)} p_0\, {}_{(e^{i\varphi} z_0)}\&_{(e^{i\varphi} z_1)}\, p_1 = \Gamma \vdash^{e^{-i\varphi} g(z_0^*, z_1^*)} \circ \vdash^{e^{i\varphi} g(z_0, z_1)} p_0\, {}_{z_0}\&_{z_1}\, p_1 \tag{9.12}$$

Let us perform the gluing operation, in Eq.(9.12), that is:

$$e^{-i\varphi} g(z_0^*, z_1^*) \circ e^{i\varphi} g(z_0, z_1) = e^{-i\varphi} g(z_0^*, z_1^*) \cdot e^{i\varphi} g(z_0, z_1) = z_0^* \cdot z_0 + z_1^* \cdot z_1 \tag{9.13}$$

where in Eq.(9.13) the definition (9.10) and the *MD* (8.4) have been used. Moreover, by using the definition of the truth-evaluation (8.5) and again (8.4), one sees that Eq.(9.12) gives back $\Gamma \vdash^{v_0+v_1=1} p_0\, {}_{z_0}\&_{z_1}\, p_1$.
We should perhaps stress the fact that ${}_{z_0}\&_{z_1}$ is a different connective from *&*, has a different definitional equation, different logical rules, as we will see in the next Section, and a different meaning. The metalinguistic origin of the connective ${}_{z_0}\&_{z_1}$ is a quantum metalanguage, which explains the mental process leading to the concept of quantum superposition.

In the physical interpretation, the non logical constants $z_0, z_1$ are the probability amplitudes $\lambda_0, \lambda_1 \in C$, the truth-values $v_0, v_1$ are the probabilities $|\lambda_0|^2$, $|\lambda_1|^2$, which sum up to 1, and the two atomic propositions $p_0, p_1$ are the two non-hermitian operators $O_0 = \lambda_0 P_0$ and $O_1 = \lambda_1 P_1$, where $P_0, P_1$ are the two projectors on the Hilbert space $C^2$.
We conclude this Section with two remarks:
i) A classical metalanguage is unable to reproduce, by the reflection principle, the logical connective for quantum superposition, which is one of the main features of Quantum Mechanics. As we have already stressed, the search of such a connective has not been pursued since now, and in fact it is absent in all kinds of Quantum logics.
ii) The connective ${}_{z_0}\&_{z_1}$ is just formed in *Lq*, where it can be used only once, because of the *MD* constraint in the quantum metalanguage. However, the connective ${}_{z_0}\&_{z_1}$ can be used three times in the logic of two qubits (*L2q*), seven times in the logic of three qubits (*L3q*) , and so on, in general $2^n - 1$ in *Lnq*, where *n* is the numbers of qubits, and $N = 2^n$ is the dimension of the Hilbert space.

## 10. The sequent calculus *Lq*

As a consequence of what has been said in the previous Section, the definitional equation for the new logical connective ${}_{z_0}\&_{z_1}$ (in the case $\Gamma \neq \emptyset$) is then:

$$\Gamma \vdash^{v_0+v_1=1} p_0\, {}_{z_0}\&_{z_1}\, p_1 \quad \underline{\text{iff}} \quad \Gamma \vdash^{v_0} p_0 \quad \underline{\text{and}} \quad \Gamma \vdash^{v_1} p_1 \qquad MD\text{: } v_0 + v_1 = 1 \tag{10.1}$$

### 10.1 The logical rules for ${}_{z_0}\&_{z_1}$

The *formation rule* of ${}_{z_0}\&_{z_1}$ is obtained from the definitional equation (10.1) from the RHS to the LHS:

$$\frac{\Gamma \vdash^{v_0} p_0 \qquad \Gamma \vdash^{v_1} p_1}{\Gamma \vdash^{v_0+v_1=1} p_0\, {}_{z_0}\&_{z_1}\, p_1} \tag{10.2}$$

The definitional equation from the LHS to the RHS gives the *implicit reflection rule* for ${}_{z_0}\&_{z_1}$ :

$$\frac{\Gamma \vdash p_0\, {}_{z_0}\&_{z_1}\, p_1}{\Gamma \vdash^{v_0} p_0} \qquad \frac{\Gamma \vdash p_0\, {}_{z_0}\&_{z_1}\, p_1}{\Gamma \vdash^{v_1} p_1} \tag{10.3}$$

By trivializing the implicit reflection rule:

$$\frac{p_0\, {}_{z_0}\&_{z_1}\, p_1 \vdash p_0\, {}_{z_0}\&_{z_1}\, p_1}{p_0\, {}_{z_0}\&_{z_1}\, p_1 \vdash^{v_0} p_0}$$

$$\frac{p_0\, {}_{z_0}\&_{z_1}\, p_1 \vdash p_0\, {}_{z_0}\&_{z_1}\, p_1}{p_0\, {}_{z_0}\&_{z_1}\, p_1 \vdash^{v_1} p_1}$$

we get the *reflection axioms* for ${}_{z_0}\&_{z_1}$ :

$$p_0\, {}_{z_0}\&_{z_1}\, p_1 \vdash^{v_0} p_0 \qquad\qquad p_0\, {}_{z_0}\&_{z_1}\, p_1 \vdash^{v_1} p_1 \tag{10.4}$$

which are axioms with partial truth-evaluations $v_0$ and $v_1$ respectively. Let us put: $p_0 \vdash^{v_0} \Delta$ , $p_1 \vdash^{v_1} \Delta'$

To get the explicit reflection rule, one must use the reflection axioms, and the quantum cut rule (defined in the previous Section):

$$\frac{p_0\ {}_{z_0}\&_{z_1}\ p_1 \vdash^{v_0} p_0 \qquad p_0 \vdash^{v_0} \Delta}{p_0\ {}_{z_0}\&_{z_1}\ p_1 \vdash^{v_0} \Delta} \qquad\qquad \frac{p_0\ {}_{z_0}\&_{z_1}\ p_1 \vdash^{v_1} p_1 \qquad p_1 \vdash^{v_1} \Delta'}{p_0\ {}_{z_0}\&_{z_1}\ p_1 \vdash^{v_1} \Delta'}$$

which leads to the *explicit reflection rule* for ${}_{z_0}\&_{z_1}$ :

$$\frac{p_0 \vdash^{v_0} \Delta}{p_0\ {}_{z_0}\&_{z_1}\ p_1 \vdash^{v_0} \Delta} \qquad\qquad \frac{p_1 \vdash^{v_1} \Delta'}{p_0\ {}_{z_0}\&_{z_1}\ p_1 \vdash^{v_1} \Delta'} \tag{10.5}$$

Let us verify that the explicit reflection rule is equivalent to the initial assertions.
To do that, let us trivialize the explicit reflection rule by putting: $\Delta \equiv p_0$ and $\Delta' \equiv p_1$. We get:

$$\frac{p_0 \vdash^{v_0} p_0}{p_0\ {}_{z_0}\&_{z_1}\ p_1 \vdash^{v_0} p_0} \qquad \frac{p_1 \vdash^{v_1} p_1}{p_0\ {}_{z_0}\&_{z_1}\ p_1 \vdash^{v_1} p_1} \tag{10.6}$$

that is, the ${}_{z_0}\&_{z_1}$ - axioms are derived from the identity axioms.
Moreover, we must verify that from the reflection axioms it is possible to get back the implicit reflection rule.
Then, let us assume $\Gamma \vdash p_0\ {}_{z_0}\&_{z_1}\ p_1$ from the definitional equation of ${}_{z_0}\&_{z_1}$ on the LHS, and let us compose it with the ${}_{z_0}\&_{z_1}$ -axioms:

$$\frac{\Gamma \vdash p_0\ {}_{z_0}\&_{z_1}\ p_1 \qquad p_0\ {}_{z_0}\&_{z_1}\ p_1 \vdash^{v_0} p_0}{\Gamma \vdash^{v_0} p_0} \qquad \frac{\Gamma \vdash p_0\ {}_{z_0}\&_{z_1}\ p_1 \qquad p_0\ {}_{z_0}\&_{z_1}\ p_1 \vdash^{v} p_1}{\Gamma \vdash^{v_1} p_1} \tag{10.7}$$

And, by omitting the axioms in the premises, we get back the implicit reflection rule:

$$\frac{\Gamma \vdash p_0\ {}_{z_0}\&_{z_1}\ p_1}{\Gamma \vdash^{v_0} p_0} \qquad \frac{\Gamma \vdash p_0\ {}_{z_0}\&_{z_1}\ p_1}{\Gamma \vdash^{v_1} p_1}$$

**10.2 General case without Meta-Data**

The Meta-Data *MD* is a constraint which forbids using the connective ${}_{z_0}\&_{z_1}$ more than once. Then, one might conclude that the logic of the qubit (*Lq*) is not properly a logic, at least not in the usual sense. Nevertheless, we think that *Lq* is a first step toward a better understanding of the qubit in logical terms.
Notice that the connective ${}_{z_0}\&_{z_1}$, used once in *Lq*, which is the logic of one qubit in $C^2$, can be used three times in the logic *L2q* of two qubits in $C^4$, seven times in the logic of three qubits *L3q* in $C^8$ and so on. In general, ${}_{z_0}\&_{z_1}$ can be used $2^n - 1$ times in the logic *Lnq* of *n* qubits in $C^{2^n}$. This sudden, discrete change in the number of times that the connective ${}_{z_0}\&_{z_1}$ can be used in *Lnq* resembles a "quantum jump" in QM.
It becomes evident that using more than once the connective ${}_{z_0}\&_{z_1}$ (quantum superposition) requires having at disposal also the connective describing the tensor product, denoted by $\otimes$, of two or more qubits. In Sect.11, the connective "times" of Linear and Basic logic, which is denoted by the same symbol $\otimes$, will be interpreted as the tensor product in the Hilbert space.
However, while the Hilbert space $C^{2^n}$ can be obtained as the tensor product of n copies of $C^2$, that is:

$$\underbrace{C^2 \otimes C^2 \otimes ....C^2}_{n\ times} = C^{2^n}$$

the logic *Lnq* of the *n* qubits of $C^{2^n}$ is not the tensor product of *n* copies of the logic of one qubit of $C^2$:

$$\underbrace{Lq \otimes Lq \otimes ....Lq}_{n\ times} \neq Lnq$$

This was shown in general by a theorem stating the impossibility of making a tensor product of (quantum) logics (Randall and Foulis, 1979).
As we said already, there is then an apparent problem with the *MD*, which constitutes a constraint on the truth-values labelling complete sequents. In fact, the requirement that the truth-values sum up to one forbids using the connective ${}_{z_0}\&_{z_1}$ more than once.
This sounds quite weird in a logical calculus, where connectives can be used for an unlimited number of times. On the other hand, the physical interpretation of truth-values as probabilities would demand the constraint to hold, because probabilities must sum up to 1.
However, let us for the moment disregard a possible physical interpretation and consider the most general case without a constraint. In this case the *MD* would reduce in assuming that the truth-evaluation of a compound proposition would be a function $H(v_0, v_1)$ of the truth-values $v_0$ and $v_1$ of atomic propositions.
Let us denote by $\Phi$ and $\Psi$ two whatsoever propositions of a formal language L, which, as before, can be asserted with degrees $z_0$ and $z_1$ respectively. Then, as before, their truth-values are $v_0$ and $v_1$ respectively. However, in this case we don't assume the *MD* constraint.
In this general case, the definitional equation of the connective ${}_{z_0}\&_{z_1}$ becomes:

$$\Gamma \vdash^{H(v_0,v_1)} \Phi {}_{z_0}\&_{z_1} \Psi \quad \underline{\text{iff}} \quad \Gamma \vdash^{v_0} \Phi \quad \underline{\text{and}} \quad \Gamma \vdash^{v_1} \Psi \qquad (10.8)$$

From the above definitional equation, from the RHS to the LHS, one gets the formation rule for ${}_{z_0}\&_{z_1}$ in the most general case:

$$\frac{\Gamma \vdash^{v_0} \Phi \qquad \Gamma \vdash^{v_1} \Psi}{\Gamma \vdash^{H(v_0,v_1)} \Phi {}_{z_0}\&_{z_1} \Psi} \qquad (10.9)$$

From the definitional equation, from the RHS to the LHS, we get the implicit reflection rule:

$$\frac{\Gamma \vdash^{H(v_0,v_1)} \Phi_{z_0} \&_{z_1} \Psi}{\Gamma \vdash^{v_0} \Phi} \qquad \frac{\Gamma \vdash^{H(v_0,v_1)} \Phi_{z_0} \&_{z_1} \Psi}{\Gamma \vdash^{v_1} \Psi} \tag{10.10}$$

By trivializing the implicit reflection rule, that is, by putting: $\Gamma = \Phi_{z_0} \&_{z_1} \Psi$, we get the $_{z_0}\&_{z_1}$ - axioms:

$$\Phi_{z_0} \&_{z_1} \Psi \vdash^{v_0} \Phi \qquad \Phi_{z_0} \&_{z_1} \Psi \vdash^{v_1} \Psi \tag{10.11}$$

Let us put: $\Phi \vdash^{v_0} \Delta$ and $\Psi \vdash^{v_1} \Delta'$ and compose with the reflection axioms, by using the quantum cut rule:

$$\frac{\Phi_{z_0} \&_{z_1} \Psi \vdash^{v_0} \Phi \quad \Phi \vdash^{v_0} \Delta}{\Phi_{z_0} \&_{z_1} \Psi \vdash^{v_0} \Delta} \qquad \frac{\Phi_{z_0} \&_{z_1} \Psi \vdash^{v_1} \Psi \quad \Psi \vdash^{v_1} \Delta'}{\Phi_{z_0} \&_{z_1} \Psi \vdash^{v_1} \Delta'}$$

which leads to the *explicit reflection rule*:

$$\frac{\Phi \vdash^{v_0} \Delta}{\Phi_{z_0} \&_{z_1} \Psi \vdash^{v_0} \Delta} \qquad \frac{\Psi \vdash^{v_1} \Delta'}{\Phi_{z_0} \&_{z_1} \Psi \vdash^{v_1} \Delta'} \tag{10.12}$$

The next step will be looking for the explicit expression of the function $H(v_0, v_1)$ such that it can reduce to the *MD* constraint $v_0 + v_1 = 1$ for the logic of the qubit *Lq*.
At this point, as we are dealing with truth-values $v_i \in [0,1]$, one might suppose that $H(v_0, v_1)$ is a T-norm, used in many-valued logics (Łukasiewicz, 1920) and Fuzzy logic (Zadeh, 1965).
A T-norm (see Hajék, 1998) is a function * from $[0,1]^2$ to $[0,1]$ which behaves exactly as the classical conjunction on the values $\{0,1\}$ and is commutative, associative, non decreasing and 1 is its unit element. For any $x, y, z \in [0,1]$, the following conditions hold:

$$x * 0 = 0$$
$$x * 1 = x$$
$$x * y = y * x$$
$$x \le y \Rightarrow x * z \le y * z$$

In Fuzzy logic, T-norms play the role of truth functions of conjunction.

There are three most important continuous T-norms:

Łukasiewicz T-norm: $x * y = \max(x + y - 1, 0)$
Gödel T-norm: $x * y = \min(x, y)$
Product T-norm: $x * y = y \cdot x$

The following theorem holds.
**Theorem:** *Every continuous T-norm is locally isomorphic to Łukasiewicz, Gödel or product T-norm.*
In Fuzzy logic, the formation rule of the conjunction & is:

$$\frac{\Gamma \vdash^{x} A \qquad \Gamma \vdash^{y} B}{\Gamma \vdash^{z = x * y} A \& B}$$

This reminds of the ${}_{z_0}\&_{z_1}$-formation rule (Eq. 10.9) in the general case , and one might suppose that $H(v_0, v_1)$ is a T-norm $v_0 * v_1$. However, as we already said, the function $H(v_0, v_1)$ should reduce to the *MD* constraint in the case of physical interest, that is, $H(v_0, v_1) = v_0 + v_1 = 1$. Unfortunately, this constraint cannot be satisfied by any of the three T-norms, because $x * y = 1$ holds only in the trivial case $x = y = 1$ in all of them. By the above theorem, then, $H(v_0, v_1)$ cannot be a T-norm.
Nevertheless, if we choose $H(v_0, v_1)$ to be a particular function of the Łukasiewicz T-norm, namely:

$$H(v_0, v_1) = 1 - T_L(1 - v_0, 1 - v_1) = 1 - \max[1 - (v_0 + v_1), 0] \qquad (10.13)$$

it can reduce to the case of physical interest.
It is worth saying a few words on a possible physical interpretation in this general case.
In this situation, there are only three possibilities:

$$H(v_0, v_1) = \begin{cases} < 1 \\ = 1 \\ > 1 \end{cases} \qquad (10.14)$$

The first possibility concerns a qumix state, that is a mixed state inside the Bloch sphere, described by a density matrix. The definition of the qumix was already given in Sect.5.
The second possibility describes the qubit state, that is a pure state on the surface of the Bloch sphere.
Finally, the third possibility concerns an entity which cannot belong to the Hilbert space, therefore it is outside the framework of Quantum Mechanics. This last case might perhaps give rise to the logic of a Quantum Field Theory of quantum information.

### 10.3 Logical description of the preparation of the optical qubit

The compound proposition $p_0\,{}_{z_0}\&_{z_1}\,p_1$ **(**whose physical interpretation is the qubit state) will be named “proposition *Q*” of the logic *Lq* of the qubit.

**Theorem**: *The compound proposition* $Q \equiv p_0\,{}_{z_0}\&_{z_1}\,p_1$ *has truth-value* $v = 1$**.**

Proof:

Let us consider the identity axioms in Eq.(10.4), which we take as premises, then, we use the ${}_{z_0}\&_{z_1}$-explicit reflection rule of Eq.(10.5) and finally, by using the formation rule for ${}_{z_0}\&_{z_1}$ of Eq. (10.2), and the *MD*, we get:

$$ {}_{z_0}\&_{z_1}\,form\;\frac{{}_{z_0}\&_{z_1}\,\exp refl\;\dfrac{p_0 \vdash^{v_0} p_0}{p_0\,{}_{z_0}\&_{z_1}\,p_1 \vdash^{v_0} p_0} \qquad \dfrac{p_1 \vdash^{v_1} p_1}{p_0\,{}_{z_0}\&_{z_1}\,p_1 \vdash^{v_1} p_1}\;{}_{z_0}\&_{z_1}\,\exp refl}{p_0\,{}_{z_0}\&_{z_1}\,p_1 \vdash^{v_0+v=1} p_0\,{}_{z_0}\&_{z_1}\,p_1} \qquad (10.15) $$

That is, proposition Q is true (with truth-value1). ▫

The logical deduction in (10.15), which is indeed quite trivial, and simply satisfies validity (a deduction is valid if the truth of its premises entails the truth of its conclusion) has nevertheless a physical interpretation which is quite important in the realization of optical quantum computers. The latter are much more robust against decoherence (entanglement with the environment, which destroys quantum superposition) with respect to other kinds of implementation of a quantum computer (for example trapped ions and the like). Then, a lot of theoretical and experimental research is at present focalized on this subject.

The operation of the connective ${}_{z_0}\&_{z_1}$, introduced in *Lq*, can be physically implemented in many different ways, the simplest one, from the logical point of view, is sketched in Fig. 10.1.

**Fig. 10.1**

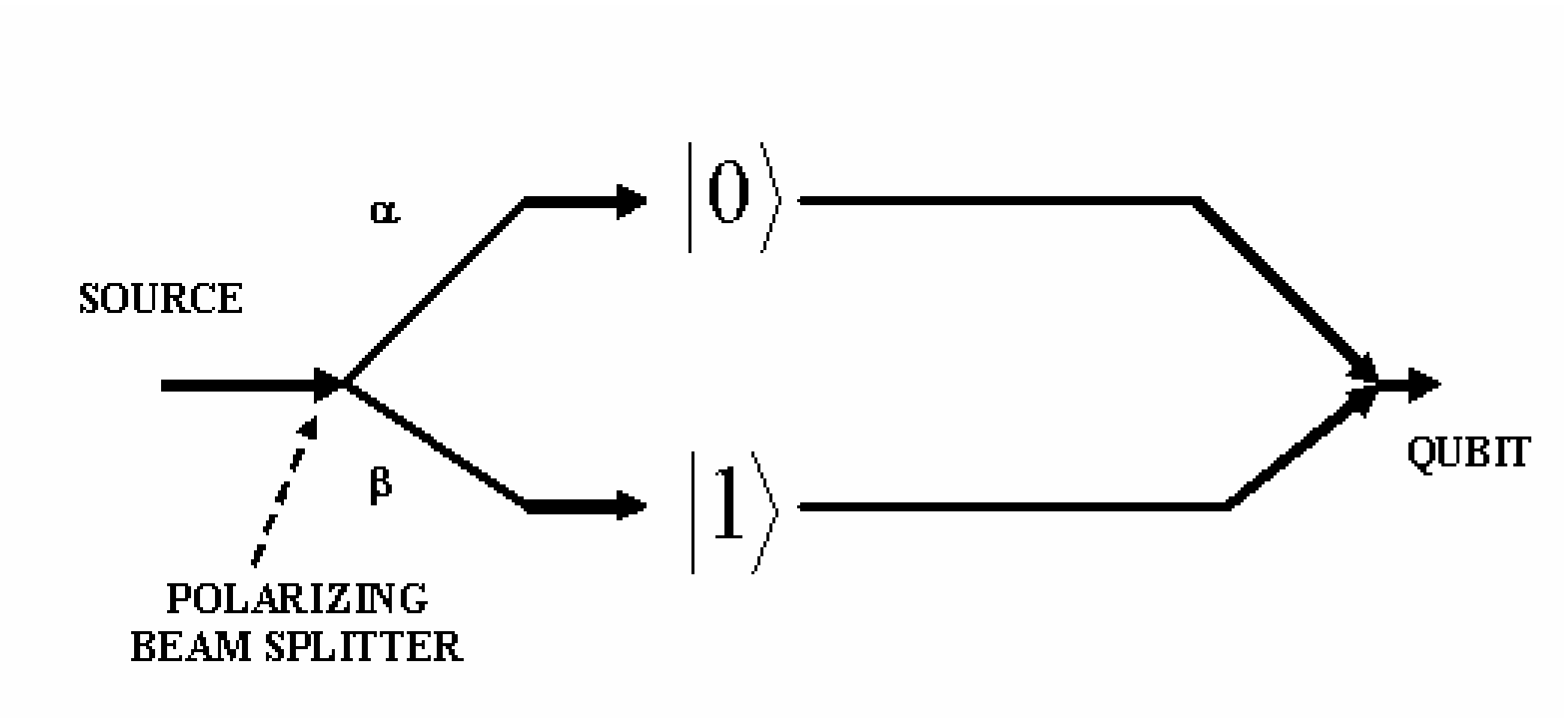

Here the polarizing beam splitter is an optical device which, applied to the light beam emitted by a coherent source (like a laser), splits it into two different beams, corresponding to two different polarization states (horizontal $|H\rangle$ and vertical $|V\rangle$ in most cases), corresponding to the two base states $|0\rangle$ and $|1\rangle$. The amplitudes of each state, here denoted by $\alpha$ and $\beta$, are complex numbers, fulfilling the relationship $\alpha^2+\beta^2=1$. The latter is a direct consequence of the fact that the beam splitter operates in a quantum regime and, therefore, the probabilities (whose total sum must give 1) are computed from the squared modules of the corresponding amplitudes. In this context $\alpha$ and $\beta$ are just corresponding to the $z_0$ and $z_1$ occurring within the logical theory *Lq*. The two beams sum up on the right hand side of Fig.10.1, owing to the presence of another beam splitter, far simpler than the one lying on the left hand side, working in a reverse direction (so as to give rise to a *recombination* of the two beams). The output of the latter is given by $\alpha|0\rangle+\beta|1\rangle$, that is, a qubit.

The $|H\rangle$ and $|V\rangle$- polarizations states with amplitudes $\alpha$ and $\beta$ of the two beams correspond to the premises in the logical deduction (10.1), that is to the identity axioms with partial truth-values $v_0$ and $v_1$. The *recombination* which gives rise to the qubit state corresponds to the step $_{z_0}\&_{z_1}$-form. of the logical deduction (10.1). The intermediate step $_{z_0}\&_{z_1}$-expl. refl. in the two branches of the logical deduction (10.1) corresponds to one of the two alternatives of absorbing the beam with polarization $|H\rangle$ ($|V\rangle$) so that the beam coming out from the second beam splitter has polarization $|V\rangle$($|H\rangle$). Finally, the Meta Data *MD*: $v_0+v_1=1$ used to get the truth-evaluation of the consequence in (10.1) corresponds to the fact that the probabilities of the occurrence of the two polarization states must sum up to 1, that is: $\alpha^2+\beta^2=1$.

We hope that this experiment might clarify the reason why we consider the qubit a compound proposition.

The concrete optical implementation of the simple arrangement shown in Fig.10.1 is rather complex. Namely the polarizing beam splitter is, in the real case, implemented through two successive stages, constituted by a simple polarizing beam splitter (typically a bi-rifrangent plate) producing the two polarized beams corresponding to the two basic states with equal amplitudes, followed by another beam splitter with two input and two output ports with suitable reflection and transmission amplitudes (for a simple introduction to the quantum theory of such beam splitters see Holbrow *et al.*, 2002). As regards the output qubit, its detection would require, in practice, two detectors (for instance avalanche photodiodes) and therefore two output lines, rather than the only one depicted in Fig.10.1. Another complication arises from the fact that, in order to minimize the effect of quantum uncertainty, one needs high photon numbers and therefore high-power lasers, a circumstance which requires, in turn, special devices and arrangements.

## 10.4 The connective ${}_{z_0^*}\vee_{z_1^*}$

We remind that in Basic logic, the connective $\vee$= "or" is the $\perp$- dual of the connective $\&$ = "and". Then, to make the dual of the sequent $\vdash A \& B$, one must make the primitive negation of the atomic propositions A and B, invert the sense of the sequent, and replace $\&$ with $\vee$:

$(\vdash A \& B)^{\perp} = A^{\perp}(\&)^{\perp}B^{\perp}\vdash = A^{\perp} \vee B^{\perp}\vdash$

To make the dual of the connective ${}_{z_0}\&_{z_1}$ we have to take into account the assertion-degrees $z_0, z_1$ and the fact that ${}_{z_0}\&_{z_1}$ is non-commutative. Therefore, we need a new kind of $\perp$-duality, denoted $\perp'$, which, besides replacing the classical $\&$ by the classical $\vee$, also exchanges $z_0$ with $z_1$. Then, we define:

$$({}_{z_0}\&_{z_1})^{\perp'} \equiv {}_{z_1}(\&)^{\perp}{}_{z_0} = {}_{z_1}\vee_{z_0} \tag{10.16}$$

The dual of the compound proposition $p_0\,{}_{z_0}\&_{z_1}p_1$ is then:

$$(p_0\,{}_{z_0}\&_{z_1}p_1)^{\perp'} = p_0{}^{\perp}({}_{z_0}\&_{z_1})^{\perp'}p_1{}^{\perp} = p_1\,{}_{z_1}\vee_{z_0}p_0 = p_0\,{}_{z_0}\vee_{z_1}p_1 \tag{10.17}$$

To make the dual of the sequent $\vdash p_0\,{}_{z_0}\&_{z_1}p_1$ we have to take into account also the fact that in *Lq* it holds the *-duality defined in Sect. 8, by which every time the sense of a sequent is inverted, the assertion-degrees are replaced by their *-duals: $z_i \to z_i{}^*$.
Then, we have:

$$(\vdash p_0\,{}_{z_0}\&_{z_1}p_1)^{\perp'} = p_0\,{}_{z_0^*}\vee_{z_1^*}p_1\vdash \tag{10.18}$$

Let us now perform the $\perp'$–*duality* on both sides of the definitional equation of ${}_{z_0}\&_{z_1}$, in the case $\Gamma = \emptyset$, given in Eq. (9.5):

$$(\vdash p_0\,{}_{z_0}\&_{z_1}p_1)^{\perp'} \quad \underline{\text{iff}} \quad (\vdash^{z_0} p_0)^{\perp'} \quad \underline{\text{and}} \quad (\vdash^{z_1} p_1)^{\perp'}$$

We get:

$$p_{0\ z_0^*}\vee_{z_1^*} p_1 \vdash \quad \underline{\text{iff}} \quad p_0 \vdash^{z_0^*} \quad \underline{\text{and}} \quad p_1 \vdash^{z_1^*} \tag{10.19}$$
$$MD(z_0, z_1): z_0^* \cdot z_0 + z_1^* \cdot z_1 = 1; \quad z_0^* \cdot z_1 + z_1^* \cdot z_0 = 0$$

Eq. (10.19) is the definitional equation of $_{z_0^*}\vee_{z_1^*}$ in the case of an empty consequent $(\Delta = \emptyset)$.
The case $\Delta = \emptyset$ enlightens better the role played by the assertion-degrees $z_0, z_1$ in the introduction of the connective $_{z_0^*}\vee_{z_1^*}$, as it was for $_{z_0}\&_{z_1}$. However, for getting a solution of the definitional equation, it is necessary to consider the case with a non-empty consequent.
When $\Delta \neq \emptyset$, the definitional equation of $_{z_0^*}\vee_{z_1^*}$ is:

$$p_{0\ z_0^*}\vee_{z_1^*} p_1 \vdash^{v_0+v_1=1} \Delta \quad \underline{\text{iff}} \quad p_0 \vdash^{v_0} \Delta \quad \underline{\text{and}} \quad p_1 \vdash^{v_1} \Delta \tag{10.20}$$
$MD$**:** $v_0 + v_1 = 1$

The above definitional equation, from the RHS to the LHS, gives the formation rule of $_{z_0^*}\vee_{z_1^*}$:

$$\frac{p_0 \vdash^{v_0} \Delta \quad p_1 \vdash^{v_1} \Delta}{p_{0\ z_0^*}\vee_{z_1^*} p_1 \vdash^{v_0+v_1=1} \Delta} \tag{10.21}$$

By solving the definitional equation from the LHS to the RHS, we get the implicit reflection rule of $_{z_0^*}\vee_{z_1^*}$:

$$\frac{p_{0\ z_0^*}\vee_{z_1^*} p_1 \vdash^{v_0+v_1=1} \Delta}{p_0 \vdash^{v_0} \Delta} \qquad \frac{p_{0\ z_0^*}\vee_{z_1^*} p_1 \vdash^{v_0+v_1=1} \Delta}{p_1 \vdash^{v_1} \Delta}$$

By trivializing the implicit reflection rule:

$$\frac{p_{0\ z_0^*}\vee_{z_1^*} p_1 \vdash^{1} p_{0\ z_0}\vee_{z_1} p_1}{p_0 \vdash^{v_0} p_{0\ z_0}\vee_{z_1} p_1} \qquad \frac{p_{0\ z_0^*}\vee_{z_1^*} p_1 \vdash^{1} p_{0\ z_0}\vee_{z_1} p_1}{p_1 \vdash^{v_1} p_{0\ z_0}\vee_{z_1} p_1}$$

We get the $_{z_0^*}\vee_{z_1^*}$-axioms:

$$p_0 \vdash^{v_0} p_{0\ z_0}\vee_{z_1} p_1 \qquad p_1 \vdash^{v_1} p_{0\ z_0}\vee_{z_1} p_1 \tag{10.22}$$

Let us put: $\Gamma\vdash^{v_0} p_0$ , $\Gamma'\vdash^{v_1} p_1$ in the premises, together with the reflection axioms. By the quantum cut rule we get:

$$\frac{\Gamma\vdash^{v_0} p_0 \qquad p_0\vdash^{v_0} p_0\,{}_{z_0}\!\vee_{z_1} p_1}{\Gamma\vdash^{v_0} p_0\,{}_{z_0}\!\vee_{z_1} p_1} \qquad \frac{\Gamma'\vdash^{v_1} p_1 \qquad p_1\vdash^{v_1} p_0\,{}_{z_0}\!\vee_{z_1} p_1}{\Gamma'\vdash^{v_1} p_0\,{}_{z_0}\!\vee_{z_1} p_1}$$

which leads to the explicit reflection rule:

$$\frac{\Gamma\vdash^{v_0} p_0}{\Gamma\vdash^{v_0} p_0\,{}_{z_0}\!\vee_{z_1} p_1} \qquad \frac{\Gamma'\vdash^{v_1} p_1}{\Gamma'\vdash^{v_1} p_0\,{}_{z_0}\!\vee_{z_1} p_1} \tag{10.23}$$

In the physical interpretation, the connective ${}_{z_0^*}\!\vee_{z_1^*}$ represents quantum superposition in the dual Hilbert space.

## 10. 5 The connective $\neg$ in *Lq*

In analogy with Basic logic, the definitional equation for the negation $\neg$ is:

$$\vdash^{z_j} \neg p_i \quad \underline{\text{iff}} \quad p_i\vdash^{z_i^*} \tag{10.24}$$

where we remind that it is: $\neg p_i \equiv p_i^{\perp} = p_j$

and $\vdash^{z_j} p_j = \left(p_i\vdash^{z_i^*}\right)^{\perp'}$

where $\perp'$, defined in Sect.8, is the generalization of the usual $\perp$.
From Eq.(9.24), one gets the $\neg$-formation rule:

$$\frac{p_i\vdash^{z_i^*}}{\vdash^{z_j} \neg p_i} \tag{10.25}$$

and the $\neg$- reflection rule:

$$\frac{\vdash^{z_i} p_i}{\neg p_i\vdash^{z_j^*}} \tag{10.26}$$

## 10.6 Structural rules

In *Lq* there are no structural rules. Therefore *Lq* is a sub-structural logic for quantum information.

In fact, our aim was to follow the guideline of Basic logic, to get a sub-structural logic of the qubit, because (structural) Quantum logics failed in describing quantum superposition. Now, Basic logic has only one structural rule, the "exchange", but even that one does not hold in *Lq*. We will explain the reason of this fact in a very intuitive way as follows.

In quantum information, there are only three physical "links" among states: superposition, tensor product, and entanglement (which is a mixture of the two).

The link "superposition" , which is an additive link, is realized by the additive conjunction $_{z_0}\&_{z_1}$ (and its $\perp' - dual$ the additive disjunction $_{z_0^*}\vee_{z_1^*}$) as we have seen in this Section.

As it will be discussed in more detail in Sect.12, the operation "tensor product" can be realized by the multiplicative conjunction $\otimes = "times"$ (and its dual, the multiplicative disjunction $\wp = "par"$). Also, the quantum correlation "entanglement" is realized by the connective @, which is a particular combination of the additive conjunction and the multiplicative disjunction. From all that it follows that a logic of quantum information having multiplicative connectives would describe at least two qubits (in $C^4$). And this is not the case of *Lq*, which is the logic of a single qubit state in $C^2$.

Now, in the exchange rule it is implicit the existence of a multiplicative connective. Let us consider, for example, "exchange left":

$$exchL \; \frac{\Gamma, A, B, \Gamma' |- \Delta}{\Gamma, B, A, \Gamma' |- \Delta}$$

and put, for simplicity: $\Gamma = \Gamma' = 0$. We get:

$$exchL \; \frac{A, B |- \Delta}{B, A |- \Delta}$$

The comma, "," in between propositions A and B is the metalinguistic link which reflects into the multiplicative connective $\otimes = "times"$:

$$\otimes - form \; \frac{A, B |- \Delta}{A \otimes B |- \Delta}$$

Roughly speaking, in sub-structural logics exchange left just says that the connective $\otimes = "times"$ is commutative, as well as exchange right says that the connective $\wp = "par"$ is commutative.

Now, in *Lq* there are no multiplicative connectives as they correspond to tensor products, which are not allowed.

In fact, let us suppose we used the connective $\otimes = "times"$ between the two atomic propositions, $p_0$ and $p_1$ of *Lq*: $p_0 \otimes p_1$, which is not anymore a proposition of *Lq*. In fact, we remind that $p_0$ and $p_1$ are interpreted as the (non-hermitian) operators $O_0$ and $O_1$, whose tensor product $O_0 \otimes O_1$ is an operator of $C^4$ (the tensor product is expressed by the same symbol $\otimes$ used for the connective "times").

However, also for the logics describing more than one qubit (for example the logic of $C^4$) the exchange rule does not hold in general. In fact, the exchange rule, as we said implies that the connective "times" (and "par") is commutative. In the physical interpretation this would mean that the tensor product of two operators of the Hilbert space is commutative, which in general is not true.
Therefore every logic of quantum information, for whatever number N of qubits, will be no-structural (it will have no structural rules).

## 10.7 Summary: the calculus *Lq*

### Logical rules:

$$_{z_0}\&_{z_1} - form \frac{\Gamma \vdash^{v_0} p_0 \qquad \Gamma \vdash^{v_1} p_1}{\Gamma \vdash^{v_0+v_1=1} p_0 \, {}_{z_0}\&_{z_1} \, p_1}$$

$$_{z_0^*}\vee_{z_1^*} - form \frac{p_0 \vdash^{v_0} \Delta \qquad p_1 \vdash^{v_1} \Delta}{p_0 \, {}_{z_0^*}\vee_{z_1^*} \, p_1 \vdash^{v_0+v_1=1} \Delta}$$

$$_{z_0}\&_{z_1} - refl \frac{p_0 \vdash^{v_0} \Delta}{p_0 \, {}_{z_0}\&_{z_1} \, p_1 \vdash^{v_0} \Delta} \qquad \frac{p_1 \vdash^{v_1} \Delta'}{p_0 \, {}_{z_0}\&_{z_1} \, p_1 \vdash^{v_1} \Delta'}$$

$$_{z_0^*}\vee_{z_1^*} - refl \frac{\Gamma \vdash^{v_0} p_0}{\Gamma \vdash^{v_0} p_0 \, {}_{z_0}\vee_{z_1} \, p_1} \qquad \frac{\Gamma' \vdash^{v_1} p_1}{\Gamma' \vdash^{v_1} p_0 \, {}_{z_0}\vee_{z_1} \, p_1}$$

$$\neg - form \frac{p_i \vdash^{z_i^*}}{\vdash^{z_j} \neg p_i}$$

$$\neg - refl \frac{\vdash^{z_i} p_i}{\neg p_i \vdash^{z_j^*}}$$

### Identity axioms:

$$p_0 \vdash^{v_0} p_0 \qquad\qquad p_1 \vdash^{v_1} p_1$$

### Cut rule:

$$\frac{\Gamma \mid\!-^{v_i} p_i \quad p_i \mid\!-^{v_i} \Delta}{\Gamma \mid\!-^{v_i} \Delta}$$

## 11. The logic of two qubits (*L2q*)

In this Section, we will introduce the logic of two qubits, *L2q* (see also Zizzi, 2006). We remind that a 2-qubits state is a vector state of the 4-dimensional complex Hilbert space $C^4 = C^2 \otimes C^2$ where $\otimes$ indicates the tensor product .
The computational basis of $C^4$ is:

$$|0\rangle \otimes |0\rangle,\ |0\rangle \otimes |1\rangle,\ |1\rangle \otimes |0\rangle,\ |1\rangle \otimes |1\rangle. \tag{11.a}$$

The four one-dimensional projectors of $C^4$ are:

$$P_{00} = P_0 \otimes P_0,\ P_{01} = P_0 \otimes P_1,\ P_{10} = P_1 \otimes P_0,\ P_{11} = P_1 \otimes P_1 \tag{11.b}$$

The most general state of $C^4$ is a linear superposition of the four basis states, namely:

$$\alpha|00\rangle + \beta|01\rangle + \gamma|10\rangle + \delta|11\rangle \tag{11.c}$$

with: $\alpha, \beta, \gamma, \delta \in C$ and $|\alpha|^2 + |\beta|^2 + |\gamma|^2 + |\delta|^2 = 1$ (11.d)

where in (11.c) the symbol of the tensor product has been omitted.
The general 2-qubits state in (11.c) is in general non-separable, that is in general it cannot be written as the tensor product of two qubits :
$\lambda_0|0\rangle + \lambda_1|1\rangle$ with $|\lambda_0|^2 + |\lambda_1|^2 = 1$ and $\lambda_0'|0\rangle + \lambda_1'|1\rangle$ with $|\lambda_0'|^2 + |\lambda_1'|^2 = 1$.
The most general state of $C^4$ in (11.c) is then said "entangled", just meaning "non-separable".
In particular, a 2-qubits state is said maximally entangled when it is one of the four Bell's states (Bell, 1987):

$$|\Phi_{\pm}\rangle_{AB} = \frac{1}{\sqrt{2}}\left(|0\rangle_A \otimes |0\rangle_B \pm |1\rangle_A \otimes |1\rangle_B\right) \tag{11.e}$$

$$|\Psi_{\pm}\rangle_{AB} = \frac{1}{\sqrt{2}}\left(|0\rangle_A \otimes |1\rangle_B \pm |1\rangle_A \otimes |0\rangle_B\right) \tag{11.f}$$

For simplicity, in the following, we will consider only maximally entangled states, that is Bell states.

### 11.1 Language of *L2q*

*Atomic propositions*: $A, B$ and their primitive negations $A,^{\perp}\ B^{\perp}$

*Logical connectives*:

The additive conjunction $_{\frac{1}{\sqrt{2}}}\&_{\frac{1}{\sqrt{2}}}$

The additive disjunction $_{\frac{1}{\sqrt{2}}}\vee_{\frac{1}{\sqrt{2}}}$

The negation $\neg$

The multiplicative conjunction $\otimes$ = "times"

The multiplicative disjunction $\wp$ = "par"

The quantum entanglement @ = "at"

The dual of @ § = "in"

.

In the following we will consider only the case where the quantum additive conjunction and the quantum additive disjunction (introduced in Sect.10 for the logic *Lq*) are labelled by equal fixed values of $z_0$ and $z_1$, precisely: $z_0 = z_1 = \frac{1}{\sqrt{2}}$.

For the sake of brevity, we will denote the commutative $_{\frac{1}{\sqrt{2}}}\&_{\frac{1}{\sqrt{2}}}$ **and** $_{\frac{1}{\sqrt{2}}}\vee_{\frac{1}{\sqrt{2}}}$ simply by the usual & an $\vee$ , with the logical rules given in Sect.7.

Also, the logical rules of the negation, the "times" and "par" are those already given in Sect.7 for Basic logic.

Then, in the following we will just introduce the new connective @ (and its dual §) with the corresponding logical rules.

**11.2 Interpretation**

In the physical interpretation *I,* the atomic propositions we are considering are the four projectors of the Hilbert space $C^4$ multiplied by the real number $\frac{1}{\sqrt{2}}$ ( instead of a complex number $z_i$ as it was in the case of *Lq***)** because we restrict ourselves to Bell' s states.

$$p_i \xrightarrow{I} \frac{1}{\sqrt{2}} P_i \qquad (i = 0,1,2,3)$$

The connective $_{\frac{1}{\sqrt{2}}}\&_{\frac{1}{\sqrt{2}}}$ is interpreted as the quantum superposition in the "cat state": $\frac{1}{\sqrt{2}}(|0\rangle + |1\rangle)$ .

The connective $\otimes$ = "times" is interpreted as the tensor product $\otimes$ in the Hilbert space $C^4 = C^2 \otimes C^2$.
The connective @ = "at" is interpreted as maximal quantum entanglement. A maximally entangled bipartite state is, for example, the Bell state:

$$\frac{1}{\sqrt{2}}(|0\rangle_A \otimes |0\rangle_B + |1\rangle_A \otimes |1\rangle_B)$$

where the labels A and B refer to two qubits $|Q_A\rangle$ and $|Q_B\rangle$. The bipartite state of two entangled qubits cannot be written as a tensor product of the two qubit states. For this reason we needed to introduce a new logical connective for describing quantum entanglement, as the connective "times" on its own, is not sufficient. The connective @ will necessarily be a particular combination of multiplicative and additive connectives.

### 11.3 The connective @ = "entanglement"

In this Subsection, we will introduce the logical connective "quantum entanglement", denoted by @. This connective, which is a function of the connectives "quantum superposition" and "par", has, nevertheless, its own definitional equation.
The connective @ is fundamental for a logical description of the quantum world, as quantum entanglement is a peculiar feature of Quantum Mechanics, with no classical analogous. Of course, here we are dealing with a complex Hilbert space which is at least four-dimensional, like $C^4$, because we need at least two qubits to get an entangled state. Therefore, we will need also multiplicative connectives, like "times" and "par" to describe the tensor product of states. In fact, as we said in the previous Subsection , the connective "times" between two propositions of $L_{2q}$ is interpreted as the tensor product in the Hilbert space. The same, of course, holds for the dual "par" in the dual Hilbert space.
In the case of the (one qubit) "cat state":

$$\frac{1}{\sqrt{2}}(|0\rangle + |1\rangle) \qquad (\lambda_0 = \lambda_1 = \frac{1}{\sqrt{2}}) \tag{11.1}$$

The connective ${}_{z_0}\&_{z_1}$ of quantum superposition, introduced in Sect.9, for $z_0 = z_1 = \frac{1}{\sqrt{2}}$, reduces to ${}_{\frac{1}{\sqrt{2}}}\&_{\frac{1}{\sqrt{2}}}$, which is commutative:

$$p_0 \; {}_{\frac{1}{\sqrt{2}}}\&_{\frac{1}{\sqrt{2}}} \; p_1 = p_1 \; {}_{\frac{1}{\sqrt{2}}}\&_{\frac{1}{\sqrt{2}}} \; p_0 \tag{11.2}$$

Due to the fact that we are restricted to the commutative case, all the inference rules for ${}_{\frac{1}{\sqrt{2}}}\&_{\frac{1}{\sqrt{2}}}$ will be the same as those for & in Basic logic.

Of course, the semantics of this logic is different from that of Basic logic, as it can be seen from the metalanguage in the definitional equation of $_{\frac{1}{\sqrt{2}}}\&_{\frac{1}{\sqrt{2}}}$ :

$$\Gamma \vdash^{v=v_0+v_1=1} p_0 \,_{\frac{1}{\sqrt{2}}}\&_{\frac{1}{\sqrt{2}}}\, p_1 \quad \underline{\text{iff}} \quad \Gamma \vdash^{v_0=\frac{1}{2}} p_0(x_0) \quad \underline{\text{and}} \quad \Gamma \vdash^{v_1=\frac{1}{2}} p_1(x_1) \qquad (11.3)$$

This is still a many-valued logic, actually it is three-valued, the possible truth-values being 0, 1, $\frac{1}{2}$.

We have introduced the "cat state", because in the logical treatment of quantum entanglement, we will consider only maximally entangled states, namely Bell's states, which are a particular quantum superposition (with a global factor $\frac{1}{\sqrt{2}}$) of bipartite states.

In what follows, we will then omit, for simplicity, this global factor in all definitions and formulas.

For simplicity, in the following, we will consider only maximally entangled states, that is, the Bell states. As we have seen, expressing the cat state $|Q\rangle_A = \frac{1}{\sqrt{2}}(|0\rangle_A + |1\rangle_A)$ in logical terms leads to the compound proposition $Q_A = p_0 \,\&\, p_1$ (where the global factor $\frac{1}{\sqrt{2}}$ has been omitted and & stands for the classical connective "and"). For brevity, we replace the notation $p_1$ with *A*, and $p_0$ (which is the primitive negation of $p_1$) with $A^{\perp}$. Therefore, we will write: $Q_A = A \,\&\, A^{\perp}$.

In the same way, the second qubit is expressed, in logical terms, by a second compound proposition $Q_B = p_0' \& p_1'$, and, as before, we replace the notation $p_1'$ with *B*, and $p_0'$ with $B^{\perp}$. Therefore, we will write: $Q_B = B \,\&\, B^{\perp}$.

Bell's states will be expressed, in logical terms, by the expression:

$$Q_A @ Q_B \qquad (11.4)$$

where @ is the new logical connective to be introduced, called "entanglement". Like all the other connectives, @ will be introduced by the reflection principle, which reflects the metalanguage into the object-language. We have at our disposal a metalanguage which comes from our knowledge of the physical structure of Bell's states. This leads us to figure out the logical structure for, say, the Bell states $|\Phi_{\pm}\rangle$ in (11.e) as :

$$(A \wp B) \,\&\, (A^{\perp} \wp B^{\perp}) \qquad (11.5)$$

Where we recall that $\wp$ is the multiplicative disjunction in Linear logic and Basic logic.
Similarly, the logical structure for the Bell's states $|\Psi_{\pm}\rangle$ in (11.f) will be:

$$(A \wp B^{\perp}) \& (A^{\perp} \wp B) \qquad (11.6)$$

In the following, we will consider only the logical expression for the states $|\Phi_{\pm}\rangle$, as the case for $|\Psi_{\pm}\rangle$ is obtained by just exchanging *A* with $A^{\perp}$ (and *B* with $B^{\perp}$).

***Definition*:**
*Two compound propositions $Q_A = A \& A^{\perp}$ and $Q_B = B \& B^{\perp}$ will be said (maximally) entangled if they are linked by the connective @ = "entanglement".*

### 11.4 Definitional equation and logical rules for @

The definitional equation for @ is:

$$\Gamma \vdash Q_A @ Q_B \quad \underline{\text{iff}} \quad \Gamma \vdash A, B \quad \underline{\text{and}} \quad \Gamma \vdash A^{\perp}, B^{\perp} \qquad (11.7)$$

where $Q_A = A \& A^{\perp}$ and $Q_B = B \& B^{\perp}$.
On the RHS of the definitional equation, we have the metalanguage, coming from our knowledge of the physical structure of Bell's states. On the LHS, instead, we have the object-language. It should be noticed that, on the RHS of the definitional equation, each of the two commas is reflected into a $\wp$, while the meta <u>and</u> is reflected into &.
Thus the connective @ is an additive as well as multiplicative connective (more exactly, additive conjunction and multiplicative disjunction), which reflects two kinds of "and" on the right: one outside the sequent (<u>and</u>) and one inside the sequent (the comma). Finally, the connective @, is a derived connective which, nevertheless, has its own definitional equation. The solution of the definitional equation for @ gives the following logical rules:

*@-form* $$\frac{\Gamma \vdash A, B \quad \Gamma \vdash A^{\perp}, B^{\perp}}{\Gamma \vdash Q_A @ Q_B} \qquad (11.8)$$

*@-impl. refl.* $$\frac{\Gamma \vdash Q_A @ Q_B}{\Gamma \vdash A, B} \qquad \frac{\Gamma \vdash Q_A @ Q_B}{\Gamma \vdash A^{\perp}, B^{\perp}} \qquad (11.9)$$

*@- axioms* $$Q_A @ Q_B \vdash A, B \qquad Q_A @ Q_B \vdash A^{\perp}, B^{\perp} \qquad (11.10)$$

*@-expl.refl.* $$\frac{A \vdash \Delta \quad B \vdash \Delta'}{Q_A @ Q_B \vdash \Delta, \Delta'} \qquad \frac{A^{\perp} \vdash \Delta \quad B^{\perp} \vdash \Delta'}{Q_A @ Q_B \vdash \Delta, \Delta'} \qquad (11.11)$$

The formation rule of @ is obtained from the @-definitional equation, from the RHS to the LHS. The implicit reflection of @ is obtained from the @-definitional

equation, from the LHS to the RHS. The @- reflection axioms are obtained by the trivialization procedure, that is, setting $\Gamma = Q_A @ Q_B$ in the implicit reflection. The @-explicit reflection rules are obtained by composition of the @- reflection axioms with the premises $A \vdash \Delta$ and $B \vdash \Delta'$, and $A^{\perp} \vdash \Delta$ and $B^{\perp} \vdash \Delta'$ respectively.

**11.5 Properties of @**

1) Commutativity of @

If the exchange rule did hold, then the connective "entanglement" would be commutative as one can easily prove.

Let us now prove $\mathrm{Q_A} @ \mathrm{Q_B} \doteq \mathrm{Q_B} @ \mathrm{Q_A}$ in the direction $\mathrm{Q_A} @ \mathrm{Q_B} \vdash \mathrm{Q_B} @ \mathrm{Q_A}$ (the proof in the opposite direction is quite the same):

$$\dfrac{\dfrac{\dfrac{A \vdash A \quad B \vdash B}{\mathrm{Q_A} @ \mathrm{Q_B} \vdash \mathrm{A,B}}\,@-\mathit{refl.}}{\mathrm{Q_A} @ \mathrm{Q_B} \vdash \mathrm{B,A}}\,\mathit{exch.\ R} \qquad \dfrac{\dfrac{A^{\perp} \vdash A^{\perp} \quad B^{\perp} \vdash B^{\perp}}{\mathrm{Q_A} @ \mathrm{Q_B} \vdash \mathrm{A}^{\perp},\mathrm{B}^{\perp}}\,@-\mathit{refl.}}{\mathrm{Q_A} @ \mathrm{Q_B} \vdash \mathrm{B}^{\perp},\mathrm{A}^{\perp}}\,\mathit{exch.\ R}}{\mathrm{Q_A} @ \mathrm{Q_B} \vdash \mathrm{Q_B} @ \mathrm{Q_A}}\,@-\mathit{form.} \tag{11.12}$$

Instead, in *Lq*, the exchange rule does not hold. Therefore, in the general case, the connective @ is not commutative.

Nevertheless, there are two particular cases in which the connective @ is commutative:

i) $B = A$, corresponding to the Bell state: $\frac{1}{\sqrt{2}}(|0\rangle \otimes |0\rangle + |1\rangle \otimes |1\rangle)$

ii) $B = A^{\perp}$, corresponding to the Bell state: $\frac{1}{\sqrt{2}}(|0\rangle \otimes |1\rangle + |1\rangle \otimes |0\rangle)$

In i) the commutativity of @ trivially follows from the fact that in this case commutativity of "times" ("par") holds.

ii) For $B = A^{\perp}$, we give the proof:

$$\dfrac{\dfrac{\dfrac{A \vdash A \quad A^{\perp} \vdash A^{\perp}}{\mathrm{Q_A} @ \mathrm{Q}_{A^{\perp}} \vdash \mathrm{A},\mathrm{A}^{\perp}}\,@-\mathit{refl.}}{\mathrm{Q_A} @ \mathrm{Q}_{A^{\perp}} \vdash \mathrm{A} \wp \mathrm{A}^{\perp}}\,\wp-\mathit{form.} \qquad \dfrac{\dfrac{A^{\perp} \vdash A^{\perp} \quad A \vdash A}{\mathrm{Q_A} @ \mathrm{Q}_{A^{\perp}} \vdash \mathrm{A}^{\perp},\mathrm{A}}\,@-\mathit{refl.}}{\mathrm{Q_A} @ \mathrm{Q}_{A^{\perp}} \vdash \mathrm{A}^{\perp} \wp \mathrm{A}}\,\wp-\mathit{form}}{\mathrm{Q_A} @ \mathrm{Q}_{A^{\perp}} \vdash (A \wp A^{\perp})\,\&\,(A^{\perp} \wp A)}\,\&-\mathit{form.} \tag{11.13}$$

Where in the conclusion in Eq.(11.13), it is:

$$(A \wp A^{\perp})\,\&\,(A^{\perp} \wp A) \equiv Q_{A^{\perp}} @ Q_A \tag{11.14}$$

Owing to the commutativity of &.

Therefore, the conclusion of Eq. (11.13) is:

$$Q_A @ Q_{A^\perp} \doteq Q_{A^\perp} @ Q_A \qquad (11.15)$$

We remark that the cases i) and ii) are the only ones of physical interest.

2) Semi-distributivity

From the definitional equation of @ with $\Gamma = \varnothing$, that is:

$$\vdash Q_A @ Q_B \quad \underline{\text{iff}} \quad \vdash A, B \quad \underline{\text{and}} \quad \vdash A^\perp, B^\perp \qquad (11.16)$$

we get:

$$(A \,\&\, A^\perp) @ (B \,\&\, B^\perp) \doteq (A \wp B) \,\&\, (A^\perp \wp B^\perp) \qquad (11.17)$$

by which we see that two terms are missing, namely $(A \wp B^\perp)$ and $(A^\perp \wp B)$, so that @ has distributivity with absorption, which we call semi-distributivity.

3) Associativity

To discuss associativity of @, a third qubit $Q_C$ is needed, and

$$Q_A @ (Q_B @ Q_C) \doteq (Q_A @ Q_B) @ Q_C \qquad (11.18)$$

cannot be demonstrated in Basic logic, as $Q_C$ acts like a context on the right.

We remind that the maximally entangled state of three qubits is the GHZ state (Greenberger *et al.*, 1990), namely: $|\Psi\rangle_{ABC} = \frac{1}{\sqrt{2}}(|0\rangle_A |0\rangle_B |0\rangle_C + |1\rangle_A |1\rangle_B |1\rangle_C)$

4) Duality

Let us define now the dual of @ :

$$(Q_A @ Q_B)^\perp \equiv \left[(A \wp B) \,\&\, (A^\perp \wp B^\perp)\right]^\perp = (A \otimes B) \vee (A^\perp \otimes B^\perp) \qquad (11.19)$$

and let us call it §, that is:

$$(Q_A @ Q_B)^\perp \equiv (Q_A \S Q_B)$$

(*vice-versa*, the dual of § is @)

The definition of the connective § is then:

$$Q_A \S Q_B \equiv (A \otimes B) \vee (A^\perp \otimes B^\perp) \qquad (11.20)$$

The connective § is to be interpreted in the dual Hilbert space.

5) Non-idempotence

**Theorem 11.1**: *The connective @ is not idempotent*

$$Q_A @ Q_A \neq Q_A \qquad (11.21)$$

The formal proof of this theorem will be given in Subsection 11.6.
Here we give an informal proof of the non-idempotence of @
We want to prove:
$Q_A @ Q_A \neq Q_A$
From the definition of @ in Eq.(11.7), by replacing $B$ with $A$ we get:
$Q_A @ Q_A \doteq (A \wp A) \& (A^{\perp} \wp A^{\perp}) \neq Q_A$
As $\wp$ is non-idempotent. In fact, to prove the idempotence of $\wp$ would require the validity of both the contraction and weakening rules.
If one makes the formal proof, one has to go both ways: to show that $A \mid\!- A \wp A$ does not hold because of the absence of the weakening rule, and that $A \wp A \mid\!- A$ does not hold because of the absence of the contraction rule.
If instead weakening and contraction did hold, then $\wp \equiv \vee$ ($\otimes \equiv \&$), and from the definition of @ , we would get:
$Q_A @ Q_A \doteq (A \vee A) \& (A^{\perp} \vee A^{\perp}) = Q_A$
Because of the idempotence of $\vee$.
In that case, from the definition of the dual of @ , namely:
$Q_A \S Q_B \doteq (A \otimes B) \vee (A^{\perp} \otimes B^{\perp})$
We will get also:
$Q_A \S Q_A \doteq (A \& A) \vee (A^{\perp} \& A^{\perp}) = Q_A$
Because of the idempotence of &.
Notice, in particular, that the formal proof that the dual of $\wp$, namely $\otimes =$"*times*" is non- idempotent, would exchange the roles of the contraction and the weakening rules used in the proof done for $\wp$ .
The fact that $\otimes$ is non-idempotent, leads to the result that the dual of @, namely §, is non-idempotent either. Then, @ (§) is non-idempotent because $\wp$ ($\otimes$) is non-idempotent.
This illustrates an obvious physical fact: self-entanglement (entanglement of a qubit with itself) is impossible as it would require a quantum clone, which is forbidden by the no-cloning theorem.
Then, one can say that the two main no-go theorems of quantum computing, namely the no-cloning and no-erase theorems are (logically) dual to each other. And the no self-entanglement "corollary" is a consequence of the first one, when entanglement is expressed in terms of §, and a consequence of the second one, when entanglement is expressed by the dual, @.

One can interpret the fact that @ (as well as its dual §) is not idempotent by the following no-go theorem in quantum computing.

**"No self-entanglement" theorem:**
*i) It is impossible to get a qubit entangled with itself.*
As the fact that @ (§) is not idempotent depends on the fact that $\wp$ ($\otimes$) is not idempotent leads to the second part of the theorem:
*ii). The impossibility of self-entanglement is a consequence of the no-erase (no-cloning) theorem.*

Schematically:

**BASIC LOGIC**

| | | | | |
|---|---|---|---|---|
| **No contraction** | **No weakening** | $\longleftrightarrow$ | **No contraction** | **No weakening** |
| $\downarrow$ | $\downarrow$ | **Symmetry** | $\downarrow$ | $\downarrow$ |
| $A \vdash A \otimes A$ | $A \otimes A \vdash A$ | $\otimes \longleftrightarrow \wp$ | $A \wp A \vdash A$ | $A \vdash A \wp A$ |
| Cannot be proved | | | | Cannot be proved |
| $\downarrow$ | | | | $\downarrow$ |
| $Q_A \S Q_A \neq Q_A$ | | **Non-idempotence of § and @** | | $Q_A @ Q_A \neq Q_A$ |

**QUANTUM COMPUTING**

| | | |
|---|---|---|
| **No-cloning** | $\longleftrightarrow$ | **No-erase** |
| $\lvert\Psi\rangle \neq \lvert\Psi\rangle \otimes \lvert\Psi\rangle$ | $\downarrow$ | $\lvert\Psi\rangle \otimes \lvert\Psi\rangle \neq \lvert\Psi\rangle$ |
| | **No self-entanglement** | |

## 11.6 Formal proof of non-idempotence of @

We try proving $Q_A @ Q_A \doteq Q_A$.
Let us try first $Q_A @ Q_A \vdash Q_A$.
There are no rules of Basic logic that we can use in the derivation, which can lead to a proof:

$$\frac{}{Q_A @ Q_A \vdash Q_A}$$

And, as the cut-elimination theorem holds in Basic logic (Faggian and Sambin, 1998) we are sure that there are no other rules leading to a proof.

Let us try now the other way around: $Q_A \mid\!- Q_A @ Q_A$.

The only rule we can use in the derivation is the @-formation rule, and there are no further rules in Basic logic, which would lead to a proof:

$$\cfrac{\overline{\quad Q_A \mid\!- Q_A, Q_A \quad}}{Q_A \mid\!- Q_A @ Q_A} @- \textit{form.}$$

And, again, because of cut-elimination, we are sure that there are no other rules leading to a proof.

For the sake of the physical interpretation, we show now that in the case the contraction and weakening rules did hold, the proof would be possible.

Let us prove first $Q_A @ Q_A \mid\!- Q_A$.

$$\cfrac{\cfrac{Q_A \mid\!- Q_A \qquad Q_A \mid\!- Q_A}{Q_A @ Q_A \mid\!- Q_A, Q_A} @- \exp l.\textit{refl.}}{Q_A @ Q_A \mid\!- Q_A} \textit{contr.} \qquad (11.22)$$

Let us prove now the other way around $Q_A \mid\!- Q_A @ Q_A$.

$$\cfrac{\cfrac{Q_A \mid\!- Q_A}{Q_A \mid\!- Q_A, Q_A} \textit{weak.}}{Q_A \mid\!- Q_A @ Q_A} @- \textit{form.} \qquad (11.23)$$

It is impossible to prove $Q_A \mid\!- Q_A @ Q_A$ in Basic logic, because the weakening rule (in the step *weak.*) does not hold. In conclusion, it is impossible to prove the idempotence of @ in Basic logic, because of the absence of the two structural rules of weakening and contraction.

### 11.7 The EPR rule

Let us consider the cut: $\cfrac{\mid\!- Q_A \quad Q_A \mid\!- A}{\mid\!- A} cut$, which corresponds, in physical terms, to measuring the qubit $|Q\rangle_A$ in state $|1\rangle_A$ .

In the same way, the cut: $\cfrac{\mid\!- Q_A \quad Q_A \mid\!- A^{\perp}}{\mid\!- A^{\perp}} cut$, corresponds to measuring the qubit $|Q\rangle_A$ in state $|0\rangle_A$ .

The cut:

$$\dfrac{\dfrac{\vdash Q_A @ Q_B \qquad Q_A @ Q_B \vdash A, B}{\vdash A, B}\; cut}{\vdash A \wp B}\; \wp - formation \tag{11.24}$$

corresponds, in physical terms, to measure the state $|1\rangle_A |1\rangle_B$. If we replace A and B in Eq.(11.24) with $A^{\perp}, B^{\perp}$, the cut corresponds to measure the state $|0\rangle_A |0\rangle_B$ It should be noticed, that, if we make a measurement of $Q_A$ (supposed entangled with $Q_B$) and get $A$ , then, by semi-distributivity of @, we have:

$$A @ Q_B \equiv A @ (B \,\&\, B^{\perp}) \doteq A \wp B \tag{11.25}$$

As it is well known in Quantum Mechanics, if two quantum systems $S_A$ and $S_B$ are entangled, they share a unique quantum state, and even if they are far apart, a measurement performed on $S_A$ influences any subsequent measurement performed on $S_B$ (the EPR "paradox"). Let us consider Alice, who is an observer for system $S_A$, which is the qubit $Q_A$, that is, she can perform a measurement of $Q_A$. There are two possible outcomes, with equal probability 1/2:
i) Alice measures 1, and the Bell state collapses to $|1\rangle_A |1\rangle_B$ .
ii) Alice measures 0, and the Bell state collapses to $|0\rangle_A |0\rangle_B$.
Now, let us suppose Bob is an observer for system $S_B$ (the qubit $Q_B$). If Alice has measured 1, any subsequent measurement of $Q_B$ performed by Bob always returns 1. If Alice measured 0, instead, any subsequent measurement of $Q_B$ performed by Bob always returns 0.
To discuss the EPR paradox in logical terms, we introduce the EPR rule:

$$\dfrac{\Gamma \vdash Q_A @ Q_B \qquad Q_A \vdash A}{\Gamma \vdash A @ Q_B}\; EPR \tag{11.26}$$

The EPR rule does cut a formula, but it is not equivalent to the cut rule (and, vice-versa, the cut rule is not equivalent to the EPR rule) as we will prove.

**Theorem 11.2**: *The EPR rule is not equivalent to the cut rule, and vice-versa.*
Proof:
Let us show first that the EPR rule is not equivalent to the cut rule. We start with the premises of the EPR rule and apply the cut rule:

$$
\text{weak.}L\;\dfrac{\dfrac{\dfrac{\Gamma\vdash Q_A @ Q_B}{\Gamma\vdash Q_A @ Q_B \quad Q_A @ Q_B \vdash A, B}\,@-\text{axiom}}{\Gamma\vdash A, B}\,\text{cut} \qquad Q_A\vdash A}{\dfrac{\dfrac{\Gamma\vdash A, B, Q_A \qquad Q_A\vdash A}{\Gamma\vdash A, A, B}\,\text{cut}}{\Gamma\vdash A, B}\,\text{contr.}R} \tag{11.27}
$$

It is impossible to demonstrate that the EPR rule is equivalent to the cut rule in Basic logic, where we don't have the structural rules of weakening and contraction. And in any logic with structural rules, the connective entanglement disappears, and the EPR rule collapses to the cut rule.

Now, we will show the vice-versa, i.e., that the cut rule is not equivalent to the EPR rule. We start with the premises of the cut rule and apply the cut rule:

$$
\dfrac{\Gamma\vdash Q_A @ Q_B \qquad \dfrac{Q_A @ Q_B\vdash A, B}{Q_A @ Q_B, Q_A\vdash A, B}\,\text{weak.}L}{\Gamma, Q_A @ Q_B\vdash A @ Q_B, B}\,EPR^C \tag{11.28}
$$

Where, in Eq.(11.28), $EPR^C$ means: "EPR rule in presence of contexts" (here $Q_A @ Q_B$ on the left and $B$ on the right). But contexts are absent in Basic logic (visibility). Furthermore, the weakening rule is not present in Basic logic. These facts lead to the conclusion that in Basic logic it is impossible to prove that the cut is equivalent to the EPR rule. Moreover, this is impossible also in sub-structural rules with contexts (like BL, BR, and BLR) because one cannot use weakening, and in structural logics because the connective entanglement disappears. In conclusion, the EPR rule is a new kind of cut peculiar to entanglement, which is possible only in *Lq*. It is a stronger rule (although less general) than the "cut over entanglement", as it uses a weaker premise to yield the same consequence.

Hence, instead of proving $Q_A @ Q_B$ in the usual way, that is, $\vdash Q_A @ Q_B$, we can just prove $Q_A$, (that is, $\vdash Q_A$), perform the usual cut (over $Q_A$): $\dfrac{\vdash Q_A \quad Q_A\vdash A}{\vdash A}\,\text{cut}$ and leave the result $A$ entangled with $Q_B$. Roughly speaking, if two compound propositions are (maximally) entangled to each other, then it is sufficient to prove only one of them. This is the logical analogue of the EPR paradox.

### 11.8 Summary: The calculus $L_{2q}$

$$
\&-\text{form}\;\dfrac{\Gamma\vdash A \qquad \Gamma\vdash B}{\Gamma\vdash A\& B} \qquad\qquad \&-\text{refl}\;\dfrac{A\vdash\Delta}{A\& B\vdash\Delta} \qquad \dfrac{B\vdash\Delta}{A\& B\vdash\Delta}
$$

$$\vee-form\frac{A\vdash\Delta \qquad B\vdash\Delta}{A\vee B\vdash\Delta} \qquad \vee-refl\frac{\Gamma\vdash A}{\Gamma\vdash A\vee B} \qquad \frac{\Gamma\vdash B}{\Gamma\vdash A\vee B}$$

$$\otimes-form\ \frac{A,B\vdash\Delta}{A\otimes B\vdash\Delta} \qquad \otimes-refl\frac{\Gamma\vdash A \qquad \Gamma'\vdash B}{\Gamma,\Gamma'\vdash A\otimes B}$$

$$\wp-form\ \frac{\Gamma\vdash A,B}{\Gamma\vdash A\wp B} \qquad \wp-refl\ \frac{A\vdash\Delta \qquad B\vdash\Delta'}{A\wp B\vdash\Delta,\Delta'}$$

$$@-form\frac{\Gamma\vdash A,B \quad \Gamma\vdash A^{\perp},B^{\perp}}{\Gamma\vdash Q_A @ Q_B} \qquad @-refl\frac{A\vdash\Delta \quad B\vdash\Delta'}{Q_A @ Q_B\vdash\Delta,\Delta'} \qquad \frac{A^{\perp}\vdash\Delta \quad B^{\perp}\vdash\Delta'}{Q_A @ Q_B\vdash\Delta,\Delta'}$$

$$\S-form\ \frac{A,B\vdash\Delta \qquad A^{\perp},B^{\perp}\vdash\Delta}{Q_A\S Q_B\vdash\Delta} \qquad \S-refl\frac{\Gamma\vdash A \qquad \Gamma'\vdash B}{\Gamma,\Gamma'\vdash Q_A\S Q_B} \qquad \frac{\Gamma\vdash A^{\perp} \qquad \Gamma'\vdash B^{\perp}}{\Gamma,\Gamma'\vdash Q_A\S Q_B}$$

$$\neg-form\frac{A\vdash}{\vdash\neg A}$$

$$\neg-refl\frac{\vdash A}{\neg A\vdash}$$

**Identity axioms:**

$A\vdash A$

**Cut rule:**

$$\frac{\Gamma\vdash A \quad A\vdash\Delta}{\Gamma\vdash\Delta}$$

**EPR rule:**

$$\frac{\Gamma\vdash \mathrm{Q_A} @ \mathrm{Q_B} \qquad \mathrm{Q_A}\vdash \mathrm{A}}{\Gamma\vdash A @ Q_B}$$

## 12. The lattices of propositions of *Lq* and *L2q*

In this Section, we will show that the algebraic structure formed by the propositions of the logic *Lq* is a lattice.

First, we will consider two propositions of the logic *Lq* (actually, an atomic proposition, and its primitive negation) represented by two non-hermitian operators of the Hilbert space $C^2$, and we will show that they form a lattice $Lq(C^2)$ which, when bounded, is orthomodular and distributive.

It would seem, then, that a logic like *Lq*, meant to describe the quantum superposition in the minimal case, that is, the one qubit state in $C^2$, turns out not to be quantum, if the term "Quantum logic" is identified with "non-distributive lattice". However, we will show that, once one considers simultaneously the logic of superposition *Lq*, and the logic of projectors, both in $C^2$, corresponding respectively to the distributive lattices $Lq(C^2)$ and $L(C^2)$, the latter discussed in Sect.2, then it is possible to get a new lattice $Lm(C^2)$, where *m* stands for *mixture*, which is non-distributive.

Thereafter, we will consider the case of $H=C^4$, relative to two pairs of non-hermitian operators, each pair associated with a qubit state. We will show that in this case, the lattice $L2q(C^4)$ is a non-distributive orthomodular lattice. It should be noticed, however, that, differently from the Birkhoff-von Neumann case, where non-distributivity is due to the non-commutativity of pairs of projectors, the lattice $L2q(C^4)$ is non-distributive although all operators commute. This circumstance is due to the fact that the operators are projectors multiplied by different complex factors corresponding to different qubit superposed states.

Apart from distributivity, all the other properties of the lattice $Lq(C^2)$, discussed in the next Subsections, also hold in the case $H=C^4$. Also, the lattice *Lq(H)* is always modular, as we will consider only finite Hilbert spaces, the most general case being

$C^{2^n} = \underbrace{C^2 \otimes \ldots\ldots \otimes C^2}_{n\ times}$ where *n* is the number of qubits.

In summary, $Lq(C^{2^n})$ is an orthomodular (and modular) lattice, which for *n=1* is distributive, while, for *n>1* is non-distributive.

### 12.1 Definition of the "meet" and "join" for the lattice $Lq(C^2)$

In this sub-section, we will define the "meet" and "join" operations within $Lq(C^2)$.

As we have seen in Sect.9, the atomic propositions of *Lq*, in the Hilbert space interpretation *H*, with $H=C^2$, are interpreted as the two non-hermitian operators:

$$O_0 = \lambda_0 P_0, \quad O_1 = \lambda_1 P_1 \qquad (12.1)$$

with $\lambda_0, \lambda_1 \in C$ and $|\lambda_0|^2 + |\lambda_1|^2 = 1$.

Now we wish to look at the algebraic structure formed by the two non-hermitian operators of Eq.(12.1). The algebraic structure we will consider is of the kind: $(S,\overset{\circ}{\wedge},\vee)$, where $S$ is a non empty set (in our case $S$ is the set of the operators $O_i = \lambda_i P_i \; (i = 0,1)$ ), $\overset{\circ}{\wedge}$ is a binary operation, which is a generalization of the meet operation for projectors, and $\vee$ is the join.

The new "meet" $\overset{\circ}{\wedge}$ is defined as follows:

$$O_0 \overset{\circ}{\wedge} O_1 = \lambda_0 P_0 \overset{\circ}{\wedge} \lambda_1 P_1 = \frac{\lambda_0 \lambda_1}{\sqrt{\lambda_0}\sqrt{\lambda_1}} P_0 \wedge P_1 \tag{12.2}$$

$$O_1 \overset{\circ}{\wedge} O_0 = \lambda_1 P_1 \overset{\circ}{\wedge} \lambda_0 P_0 = \frac{\lambda_1 \lambda_0}{\sqrt{\lambda_1}\sqrt{\lambda_0}} P_1 \wedge P_0 \tag{12.3}$$

$$O_0 \overset{\circ}{\wedge} O_0 = \lambda_0 P_0 \overset{\circ}{\wedge} \lambda_0 P_0 = \frac{\lambda_0{}^2}{\sqrt{\lambda_0}\sqrt{\lambda_0}} P_0 \wedge P_0 \tag{12.4}$$

$$O_1 \overset{\circ}{\wedge} O_1 = \lambda_1 P_1 \overset{\circ}{\wedge} \lambda_1 P_1 = \frac{\lambda_1{}^2}{\sqrt{\lambda_1}\sqrt{\lambda_1}} P_1 \wedge P_1 \tag{12.5}$$

Then, the definition of the meet can be given, for the atomic formulas $O_i$, in the general form:

$$O_i \overset{\circ}{\wedge} O_j \overset{def}{=} \frac{1}{\sqrt{\lambda_i \lambda_j}} O_i \wedge O_j \qquad (i, j = 0,1) \tag{12.6}$$

where:

$$O_i \wedge O_j = \lambda_i \lambda_j P_i \wedge P_j = \lambda_i \lambda_j P_i \cdot P_j = O_i \cdot O_j \tag{12.7}$$

Notice that for $i \neq j$, Eqs.(12.6) and (12.7) give the orthogonal law:

$$O_i \overset{\circ}{\wedge} O_j = 0 \tag{12.8}$$

The definition of the join is the usual one, that is, the sum minus the meet of the operators $O_i$:

$$O_i \vee O_j \overset{def}{=} O_i + O_j - O_{i\,\frac{1}{\sqrt{\lambda_i}}} \wedge {}_{\frac{1}{\sqrt{\lambda_j}}} O_j \qquad (i, j = 0,1) \tag{12.9}$$

In the case $i \neq j$, Eq.(12.9), owing to Eq.(12.8), trivially gives:

$$O_i \vee O_j = O_i + O_j = \lambda_i P_i + \lambda_j P_j \tag{12.10}$$

The above definitions of the meet and join allow to recover the idempotent laws:

$$O_i \,{}_{\frac{1}{\sqrt{\lambda_i}}}\wedge_{\frac{1}{\sqrt{\lambda_i}}}\, O_i = O_i \qquad O_i \vee O_i = O_i \tag{12.11}$$

Also, the commutative laws hold:

$$O_i \vee O_j = O_j \vee O_i \qquad O_i \,{}_{\frac{1}{\sqrt{\lambda_i}}}\wedge_{\frac{1}{\sqrt{\lambda_j}}}\, O_j = O_j \,{}_{\frac{1}{\sqrt{\lambda_j}}}\wedge_{\frac{1}{\sqrt{\lambda_i}}}\, O_i \tag{12.12}$$

as well as the absorption (saturation) laws:

$$O_i \vee (O_i \,{}_{\frac{1}{\sqrt{\lambda_i}}}\wedge_{\frac{1}{\sqrt{\lambda_j}}}\, O_j) = O_i \tag{12.13}$$

To get a bounded lattice, we define the "unity" $\hat{I}$ as:

$$\hat{I} = O_0 + O_1 = \lambda_0 P_0 + \lambda_1 P_1 \tag{12.14}$$

and the 0 as:

$$0 = O_0 O_1 \tag{12.15}$$

The bounded lattice *Lq* is then: $(S, \overset{\circ}{\wedge}, \vee, \hat{I}, 0)$.

## 12.2 Complementarity of *Lq*($C^2$)

By the definitions of the meet $\overset{\circ}{\wedge}$ and the join $\vee$ given above, and by the use of Eqs. (12.8) and (12.10), we see that this lattice is complemented:

$$O_0 \overset{\circ}{\wedge} O_1 = 0 \tag{12.16}$$

$$O_0 \vee O_1 = \hat{I} \tag{12.17}$$

$$0 \overset{\circ}{\wedge} \hat{I} = 0 \tag{12.18}$$

$$0 \vee \hat{I} = \hat{I} \tag{12.19}$$

When the new meet $\overset{\circ}{\wedge}$ is performed between an atomic proposition $O_i$ and the "unity" $\hat{I}$, it should give $O_i$, as usual. One can recover this result, by using the definition of the "meet" given above, which automatically implies the distributive law of the meet with respect to the sum:

$$O_i \overset{\circ}{\wedge}(O_i + O_j) = O_i \tag{12.20}$$

where, in (12.20), the idempotent law (12.11) and the orthogonal law (12.8) have been used.
Therefore, it holds:

$$O_i \overset{\circ}{\wedge} \hat{I} = O_i \tag{12.21}$$

By the definition of the meet $\overset{\circ}{\wedge}$, it also holds:

$$O_i \overset{\circ}{\wedge} 0 = 0 \tag{12.22}$$

The partial order relation $\leq$ is associated to the above operations of meet and join as:
$x \leq y$ iff $x \wedge y = x$
$x \leq y$ iff $x \vee y = y$
In our case, the above relations read as:

$$O_i \leq \hat{I} \text{ iff } O_i \overset{\circ}{\wedge} \hat{I} = O_i \tag{12.23}$$

which is satisfied because of Eq.(12.21).

$$O_i \leq \hat{I} \text{ iff } O_i \vee \hat{I} = \hat{I} \tag{12.24}$$

It is easy to show that the relation in Eq.(12.24) is satisfied. In fact, it holds:

$$O_i \vee \hat{I} = O_i \vee (O_0 + O_1) = O_i + O_0 + O_1 - O_i \overset{\circ}{\wedge}(O_0 + O_1) = O_i + O_0 + O_1 - O_i = O_0 + O_1 = \hat{I} \tag{12.25}$$

where the definition of $\hat{I}$ in Eq.(12.17) and Eq.(12.21) have been used.
In a similar way it is also trivial to show that:

$$0 \leq \hat{I} \text{ and } 0 \leq O_i . \tag{12.26}$$

## 12.3 Orthocomplementarity of $Lq(C^2)$

At this point we can say that the lattice $Lq(C^2)$ is complemented, and the partial order relation we explicated above will allow us to show that this lattice is also orthocomplemented.
Then we define an order-reversing involution $\perp$ that maps each element to a complement:

$$O_i^{\perp} = \hat{I} - O_i \tag{12.27}$$

That is, explicitly:

$$O_0^{\perp} = O_1, \; O_1^{\perp} = O_0 \tag{12.28}$$

Moreover, it holds trivially:

$$\hat{I}^{\perp} = 0 \tag{12.29}$$

$$0^{\perp} = \hat{I} \tag{12.30}$$

Then, it is straightforward to see that the complemented lattice $Lq(C^2)$ is in fact orthocomplemented, as the following axioms are satisfied:
Complement law:

$$O_i^{\perp} \vee O_i = \hat{I} \qquad O_i^{\perp} \overset{\circ}{\wedge} O_i = 0 \tag{12.31}$$

$$\hat{I}^{\perp} \vee \hat{I} = \hat{I} \qquad \hat{I}^{\perp} \overset{\circ}{\wedge} \hat{I} = 0 \tag{12.32}$$

$$0^{\perp} \vee 0 = \hat{I} \qquad 0^{\perp} \overset{\circ}{\wedge} 0 = 0 \tag{12.33}$$

Involution:

$$(O_i^{\perp})^{\perp} = O_i \tag{12.34}$$

$$(\hat{I}^{\perp})^{\perp} = \hat{I} \tag{12.35}$$

$$(0^{\perp})^{\perp} = 0 \tag{12.36}$$

Order-reversing:

$$O_i \le \hat{I} \text{ implies } \hat{I}^{\perp} \le O_i^{\perp} \tag{12.37}$$

$$0 \le O_i \text{ implies } O_i^{\perp} \le 0^{\perp} \tag{12.38}$$

$0 \leq \hat{I}$ implies $\hat{I}^{\perp} \leq 0^{\perp}$ (12.39)

## 12.4 Orthomodularity of $Lq(C^2)$

As it was already recalled in Sect.2, an orthocomplemented lattice such that for any two elements the orthomodular law holds, is called orthomodular.
The orthomodular law is:
if a $\leq$ c, then $a \vee (a^{\perp} \wedge c) = c$

In our case we must verify that, since $O_i \leq \hat{I}$, it holds: $O_i \vee (O_i^{\perp} \overset{\circ}{\wedge} \hat{I}) = \hat{I}$. In fact, we have:

$$O_i \vee (O_i^{\perp} \overset{\circ}{\wedge} \hat{I}) = O_i \vee O_i^{\perp} = \hat{I} \qquad (12.40)$$

where Eq.(12.21) and the first equation in Eq.(12.31) have been used.
Also, we must verify that, since $0 \leq O_i$, it holds: $0 \vee (0^{\perp} \overset{\circ}{\wedge} O_i) = O_i$. In fact, we have:

$$0 \vee (0^{\perp} \overset{\circ}{\wedge} O_i) = 0 \vee (\hat{I} \overset{\circ}{\wedge} O_i) = 0 \vee O_i = 0 + O_i - 0 \overset{\circ}{\wedge} O_i = O_i . \qquad (12.41)$$

Finally, we must verify that, since $0 \leq \hat{I}$, it holds: $0 \vee (0^{\perp} \overset{\circ}{\wedge} \hat{I}) = \hat{I}$. In fact, we have:

$$0 \vee (0^{\perp} \overset{\circ}{\wedge} \hat{I}) = 0 \vee (\hat{I} \overset{\circ}{\wedge} \hat{I}) = 0 \vee \hat{I} = \hat{I} \qquad (12.42)$$

where Eq.(12.33) has been used.
At this point we can say that the lattice $Lq(C^2)$ is orthomodular.

## 12.5 Modularity of $Lq(C^2)$

As we already said in Sect.8, when the Hilbert space *H* is finite, the lattice of projectors is modular. In our case, the Hilbert space under consideration, namely $C^2$, is finite. However, the lattice $Lq(C^2)$ is not simply a lattice of projectors on *H*, in fact it is a lattice of non-hermitian operators on *H*. Nevertheless, the modular law, as we will show below, is satisfied also in our case.
Modular law: $a \leq c \rightarrow a \vee (b \wedge c) = (a \vee b) \wedge c$
In our case (for $a = O_i$, $b = O_j$, $c = \hat{I}$, and $i \neq j$) the modular law reads:

$$O_i \leq \hat{I} \rightarrow O_i \vee (O_j \overset{\circ}{\wedge} \hat{I}) = (O_i \vee O_j) \overset{\circ}{\wedge} \hat{I} \quad \text{for} \quad i \neq j \qquad (12.43)$$

The term on the LHS of Eq.(12.43) gives:

$$O_i \vee (O_j \overset{\circ}{\wedge} \hat{I}) = O_i \vee O_j = \hat{I}$$

and the RHS of Eq.(12.43) gives:

$(O_i \vee O_j) \overset{\circ}{\wedge} \hat{I} = \hat{I} \overset{\circ}{\wedge} \hat{I} = \hat{I}$

Therefore the modular law is satisfied in this case.

For $a = 0$, $b = O_i$, $c = O_j$, the modular law reads:

$$0 \le O_j \rightarrow 0 \vee (O_i \overset{\circ}{\wedge} O_j) = (0 \vee O_i) \overset{\circ}{\wedge} O_j \qquad (12.44)$$

The term on the LHS of Eq.(12.44) gives:

$0 \vee (O_i \overset{\circ}{\wedge} O_j) = 0 \vee 0 = 0$

and the RHS of Eq.(12.44) gives:

$(0 \vee O_i) \overset{\circ}{\wedge} O_j = O_i \overset{\circ}{\wedge} O_j = 0$

Therefore, the modular law holds also in this case.

One more case is obtained for: $a = 0$, $b = O_i$, $c = \hat{I}$, for which the modular law reads:

$$0 \le \hat{I} \rightarrow 0 \vee (O_i \overset{\circ}{\wedge} \hat{I}) = (0 \vee O_i) \overset{\circ}{\wedge} \hat{I} \qquad (12.45)$$

The term on the LHS of Eq.(12.45) gives:

$0 \vee (O_i \overset{\circ}{\wedge} \hat{I}) = 0 \vee O_i = O_i$

and the term on the RHS gives:

$(0 \vee O_i) \overset{\circ}{\wedge} \hat{I} = O_i \overset{\circ}{\wedge} \hat{I} = O_i$

Then, the modular law is satisfied also in this case.

The last case is given for: $a = 0$, $b = \hat{I}$, $c = O_i$

$$0 \le O_i \rightarrow 0 \vee (\hat{I} \overset{\circ}{\wedge} O_i) = (0 \vee \hat{I}) \overset{\circ}{\wedge} O_i \qquad (12.46)$$

The term on the LHS of Eq.(12.46) gives:

$0 \vee (\hat{I} \overset{\circ}{\wedge} O_i) = 0 \vee O_i = O_i$

and the term on the RHS gives:

$(0 \vee \hat{I}) \overset{\circ}{\wedge} O_i = \hat{I} \overset{\circ}{\wedge} O_i = O_i$

Then, the modular law holds also in this last case.

### 12.6 Distributivity of $Lq(C^2)$

It is easy to show, by direct substitution of the elements of $Lq(C^2)$ into the distributive laws quoted in Sect.2, and by the use of the meet and join defined in Subsection 12.1, that this lattice is distributive.

Therefore, the algebraic structure of the logic of quantum superposition in $C^2$, in terms of non-hermitian operators (weak measurements) is Boolean. Moreover, as we already showed in Sect.2, the same is true when this logic is given in terms of projectors of $C^2$. Nevertheless, when the two kinds of operators are taken together, with a specific partial order relation, to form a unique lattice, then, as we will show in the next Subsection, this new lattice is non-distributive.
The Hasse diagram of $Lq(C^2)$ is showed in Fig. 12.1.

**Fig.12.1**
**The lattice Lq($C^2$)**

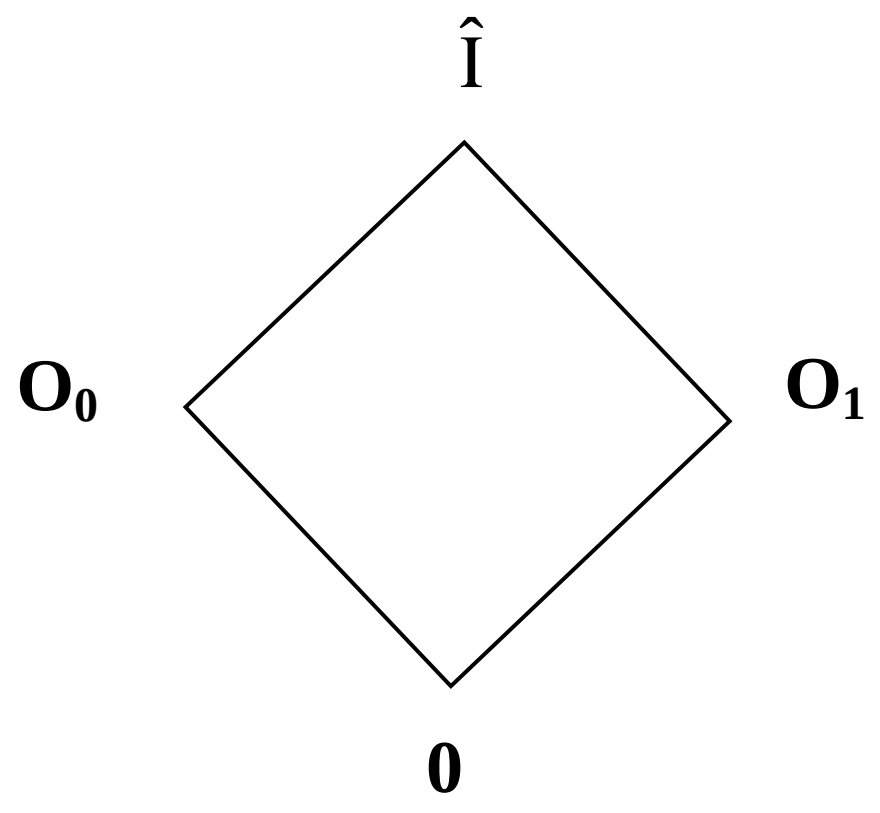

### 12.7 The lattice $L_m(C^2)$

In this Section, we will consider the lattice $L_m(C^2)$, whose underlying set consists of the two non-hermitian operators of *Lq*, and the two one-dimensional projectors of the Birkhoff-von Neumann Quantum logic, both in the case of $C^2$.
Before discussing the choice of the partial order relation for this lattice, we make some useful considerations.
We recall that the probability *p(i)* of getting the basis state $|\Psi_i\rangle$ from the superposed state $|\Psi\rangle = \sum_i \lambda_i |\Psi_i\rangle$ upon a projective measurement $P_i$ is given by:

$p(i) = \langle\Psi|P_i|\Psi\rangle = |\lambda_i|^2$

This equals the expectation value of the Hermitian operator $O_i^\dagger O_i$ (where $O_i = \lambda_i P_i$) on the basis state $|\Psi_i\rangle$, namely:

$$\langle\Psi_i|O_i^\dagger O_i|\Psi_i\rangle = |\lambda_i|^2 \qquad (i = 0,1) \qquad (12.47)$$

Instead, the e.v. of the one-dimensional projector $P_i$ on the basis state $|\Psi_i\rangle$, which is its eigenstate with eigenvalue 1, is, of course, 1:

$$\langle\Psi_i|P_i|\Psi_i\rangle = 1 \qquad (12.48)$$

This can be interpreted as the probability of getting the state $\left|\Psi_i\right\rangle$ once the projective measurement $P_i\left|\Psi\right\rangle$ has been performed.

Our claim is that the expectation values (12.47) and (12.48) are in fact "probabilities" of getting the eigenstates of the operators $O_i$ and $P_i$ respectively, and therefore they can be given the role of the "measure" $\mu$ of the range where those operators "project".

Then, we will set a relation of range-measure between the ranges of the two operators $O_i$, that is, between *ran*($O_0$) and *ran* ($O_1$), and also between the range of each operator $O_i$, and the range of the corresponding projector $P_i$, that is, between *ran*($O_i$) and ran($P_i$) for *i= 0,1*.

Let us now define:

$$\mu(ran(O_i)) \equiv \left\langle \Psi_i \right| O_i^{\dagger} O_i \left| \Psi_i \right\rangle ; \qquad \mu(ran(P_i)) \equiv \left\langle \Psi_i \right| P_i \left| \Psi_i \right\rangle = 1 \tag{12.49}$$

Therefore we get:

$$\mu(ran(O_i)) \le \mu(ran(P_i)) \text{ for every } i. \tag{12.50}$$

From the above relation it follows:

$$ran(O_i) \subseteq ran(P_i) \text{ for every } i. \tag{12.51}$$

At this point, following the procedure adopted by Birkhoff-von Neumann described in Sect.2, we will introduce a partial order relation between two operators, induced by the relation of set theoretical inclusion between their ranges. It should be noticed that in our case, the set theoretical inclusion of the ranges is induced in turn by the range-measure $\mu$.

We consider the poset: $P = \{S, \le\}$where the underlying set *S* consists of the two non-hermitian operators $O_i$ and of the two one-dimensional projectors $P_i$ of $C^2$, and the partial order relation $\le$ is defined as:

$$ran(O_i) \subseteq ran(P_i) \rightarrow O_i \le P_i \tag{12.52}$$

This allows to define the meet and join operations:

$$\text{if } O_i \le P_i \text{ then } O_i \wedge P_i = O_i \tag{12.53}$$

$$\text{if } O_i \le P_i \text{ then } O_i \vee P_i = P_i \tag{12.54}$$

The lattice $L_m(C^2) = \{S, \wedge \vee\}$ becomes bounded through the introduction of the unity (the identity operator $I_2$ of $C^2$), and 0 **(**the null operator of $C^2$**).**

The Hasse diagram of the bounded lattice $L_m(C^2) = \{S, \wedge, \vee, \mathrm{I}_2, 0\}$ is given in Fig. 12.2.

**Fig. 12.2**
**The lattice $L_m(C^2)$**

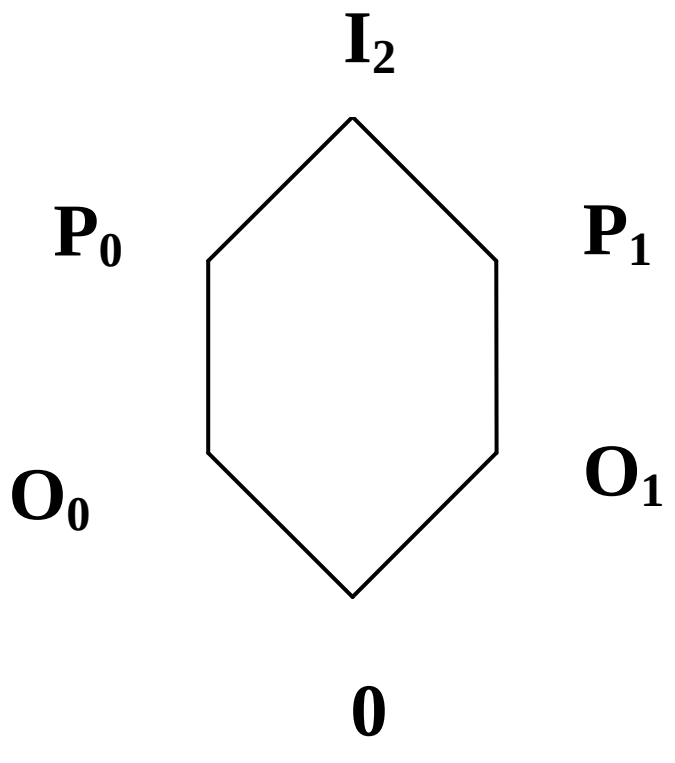

It is straightforward to show that the lattice $L_m(C^2)$ is non-distributive. In fact, let us consider the distributive law: $a \wedge (b \vee c) = (a \wedge b) \vee (a \wedge c)$ in the case:
$a = P_0 \quad b = O_0 \quad c = O_1$.
The LHS of the distributive law gives:

$$P_0 \wedge (O_0 \vee O_1) = P_0 \wedge \mathrm{I}_2 = P_0 \tag{12.55}$$

while the RHS gives:

$$(P_0 \wedge O_0) \vee (P_0 \wedge O_1) = O_0 \vee 0 = O_0 \tag{12.56}$$

As the results on the RHS of Eqs.(12.55) and (12.56) are different, we conclude that the distributive law is not satisfied in this case.
Therefore, the lattice $L_m(C^2)$ is non-distributive.

### 12.8 Non-distributivity of the lattice $L2q(C^4)$

We consider the case of two qubits, in $C^4$, described by the two pairs of non-hermitian operators: $(O_0 = \lambda_0 P_0,\ O_1 = \lambda_1 P_1)$ and $(O_0' = \lambda_0' P_0,\ O_1' = \lambda_1' P_1)$.
The bounded lattice $L2q(C^4)$ is:

$$Lq(C^4) = \{S, \leq, \mathrm{I}_4, 0\}$$

Where S is the underlying set of the four non-hermitian operators:

$S = \left\{ O_0, O_0', O_1, O_1' \right\}$

The partial order relation $\leq$ for the pair ($O_0$', $O_0$) is defined as:

$$\mu(ran(O_0')) \leq \mu(ran(O_0)) \rightarrow ran(O_0') \subseteq ran(O_0) \rightarrow O_0' \leq O_0 \qquad (12.57)$$

or, what is the same:

$$O_0' \leq O_0 \text{ iff } |\lambda_0'|^2 \leq |\lambda_0|^2 \qquad (12.58)$$

From the above relation, and from the constraints:

$$|\lambda_0|^2 + |\lambda_1|^2 = 1, \ |\lambda_0'|^2 + |\lambda_1'|^2 = 1. \qquad (12.59)$$

it follows:

$$O_1' \leq O_1 \qquad (12.60)$$

The unity $I_4$ is the identity operator of $C^4$, and 0 is the null operator of $C^4$.
Eqs.(12.58) and (12.60) allow to define the meet and join operations:

$$O_0' \leq O_0 \rightarrow O_0' \wedge O_0 = O_0' \qquad O_0' \leq O_0 \rightarrow O_0' \vee O_0 = O_0 \qquad (12.61)$$

$$O_1 \leq O_1' \rightarrow O_1 \wedge O_1' = O_1 \qquad O_1 \leq O_1' \rightarrow O_1 \vee O_1' = O_1' \qquad (12.62)$$

The Hasse diagram of the lattice ***L2q***(***$C^4$***) is given in Fig.12.3.

**Fig. 12.3**
**The lattice *L2q*($C^4$)**

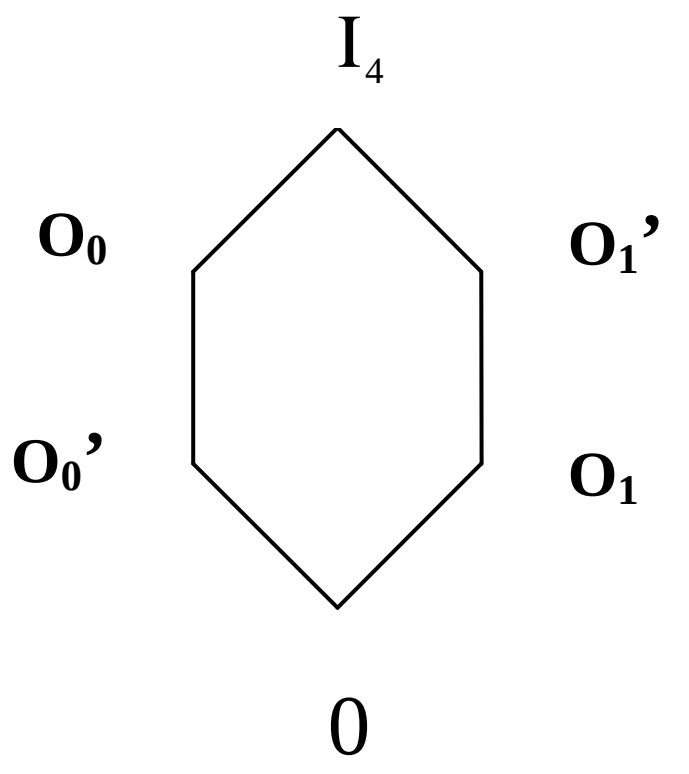

Let us consider now the distributivity law:

$a \wedge (b \vee c) = (a \wedge b) \vee (a \wedge c)$
in the case: $a = O_0$, $b = O_0'$, $c = O_1'$
The LHS of the distributivity law gives:

$$O_0 \wedge (O_0' \vee O_1') = O_0 \wedge \mathrm{I}_4 = O_0 \qquad (12.63)$$

while the RHS gives:

$$(O_0 \wedge O_0') \vee (O_0 \wedge O_1') = O_0' \vee 0 = O_0' \qquad (12.64)$$

As the results on the RHS of Eqs.(12.63) and (12.64) are different, it follows that the distributive law is not satisfied.
Therefore, we conclude that the lattice $L2q(C^4)$ is non-distributive.

## 13. Quantum Metalanguage and Constructivism

The study of the “natural” logic of mental processes is very important for the development of a constructivist approach to logic, by which logic is a by-product of the mind.

In this regard, there are two different points of view, the microscopic and the macroscopic one.

According to the latter, one might start from the phenomenology of the “thinking processes”, as studied by Cognitive Science where, however, a unified theory of these processes is still missing. Also the so-called “philosophy of mind” (see, in this regard, Pylyshyn, 1984; Scheutz, 2002; Kim, 2006; Braddon-Mitchell and Jackson, 2007; Bechtel, 2008; McLaughlin *et al*., 2009) in general takes the macroscopic point of view. The latter appear to be adopted even by constructivist approaches, like dynamical constructivism (see Sambin, 2002), which seem to rely on the cognitive and/or social interpretation of logical and mathematical activities, in agreement with the arguments put forward by researchers dealing with the philosophy of mind.

Within the macroscopic point of view, we believe that Basic logic (see, for instance, Sambin *et al*., 2000) is the best candidate for a “natural” logic of mind, in a foundational sense, that is, a logic which can describe the basic mental operations. In fact Basic logic, like linear logic, is a sub-structural logic because within it some of the structural rules of sequent calculus, namely the contraction and the weakening rules, do not hold (while the exchange rule is still holding).

In so far, Basic logic has been considered as the weakest logic, even more than linear logic as the latter can be obtained from Basic logic by allowing free contexts on both sides of the sequents. Therefore, Basic logic stands for the platform of all other logics. The latter can be considered as its extensions, which are too abstract and structured to allow a representation of the most fundamental thought processes.

Instead the former point of view (the microscopic one) assumes that the macroscopically observed phenomenology of mental processes is nothing but a by-product of the activities of a huge number of microscopic components, reciprocally interacting, often in a cooperative way. In other words, by resorting to a metaphor based on physics, mental activities might be compared to thermo-dynamical processes described through macroscopic variables like pressure or temperature which, however, are due to statistical effects of the microscopic interactions occurring between atoms or molecules. It is, however, very difficult to individuate what are the “atoms” underlying mental activity. The supporters of the Connectionist approach in psychology often speak of “micro cognitive units”, without, however, being able to make this concept more precise. Here we prefer to identify the microscopic entities supporting the existence of mental processes with the basic components of biological matter constituting the brain, which can be observed and studied. More precisely, as the behaviour of these components appears to be ruled by quantum-like laws, we rely on the quantum processes occurring in the brain (on this topic see Ricciardi and Umezawa, 1967; Stuart *et al*., 1979; Stapp, 1993; Jibu and Yasue, 1995; Tuszyński, 2006; Abbott *et al*., 2008), as formalized within quantum theory.

The two points of views are not incompatible, they just concern different contexts. However, the microscopic point of view, which is based on a mathematical formalism (that of quantum mechanics and quantum field theory), seems to be better suited to a reconstruction of a logic, based on a bottom-up strategy.

In this thesis we adopt the microscopic point of view and, in particular, we focus on quantum metalanguage which is considered, here, as an "emergent" feature of brain quantum processes. In this regard we stress that, within a microscopic approach, which deals with at least two description levels (the microscopic and the macroscopic one), the reference to the concept of "emergence" is mandatory. Namely it refers to all processes which allow the occurrence of a macroscopic phenomenology, just with the specific observed features, from the dynamics and the interactions taking place at the microscopic level. Without entering into technical details, for sake of brevity, about the mathematical theories of emergence (often relying on the methods of Statistical Physics), we limit ourselves to mention that a physically grounded description of emergence is possible only within the context of quantum theories. We do not deal here with particular models of emergence, like the ones describing the arising of the metalanguage from the interactions between microscopic brain components, as these models are not the subject of this thesis.

As a consequence of the above, we are led to identify the metalanguage, viewed as a macroscopic entity emerging from the microscopic processes occurring within the brain, with a quantum metalanguage. This new kind of metalanguage should lead to a new quantum logic, which describes the fundamental aspects of mind functioning at the quantum level. We remark here that the term "logic" is used in this context as a sort of label to denote a formalized description of some basic mental operations. Namely this description, as seen in previous Sections of this thesis, implies a number of strange features which could prevent some logicians from using the term "logic". On the other hand, these features are nothing but a consequence of the fact that even quantum mechanics is endowed with "strange" and unusual features, at least according to the common sense. This new quantum logic, which is sub-structural, has the three features which characterize Basic logic, that is, symmetry, reflection, and visibility. Furthermore, this logic does not allows the exchange rule, therefore it is even weaker than Basic logic. We call this logic quantum because it describes quantum information, and because its algebraic structure is (a part from the trivial case of a single qubit) a non distributive lattice, but it is different from standard quantum logics, which instead are structural.

The fact that this sub-structural quantum logic is weaker than Basic logic can be seen as related to a bottom-up construction of mental operations from the microscopic level to the macroscopic one.

As regards metalanguage, we remark that it is quite an old subject in logic, which rests at the roots of semantics and model theory. It is well known that propositions can be viewed as the primary bearers of truth and falsity. If we deal with a language (the object-language) denoted by *L*, then the definition of its propositions must be made within another language known as the metalanguage, *M*. The metalanguage should contain a copy of the object language (so that anything one can say in *L* can be said in *M* too), and *M* should also be able to talk about the sentences of *L* and their syntax. We recall that Tarski (see Tarski, 1944; a complete view of his approach can be found in the partial collection of his papers Tarski, 1983) allows *M* to contain notions from set theory, and a predicate symbol *True*. The main purpose of the metalanguage is therefore to formalise what is being said about the object-language.

As we said above, *M* can be viewed as an emergent feature of human mind or, speaking in rough terms, of human thought, if a constructivist approach to mathematics and logic is adopted. There are many different approaches to constructivism (see in this regard the references quoted in the Introduction of this thesis). We quote here constructivist (or intuitionist) logic, which is the logic system developed by Heyting to provide a formal basis for Brouwer's programme of intuitionism. The system preserves justification, rather than truth, in derivations (see Heyting, 1971).

But the concept of metalanguage is also very useful in computer science, where it stands for the language of control (there is an extensive literature on the use of metalanguage in computer science; here we will limit ourselves to quote Melhan, 1993; Nürnberg, 2003; Hicks, 2004). We refer to the metalanguage as a control over the object-language, as meant within computer science (with particular attention to quantum computing).

In the schematic description of control processes an important role is played by representation of the controlled system in the controlling device (see, for instance, Turchin *et al.*, 1996, Joslyn, 2001). When the controlled system is some object-language *L*, the language used to make the representations in the controlling system, let it be *L'*, is referred to as a metalanguage (see, in this regard, Turchin, 1999). Control is the operation mode of a *control system* which includes two subsystems: the controlling one (a controller) *C*, and the controlled one, *S*. They interact, but there is a difference between the action of *C* on *S*, and the action of *S* on *C*. The controller *C* may change the state of the controlled system *S* in any way, including the destruction of *S*. In general, the converse is not true (that is, the controlled system cannot destroy or modify the controlling system). However this feature depends on the complexity of the system. Namely we cannot exclude that, within suitably complex systems, it could occur, just through processes of emergence, a sort of “self-organization” leading to a blockage of controlling action in the case the latter is against the very “scope” of the system. This can happen when the control is introduced by the external world, like, for instance, in social systems.

Of course, when we talk about an internal control as in the case we consider in this thesis, that is the control arising in thought processes, things are different. The worse it can happen, is that the goal of the internal control can be modified due to new information acquired by the system, all that leading to a more sophisticated scope.

The characteristic of a quantum system, from the perspective of control theory, is that measurement causes its state to change. This property is referred to as *back action*. In quantum mechanics the effect of back action is to introduce a type of quantum noise into the measurement of the system's state. However, it is possible to think of quantum control in a context that is similar to classical control, but with an additional noise source due to the measurement back action.

Observation plays a central role in feedback control: there can be no feedback without checking up on the state of the system to ensure that it is doing the right thing. However, observation induces unavoidable randomness due to the projection postulate. Fortunately, quantum systems can be measured in many different ways, some of which are less intrusive than others. In the worst case scenario, a projective measurement instantaneously and completely collapses the quantum system into an eigenstate of the measurement operator. This drastic, sudden change in the quantum state appears to be a poor candidate for feedback control.

Instead, weak measurements of the quantum system result in less drastic fluctuations and only gradually reduce the system to a measurement eigenstate.

We exploit the very fortunate case where these two aspects (observation and control) of metalanguage coincide, namely the quantum case. Within this context we must resort to a quantum model of mind, and we use the expression *quantum mind* to denote all quantum features of mental operations. As said before, we consider the quantum mind as an emergent feature of quantum processes in the brain. The constructivist approach to logic and the quantum model of mind are strictly related. In fact, in the constructivist approach to logic, the latter can be viewed as an *invention* of the mind (see Sambin, 2002, p.270). In this context, quantum metalanguage belongs to the quantum mind, arising from a quantum process in the brain, which we call meta-thought. The latter is the process of thinking about our own thought. It should tell us whether a certain metalanguage is correctly selected to reflect, by the reflection principle (see, e.g., Sambin *et al*., 2000) into the corresponding object-language. Although meta-thought is a quantum process, it does not discriminate between the classical and the quantum case, its only requirement being that no contradiction arises between metalanguage and object-language. Meta-thought allows a classical metalanguage for a classical object-language (for example that produced by a classical Turing machine) as well as a quantum metalanguage for a quantum object-language (for example the one produced by a quantum computer).

Let us now consider in more detail the brain quantum processes. In this regard we remark that the brain can be seen as a quantum open dissipative system. The attribute "open" denotes the fact that it is an open system in the usual thermo-dynamical sense because its components are always changing, owing both to the input of new components (typically macromolecules) from outside and to the disappearance of components already present (like macromolecules or even whole cells). The attribute "dissipative" denotes the fact that most processes, at the basic level, do not fulfil the energy conservation principle in the usual sense. Among these processes we can quote the diffusion, inelastic scattering between molecules, energy transfer towards the intracellular or extra-cellular liquid. As it is well known, the description of dissipative processes within quantum mechanics is a difficult problem. We remind that the traditional quantum mechanical formalism has been built to describe systems characterized by the presence of an Hamiltonian and whence fulfilling the energy conservation principle. Within this context one traditionally makes use of hermitian operators with real eigenvalues. Some simple cases of dissipative systems, like the damped harmonic oscillator, could, however, be cast under an Hamiltonian form. Nevertheless a quantization of these systems, performed through usual methods, could give rise to troubles with the uncertainty principle, whose form should vary with time. Furthermore, it would be difficult to interpret the non-hermitian operators occurring in these strange cases, as they should be associated to complex eigenvalues. These difficulties, as known from the history of theoretical physics, have been circumvented by Celeghini, Rasetti and Vitiello who showed (see Celeghini *et al.*, 1992) that, by introducing, beside the dissipating system, its *double*, that is a sort of mirror system absorbing just the energy dissipated by the original system, it was possible to deal again with an Hamiltonian energy-conserving system, including both the system and its double. This "doubling" formalism allowed an easy interpretation of the imaginary parts of eigenvalues of non-hermitian operators, associated to the energy transfer between the system and its "double". Looking only at the original dissipating system these transfers give rise to a non-unitary dynamics which, as everyone knows, can be allowed only within the context of Quantum Field Theory (see, for instance, Umezawa, 1993). This entails that, really, when we deal with an open dissipative quantum system, we are beyond the context of traditional Quantum Mechanics, but rather within Quantum Field Theory.

Meta-thought is a quantum dissipative process. The stages of such a process are described by logical propositions which are interpreted as non-hermitian operators. The resulting metalanguage is formed by such asserted propositions, carrying along assertion-degrees. The states of the model satisfying the theory are generalized coherent states, very robust against decoherence. Therefore, the quantum metalanguage is a form of quantum control on decoherence. This is a non-destructive quantum control, like that performed by a weak measurement (see, on this topic, Aharonov *et al.*, 1988; Aharonov and Botero, 2005; Oreshkov and Brun, 2005).

In our case, the non-hermitian operators $O_i \quad (i = 1,2....n)$, which play the role of weak measurements, are given by the multiplication of a complex number $\lambda_i \; (i = 1,2....n)$, which is a probability amplitude, times a one-dimensional projector operator $P_i$ on the finite-dimensional Hilbert space *H* of dimension *n*, namely: $O_i = \lambda_i P_i$ with the constraint: $\sum_i^n |\lambda_i|^2 = 1$.

We remind here that the non-hermitian operators $O_i$ are the interpretation in the Hilbert space *H*, of the n atomic propositions $p_i$ of the quantum object-language *Lq*, already introduced in Sect.8. Moreover, an asserted atomic proposition $p_i$, which belongs to the quantum metalanguage, is denoted by the (weighted) sequent: $\vdash^{z_i} p_i$, where the "assertion-degrees" $z_i$ are interpreted as the probability amplitudes $\lambda_i$ in a general superposed state of the Hilbert space *H*. The atomic assertions $\vdash^{z_i} p_i$ are interpreted as $O_i|\Psi_i\rangle$, where the $|\Psi_i\rangle$ form an orthonormal basis for *H*. Roughly speaking, the interpretation of the propositions $p_i$ is necessary only if one wishes to have an algebraic picture (for example the lattice of propositions $p_i$) while the interpretation of atomic assertions requires a model (in our case the basis vectors $|\Psi_i\rangle$ of *H*). Then understanding the meaning (semantics) of the metalanguage being used requires a model (possibly of the real world). This is the process of meta-thought. When the model is QM, the process of meta-though must be quantum, which brings us back to the concept of quantum mind.
Dissipative quantum systems have been studied (see Vitiello, 1995; his work fits in with the general framework described in Umezawa, 1993) in terms of unitarily non equivalent representations, once the assumption is taken that the rest of the universe plays the *role* of the environment. The whole system then is closed, and the total energy is constant. In the case of the quantum brain system, the mental representation of the rest of the universe can be described as deriving from the general mechanism known as *doubling* and mentioned above (for a general overview see Vitiello, 2001).
As already illustrated, the simplest example of *doubling* can be found in the theory of the damped harmonic oscillator, where, to implement a canonical quantization scheme, one must double the phase-space dimension, in order to deal with an effective isolated system. The new degrees of freedom introduced in this way, are assumed to represent the bath, which absorbs the energy dissipated by the oscillator (see Celeghini *et al*., 1992).
In QM, however, standard canonical commutation relations are not preserved in time because of the appearance of a damping term. This problem is overcome within QFT, which allows the existence of unitarily non equivalent representations, that is, the Hilbert space *H* at time t is different from the Hilbert space *H'* at time t'.
In this framework, the imaginary part of our non-hermitian operator, which is just the damping term appearing in the CR at a given time, can be interpreted as the transition operator among different unitarily non equivalent representations, performing a quantum tunnelling.
The propositions are then asserted in different ways on model states at different fixed times, and the assertion-degree is, in the physical interpretation, a complex number, whose phase is the damping term. This means that part of the assertion-degree is absorbed by the *double* (the same proposition expressed by the environment).
We find that dissipation is lowered by the presence of generalized coherent states, that is, by the quantum control (we stress that generalized coherent states have been introduced in quantum theories years ago; see, for instance, Perelomov, 1986; a related work is the one of Ali *et al*., 2000).
In summary, we believe that quantum metalanguage emerges from quantum brain processes, and acts as a quantum control against dissipation: in this way, some quantum aspects of our thought are not lost. These might generate the most creative mental attitudes, one of which, for example, is the ability to handle a constructive approach to mathematics, in particular to logic.
In particular in this thesis we have studied the quantum metalanguage in the framework of sequent calculus, and exploited the reflection principle, where the metalanguage reflects into the object-language by the definitional equation. We found that the quantum metalanguage allows introducing the new logical connective of quantum superposition. This new logical connective quantum “and” is labelled by non logical constants, which are, in the interpretation, the complex probability amplitudes of the resulting quantum superposition.

## 14. Quantum Metalanguage and Interpretations of Quantum Mechanics

In this Section we will take into account the contribution of two philosophical schools (namely, logical positivism, and scientific realism) underlying the two main interpretations of Quantum Mechanics (QM): the instrumentalist and the realist interpretations, respectively. We will follow the classification introduced by de Muynck (see de Muynck, 2002). Thereafter, we will discuss which are the logics associated to these interpretations, and how the corresponding metalanguages reflect into the logical languages. In effect, the choice of the metalanguage is not absent from a personal point of view of the physical world. We think that a metalanguage already involves an interpretation of the physical theory under consideration. On the other hand, the mathematical formalism of the physical theory encompasses a logical language. If the chosen metalanguage does not reflect properly into the logical language, by abstraction, some inconsistencies and paradoxes (like the measurement problem in QM) can arise.

### 14.1 Logical positivism

Logical positivism is a school of philosophy which developed in the 20's from the Vienna Circle. It combines empiricism, the idea that observational evidence is indispensable for knowledge of the world, with a version of rationalism incorporating mathematical and logic-linguistic constructs and deductions in epistemology.

The tenets of logical positivism are in opposition to all metaphysics, especially ontology and synthetic a priori propositions. There is the rejection of metaphysics not as wrong but as having no meaning. A 1929 pamphlet written by Hahn, Neurath, and Carnap (see Hahn *et al.* 1929) summarized the doctrines of logical positivism.

### 14.2 Scientific realism

Scientific realism has developed mostly as a reaction to Logical positivism. According to scientific realism, an ideal scientific theory has: i) a semantic commitment to realism, ii) an ontological commitment, iii) an epistemological commitment.

i) The claims the theory makes are either true or false, depending on whether the entities talked about by the theory exist and are correctly described by the theory.
ii) The entities described by the scientific theory exist objectively and independently from the human mind.
iii) There are reasons to believe in some significant portion of what the theory says.

### 14.3 On the interpretations of Quantum Mechanics

The two main interpretations of QM are:

i) The Copenhagen (or standard) interpretation based on logical positivism and instrumentalism (provided we have only one interpretation of this kind, a circumstance now questioned by a number of authors; see, e.g., Howard, 2004;

Camilleri, 2007; Gomatam, 2007). Instrumentalism denies that theories are truth-valuable; instead, they should be treated like a black box into which you feed observed data, and through which you produce observable predictions. This requires a distinction between theory and observation.

ii) The realist interpretation, whose best example is the Bohm interpretation (for an overview see Bohm and Hiley, 1993; a collection of contributions devoted to solve the problems raised by this interpretation is contained in Cushing *et al*., 1996).

Most physicists don't think quantum mechanics *needs* interpretation. More precisely, they think it only requires a minimal instrumentalist interpretation. Minimal instrumentalism is the view that concepts and theories are merely useful instruments whose worth is measured not by whether the concepts and theories are true or false (or correctly depict reality), but by how effective they are in explaining and predicting phenomena. All that can be synthesized by the famous phrase “Shut up and calculate!”, first appeared in a paper written by N.D. Mermin (see Mermin, 1989). This methodological viewpoint often contrasts with scientific realism, which defines theories as being more or less true. However, minimal instrumentalism is more of a pragmatic approach to science, information and theories than an ontological statement.

## 14.4 The Copenhagen Interpretation

The Copenhagen interpretation, which is rooted in the philosophy of positivism, claims that quantum mechanics deals only with the probabilities of observable quantities; all other questions are considered metaphysical. This interpretation regards the wave-function as a mathematical tool used in the calculation of probabilities with no physical existence. Wave-function collapse is therefore a meaningless concept: the wave-function describes only a specific experiment, the outcome of which transcends the capability of Quantum Mechanics, while assessment of probabilities, the domain handled by Quantum Mechanics, requires a set of experiments. The Copenhagen Interpretation holds that any wave-function is only an abstraction, or is at least non-committal about its being a discrete entity or a discernible component of some discrete entity. The measurement problem in Quantum Mechanics is the unresolved problem of how wave-function collapse occurs. The inability to observe this process directly has given rise to different interpretations of Quantum Mechanics, and poses a key set of questions that each interpretation must answer. The wave-function in Quantum Mechanics evolves according to the Schrödinger equation into a linear superposition of different states, but actual measurements always find the physical system in a definite state. Any future evolution is based on the state the system was discovered to be in when the measurement was made.

## 14.5 The realist Interpretation

The realist interpretation of QM was based on scientific realism while, during what is perhaps the most notable example of revolution in science (the development of Quantum Mechanics in the 20’s) the dominant philosophy of science was logical positivism. The alternative to the Copenhagen School was the realist Bohm interpretation of Quantum Mechanics. The Bohm interpretation of Quantum Mechanics is an interpretation postulated by David Bohm in which the existence of a non-local universal wave-function allows distant particles to interact instantaneously. The interpretation generalizes Louis de Broglie's pilot wave theory of 1927 (the whole original contribution of de Broglie to the 1927 Solvay Conference has been reprinted in Bacciagaluppi and Valentini, 2006; books devoted to this theory are de Broglie, 1956; Valentini, 2006), which posits that both wave and particle are real. The wave-function 'guides' the motion of the particle, and evolves according to the Schrödinger equation. The interpretation is deterministic (unlike the Copenhagen interpretation). It says that the state of the universe evolves smoothly through time, without the collapsing of wave-functions when a measurement occurs. However, it does this by assuming a number of hidden variables, namely the positions of all the particles in the universe, which, like probability amplitudes in other interpretations, can never be measured directly. The Bohm interpretation tries to solve the measurement problem: this interpretation contains not only the wave-function, but also the information about the position of the particle(s). The role of the wave-function is to create a "quantum potential" that influences the motion of the "real" particle in such a way that the probability distribution for the particle remains consistent with the predictions of orthodox Quantum Mechanics. According to the Bohm interpretation combined with the von Neumann theory of measurement in quantum mechanics, once the particle is observed, other wave-function channels remain empty and thus ineffective, but there is no true wave-function collapse. Decoherence provides that this ineffectiveness is stable and irreversible, which explains the apparent wave function collapse.

### 14.6 The transcendental Interpretation

Within this thesis, we have developed a new approach to possible interpretations of QM, which stands in between realism and instrumentalism. In fact, like in realism, we maintain that a physical (quantum-computational) system , before any measurement, has an objective reality, which is nothing else than (quantum) information. In this first stage, we consider the quantum (computational) process, with reference to possible empirical events.

At the moment of measurement, however, the quantum system loses its objective reality (there is a loss of quantum information) and becomes dependent on the external world’s judgements, by which only a partial information is retained. This second stage is linked to instrumentalism in the sense that we get the events (outputs), and we cannot anymore talk about any objective reality, but because we destroyed it, not because it did not exist before. However, for any practical purpose, an instrumentalist would say that the event is real, not the process.

In the first stage, the quantum system is not measured, therefore it cannot be described in terms of experimental propositions, like projectors. To describe a quantum system before measurement, we introduced a new notion of assertion, by which propositions are asserted with an assertion degree, related to a probability amplitude. We called these propositions transcendental as their propositional status, as truth-bearers, is like the one of Kant’s *transcendental* arguments (as introduced within I. Kant’s *Critique of Pure Reason*; see for a recent edition Kant, 1993). More precisely, transcendental arguments attempt to prove a proposition starting from the fact that this proposition is the precondition of some other well-established propositions (the precondition being in our case only partially asserted). The utility of this approach is that it leads to a quantum metalanguage, which reflects properly into the object-language of a quantum computer, allowing a quantum control. We remark that a first reference to the role of Kant’s arguments within physical theories has been already made by Popper (see Popper, 1972). The importance of a transcendental approach to Quantum Mechanics has been held, in more recent times, by M. Bitbol (see Bitbol, 1998).

## 14.7 The mathematical counterparts

In the philosophy of mathematics, constructivism asserts that it is necessary to "construct" a mathematical object to prove that it exists. When one assumes that an object does not exist and derives a contradiction from that assumption, one still has not found the object and therefore not proved its existence, according to constructivists.

Intuitionism is one kind of constructivism. Intuitionism maintains that the foundations of mathematics lie in the individual mathematician's intuition. Constructivism is not based on this view of intuition, and is consonant with an objective view of mathematics.

Constructivist mathematics uses intuitionist logic, which is essentially classical logic without the law of the excluded middle. This is not to say that the law of the excluded middle is denied entirely; special cases of the law will be provable. It is just that the general law is not assumed as an axiom. The algebraic structure underlying intuitionist logic is Heyting algebra. Constructivism in mathematics corresponds to the transcendental approach to QM.

On the other hand, mathematical realism, like realism in general, holds that mathematical entities exist independently of the human mind. Thus humans do not invent mathematics, but rather discover it, and any other intelligent beings in the universe would presumably do the same.

Mathematical realism corresponds to naïve realism in QM.

Many mathematicians have been mathematical realists; they see themselves as discoverers of naturally occurring objects. For example, Gödel believed in an objective mathematical reality that could be perceived in a manner analogous to sense perception (for exhaustive information about realism and empiricism in mathematics see Maddy, 1997).

Platonism is the form of realism that suggests that mathematical entities are abstract, have no spatiotemporal or causal properties, and are eternal and unchanging. This is often claimed to be the view most people have of numbers. The term *Platonism* is used because such a view is seen to parallel Plato's belief in a "World of Ideas" .

Empiricism is a form of realism that denies that mathematics can be known a priori at all. It says that we discover mathematical facts by empirical research, just like facts in any of the other sciences. An important early proponent of this view was John Stuart Mill (the main book of Mill on this topic has been recently reprinted; see Mill, 2002). Contemporary mathematical empiricism, formulated by Quine and Putnam, is primarily supported by the *indispensability argument*: mathematics is indispensable to all empirical sciences, and if we want to believe in the reality of the phenomena described by the sciences, we ought also believe in the reality of those entities required for this description. Putnam strongly rejected the term "Platonist" as implying an overly-specific ontology that was not necessary to mathematical practice in any real sense. He advocated a form of "pure realism" that rejected mystical notions of truth . Empiricism in mathematics corresponds to instrumentalism in physics (or to the Copenhagen school).

### 14.8 The Kochen-Specker theorem

The Kochen-Specker theorem (Kochen and Specker, 1967) asserts the non-existence of valuations in quantum theory, subject only to the rather plausible requirement that the value of a function of a physical quantity should be the result of applying that function to the value of the quantity. In symbols, if V is a putative value function, and if f is a real-valued function of real numbers then, if A is any physical quantity, the requirement is:

$$V (f (A)) = f (V (A))$$

When applied to propositions, the theorem asserts the non-existence of any consistent assignment of true-false values to the propositions in quantum theory. However, although the theorem forbids any absolute assignment of truth values, it does not exclude truth-values that are contextual . Here, ‘contextual’ means that the truth-value given to a proposition depends on which other compatible (meaning ‘simultaneously measurable’) propositions are given values at the same time. Also, it does not excludes many-valued truth values (or truth-degrees).

### 14.9 Metalanguages for Interpretations of Quantum Mechanics

Given a quantum observable associated with an hermitian operator *A*, and an eigenvalue $a \in R$ of *A*, there are two main conceptual attitudes:

i) The instrumentalist approach (in terms of standard quantum logic)

ii) The (naïve) realist approach (in terms of Boolean logic)

As the realist approach is restricted, by the Kochen- Specher theorem, to be formulated only in terms of many-valued logics, which include probabilistic and fuzzy logics, we will consider here two main sub-cases of the realist approach:.

iii) The probabilistic approach (in terms of probabilistic logic)

iv) The fuzzy approach (in terms of fuzzy logic)

Furthermore, we will discuss also a third sub-case, which we have developed within this thesis:

v) The transcendental approach (in terms of the many-valued generalization of Basic logic).

**14.10 The instrumentalist approach.**

The instrumentalist approach can be summarized by the following sentence: “ *If* a *measurement* of *A* is performed, the probability that the *result* will be *a*, is *p*”.

In the instrumentalist approach, one adopts a quantum reasoning (standard quantum logic) to describe a quantum system. But the metalanguage, which is classical, reflects badly into the quantum object language, giving rise to the infamous measurement problem.

The propositions are, in this case, projective measurements. There are truth-values, which are indeterminate. The classical metalanguage corresponds, in the particular case of quantum information, to a classical control on a quantum computer.

The classical metalanguage, which would stand on the RHS of the definitional equation for the connective ${}_{z_0}\&_{z_1}$ = “quantum superposition”:

$$\vdash p_0 \;{}_{z_0}\&_{z_1}\; p_1 \qquad \underline{\text{iff}} \qquad \vdash p_0 \qquad \underline{\text{and}} \qquad \vdash p_1$$

reflects badly into the object-language. In fact, while in the RHS of the definitional equation the assertion-degrees $z_0$ and $z_1$ are absent, they appear in the LHS. The interpretation of this metalinguistic inconsistency is that a classical control is inefficient on a quantum computer.

**14.11 The realist approach**

The (naïve) realist approach can be summarized by the following sentence: “ The physical quantity *A has* a value, and the probability that this value *is a*, is *p*”.

In the naïve realist approach, one adopts a classical reasoning to describe a quantum system. Both the metalanguage and the object-language are classical: the measurement problem does not arise, but just because of the classical character of the object-language. In computer science, the classical metalanguage corresponds to a classical control of the classical object-language of a classical computer. Furthermore, the resulting Boolean logic, which is bivalent, that is has two truth values $\{0,1\}$, is a wrong result. In fact, it is forbidden by the Kochen-Specker theorem, which states that it is not possible to assign truth values to propositions (projectors) in Quantum Logic, because of the non distributive property.

The classical control is the classical metalanguage on the RHS of the definitional equation of the classical & = “and” of the classical object-language:

$$\vdash p_0 \,\&\, p_1 \qquad \underline{\text{iff}} \qquad \vdash p_0 \qquad \underline{\text{and}} \qquad \vdash p_1$$

**14.12 The probabilistic approach**

The probabilistic approach can be summarized as follows. The proposition: "The value of *A* is *a*" has truth value equal to the probability *p* that the value is *a*. As the probability's range is [0,1] this approach leads to a many-valued logic (see, e.g., Malinowski, 1993).

## 14.13 The fuzzy approach

The fuzzy approach can be summarized as follows. The sentence: "The observable *A* has a value *a*" has truth-degree $x \in [0,1]$.

In fuzzy logic, the truth values for conjunctive and disjunctive connectives are given by T-norms (see Hajék, 1998; Klement *et al.*, 2000) and co-T-norms respectively. However, the choice of the adequate T-norms, is a procedure case-dependent, and there are difficulties in finding a unifying principle, and to generalize T-norms to the quantum case. The premises being fuzzy, the metalanguage is fuzzy, that is non-classical. However, it is not quantum and cannot reflect properly into a quantum object-language (this explains the difficulties with quantum T-norms). The fuzzy metalanguage corresponds to a fuzzy control for a classical computer.

$$\Gamma \vdash^{x \circ y} p_0 \,\&\, p_1 \qquad \underline{\text{iff}} \qquad \Gamma \vdash^{x} p_0 \qquad \underline{\text{and}} \qquad \Gamma \vdash^{y} p_1$$

where $x, y \in [0,1]$ and $x \circ y$ is a T-norm.

Notice that, in the case of the qubit, $x = v_0 \quad y = v_1$, but the MD $v_0 + v_1 = 1$ cannot be satisfied by any T-norm (Łukasiewicz, Gödel or product T-norm).

## 14.14 The transcendental approach

The transcendental approach can be summarized as follows: The sentence :"The observable *A* has the value *a*" is asserted with a certain degree of assertion which is a probability amplitude. The transcendental and the probabilistic approaches seem almost identical at a first sight. In fact, in both cases the truth-degree *v* results to be equal to the probability *p*. However, while in the probabilistic approach this equality is given a priori, in the transcendental approach it is a consequence of having assertion-degrees. And the latter are the necessary condition to get an adequate quantum metalanguage. Moreover, in the transcendental approach truth-degrees concern non-experimental propositions, that is propositions before a measurement and, of course, probabilities concern experimental propositions (measurements). Instead, in the probabilistic approach, truths and probabilities refer to the same kind of propositions: the experimental ones. The transcendental approach is not a realist approach either: it has a realist flavour but the presence of the quantum cut rule (projective measurement in the interpretation) leads to a kind of mixture of empiricist-probabilistic approaches.

In the transcendental approach, both the metalanguage and the object-language are quantum.

The quantum metalanguage corresponds to a quantum control for a quantum object-language. Starting from the quantum metalanguage, we get the definitional equation for the connective of the intrinsic quantum superposition of the qubit by the definitional equation:

$$. \mid\!- p_0 \,{}_{z_0}\&_{z_1}\, p_1 \quad \underline{\text{iff}} \quad \mid\!-^{z_0} p_0 \quad \underline{\text{and}} \quad \mid\!-^{z_1} p_1$$

*MD*: $z_0^* \cdot z_0 + z_1^* \cdot z_1 = 1$

**14.15 Ontology of processes and quantum metalanguage**

Ontology is the study of reality, and of its categories. Originally, ontology was part of metaphysics but nowadays is very important in philosophy of science, in particular for the foundations of quantum mechanics. A modern, theoretical scientific approach to ontological questions is formal ontology. The list of references dealing with this subject is very long; we will limit ourselves to quote Cocchiarella (1996); Zalta (2000); Hofweber (2004); Merriks (2007). An important paper criticizing the usual approach to ontology is Brink& Rewitzky (2002).

Ontology concerns questions regarding what entities exist or can come to existence. Entities have distinct, separate existences, although it need not to be a material existence. For example, abstractions can be regarded as entities. A very important abstraction, in our case, is the passage from a metalanguage to a formal (object) language. What is real, in this case, the (quantum) metalanguage or the (quantum) object language, or just the process of abstraction passing from the former to the latter? In an ontology of events, one would answer that what is real is the object-language. In an ontology of processes the answer would be: the abstraction from the metalanguage to the object language. We agree with the second answer: this is in fact the case of an ontology of processes. Aristotle described ontology as “the science of being in the capacity of being”. In his words, we see that the process is evident. If he stopped the sentence at “the science of being” it would have been an ontology of events. Instead, the last piece of his sentence “in the capacity of being” shows that what he had in mind was the possibility, and the very process of becoming to existence.

In philosophy, events are objects in time, or instantiations of properties in objects, or both, depending on different philosophical definitions. In any case, the concept of event is based on the presence of suitable variables (supposed here numerical and measurable for simplicity) which can describe the state of the world. An event corresponds to the occurrence of the fact that these variables acquire some well definite values at a certain instant of time (in an appropriate space-time reference frame). An ontology of events presupposes not only that all phenomenological observations are observations of events, but also that all these observations can be explained or foreseen exclusively in terms of correlations among events. In many domains of Physics, an ontology of events seems to be the only possible one, or at least the most convenient. The typical case is the scattering of particles. All that we can practically observe are the events before and after the scattering. All that happens in the meanwhile is unknowable, and the only theories we can make concern the correlations among input and output events. This way of reasoning is sufficient for many practical purposes.

Let us now come to processes. A process is a temporal sequence of events. A process is ruled on by some dynamical law which characterizes the process itself. For example, a calculation is ruled on by the implemented algorithm. There are different kinds of processes, from the deterministic ones, to the probabilistic ones, to those totally random.

An ontology of processes does not deny that observations are about events, but hold that events are explained only in terms of the underlying process, and that the descriptions of events and processes are somehow inseparable. The expression “ontology of processes” has been borrowed from information science, where it has been introduced within the context of space-temporal databases (see, for instance, Kuhn (2001); Franck (2003)).

In an ontology of processes, even if the descriptions of processes are only mental, they admit an interpretation of what is observed in the events, because they relate these events to others observed in other domains, or on other levels of experience.

Then, for example, the knowledge of an algorithm allows to explain why a certain initial event (the appearance of a certain symbolic structure) leads to some final event (the appearance of another symbolic structure). But also, the knowledge of the algorithm allows the interpretation of the final event as the result of a computation. Therefore, an ontology of processes is necessary to describe computation, but it is not so important if we want to study, for example, the scattering among particles, where the probabilistic correlations might suffice to understand what happens physically. Process theory is a form of scientific research study in which events or occurrences are said to be result of certain input states leading to a certain outcome (output) state following a set process. For example, in computing, the set process can be the running program, or a task to be executed.

An ontological question, which is very important in philosophy of science is : “What it means to say that a physical object is real?”. In quantum physics, this is a very controversial question, about which the debate goes on, since the origin of QM. The question about the reality of a quantum state is in fact at the core of the foundations of QM.

Among ontological approaches, the most significant are realism and empiricism. In the realist approach, the idea is that facts are out there just waiting to be discovered. In the empiricist approach, instead, the idea is that we cannot say anything about the existence of the facts of the world until we make observations. The former approach, when applied to the foundations of quantum mechanics, originates the realist interpretation, while the latter originates the line of thought of the Copenhagen school. As the Copenhagen interpretation relies on measurements and output data, is in fact an ontology of events.

Within this thesis, we have adopted a conceptual framework closer to realist interpretations which use an ontology of processes. However, our choice was not made *a priori*, but was dictated by the necessity to recover correctly the object language by reflection of an appropriate (quantum) metalanguage. Now, a natural question arises: which is the relation between an ontology of processes and constructive mathematics?

On one side, an ontology of processes leads to a realist approach, and this could be interpreted, in mathematics, as Platonism: mathematic entities exists out there, in the platonic world, just waiting to be discovered. On the other hand, an ontology of events, leads to an empiricist approach, which, in mathematics corresponds to constructivism: a sentence is true if it can be proved.

Logical, or mathematical proofs correspond to experiments in Physics: provable in mathematics is the correspondent of measurable in physics. If things would stand really like that, it would be very unpleasant, because one would like to reconcile a realist interpretation of quantum mechanics with a constructivist approach to mathematics. Actually, there is a little delicate argument which should not be ignored. The "process" we consider is stopped at a certain point. Thereafter, we consider only events. It is a mixture of the two approaches. In fact, the ontology of processes regards only the abstraction of the reflection of a quantum metalanguage into a quantum object-language. This concerns the internal logic of quantum information. When a measurement is performed, that is, when one is concerned with experimental propositions, the ontology is an ontology of events. Translated in mathematical terms, one might make the abstraction that there are some pre-theorems which become theorems by proof.

An example of a pre-theorem is given by quantum computing, as we will illustrate in what follows.

Let us consider a theorem T to be proven by a quantum computer (let us suppose that the task of proving T is too hard and long lasting for a human agent or a classical computer). At the end of the computation, we will know whether T is true or false, (by a measurement giving 1 for true or 0 for false) but we will not be able to follow all the intermediate steps of the proof . In fact any action (measurement) made from outside would break quantum superposition, and stop the quantum computational process (the proof). Then, if theorem T will result true, it will look like a Gödel sentence to an external observer, that is, true but not provable. Instead, for an *internal observer* this would not be the case. The internal observer is performing the proof through a quantum computation. For him, theorem T is true and provable.

In a sense, a pre-theorem contains an infinite amount of information, which is not accessible from outside. When the theorem is proved, we get only a bit of information: 1= true or 0 = false. A quantum metalanguage might in principle give the means to reconstruct the proof, as it reflects properly in the internal logic without destroying the whole process.

In this sense, we have a kind of quantum Platonism in a logical approach to quantum information, but then it is converted into a constructive approach.

An ontology of processes for partially asserted propositions, together with the usual ontology of events for experimental propositions, can treat the dualism between the realist interpretation of quantum mechanics and the Copenhagen interpretation, as well as the dualism between Platonism and constructivism in mathematics.

What we have used here is a sort of dialectic approach, like the one developed by Hegel (a useful discussion about the implications of Hegel’s approach is Williams, 1989). In our case, dialectic stands in the correlation between quantum metalanguage and quantum-object language. Differently from the classical case, where the metalanguage *speaks about* the object-language, here the quantum metalanguage *discusses with* the object-language, by means of a quantum control, which operates in terms of weak measurements.